%% file: main.tex
\RequirePackage{fix-cm}
\documentclass[10pt]{article}
\usepackage[a4paper,left=17mm,right=17mm,top=21mm,bottom=22mm]{geometry}
\usepackage{fontspec}
\usepackage{graphicx,booktabs,tabularx,caption,hyperref,xcolor,placeins,float,amsmath,amssymb}
\definecolor{ink}{HTML}{172A3A}
\definecolor{linkink}{HTML}{12685E}
\hypersetup{colorlinks=true,linkcolor=linkink,urlcolor=linkink,pdftitle={S²-MAS: A Society-Scale Multi-Agent Simulation Framework with a 500-Million-Persona Resource},pdfauthor={Yang Li, Yike Wang, Di Wan, Chujun He, Zongsi Xu, Yitao Yu, Shuai Zhong, Minghao Pan, Quanfeng Lu, Xiyu Chen, Sheng Zhong, Deng Binjie, Yichen Guo, Zhenwei Guo, Hou Jun, Kunhan Guo, Weihao Lin, Tuo Zhou, Hai Liu, Chenwei Wu, Jianyuan Zhong, Xilin Dai, Ye Luo}}
\DeclareCaptionLabelSeparator{bar}{\space|\space}
\DeclareCaptionFont{paperfont}{\fontsize{8.5}{10.3}\selectfont}
\makeatletter
\renewcommand\section{\@startsection{section}{1}{\z@}{-2.2ex plus -1ex minus -.2ex}{.9ex plus .2ex}{\normalfont\large\mdseries}}
\renewcommand\subsection{\@startsection{subsection}{2}{\z@}{-1.7ex plus -.5ex minus -.2ex}{.6ex plus .2ex}{\normalfont\normalsize\mdseries}}
\makeatother
\begin{document}

\begin{center}
{\LARGE S\textsuperscript{2}-MAS: A Society-Scale Multi-Agent Simulation\\Framework with a 500-Million-Persona Resource\par}
\vspace{7pt}
{\fontsize{10.5}{13}\selectfont
Yang~Li\textsuperscript{1,*}, Yike~Wang\textsuperscript{1,*}, Di~Wan\textsuperscript{1}, Chujun~He\textsuperscript{2}, Zongsi~Xu\textsuperscript{1}, Yitao~Yu\textsuperscript{1}\par
Shuai~Zhong\textsuperscript{1}, Minghao~Pan\textsuperscript{10}, Quanfeng~Lu\textsuperscript{1}, Xiyu~Chen\textsuperscript{1}, Sheng~Zhong\textsuperscript{3}, Deng~Binjie\par
Yichen~Guo\textsuperscript{7}, Zhenwei~Guo\textsuperscript{4}, Hou~Jun\textsuperscript{1}, Kunhan~Guo\textsuperscript{5}, Weihao~Lin\textsuperscript{2}, Tuo~Zhou\textsuperscript{1}\par
Hai~Liu\textsuperscript{6}, Chenwei~Wu\textsuperscript{8}, Jianyuan~Zhong\textsuperscript{9}, Xilin~Dai\textsuperscript{9}, Ye~Luo\textsuperscript{1,11,\textdagger}\par
}\vspace{5pt}
{\fontsize{8.4}{10.2}\selectfont
\textsuperscript{1}The University of Hong Kong\par
\textsuperscript{2}Faculty of Business and Economics, The University of Hong Kong\par
\textsuperscript{3}HKU Business School Shenzhen Research Institutes, The University of Hong Kong\par
\textsuperscript{4}Risk Management Institute, National University of Singapore\par
\textsuperscript{5}School of Computing and Data Science, The University of Hong Kong\par
\textsuperscript{6}School of Artificial Intelligence, Jiangxi Science and Technology Normal University\par
\textsuperscript{7}Nanyang Technological University\par
\textsuperscript{8}University of Michigan, Ann Arbor\par
\textsuperscript{9}The Chinese University of Hong Kong\par
\textsuperscript{10}California Institute of Technology\par
\textsuperscript{11}Shenzhen Loop Area Institute\par
\vspace{3pt}\textsuperscript{\textdagger}Corresponding author: Ye Luo\par
Research contact: \href{mailto:yl9919@connect.hku.hk}{\nolinkurl{yl9919@connect.hku.hk}}\par
}\vspace{5pt}
\vspace{7pt}
{\small 28 October 2026\par}
\end{center}
\vspace{3pt}
\begin{center}{\normalsize Abstract}\end{center}
\begingroup
\setlength{\parindent}{0pt}
\setlength{\parskip}{0pt}
\begin{minipage}{\textwidth}\small
Simulating economic responses at society scale requires decision rules that respond to household circumstances and accounting rules that track their consequences. We introduce S\textsuperscript{2}-MAS, a multi-agent simulation framework with a resource of 500 million synthetic household personas. It combines trainable candidate-scoring policies, reusable financial scenarios and explicit state transitions. Trained with SFT + RLCD on 2.93 million examples, YN-mini-MLX achieves 74.55\% accuracy and 38.37\% question-macro F1 on held-out consumer financial-choice tasks, improving on the training-question majority baseline by 2.78 and 6.21 percentage points. A completed application evaluates 12 million distinct households in two purchase-funding conditions: 24 million one-shot choices feed 288 million monthly ledger states in 3 h 59 min on 14 GPUs. Jointly increasing purchase goals and approved limits produces heterogeneous consumption, cash and repayment paths, with much of the consumption variation arising from transitions between deferral and purchase. In a separate 1,060-household retrospective consumption experiment, standalone YN-master-9B has 3.42\% lower annual MAE than the deterministic fully rational benchmark, alongside a 19.7\% lower predicted population mean than observed. Information-timing tests show how year labels and later outcome information change decisions. Together, these experiments establish an executable alternative for studying heterogeneous household behavior with learned policies and explicit financial accounting.
\end{minipage}
\endgroup
\par\vspace{10pt}

\section*{1. Introduction}

\begin{figure}[!htbp]

\begin{minipage}{\textwidth}

\centering

\includegraphics[width=\textwidth,height=.66\textheight,keepaspectratio]{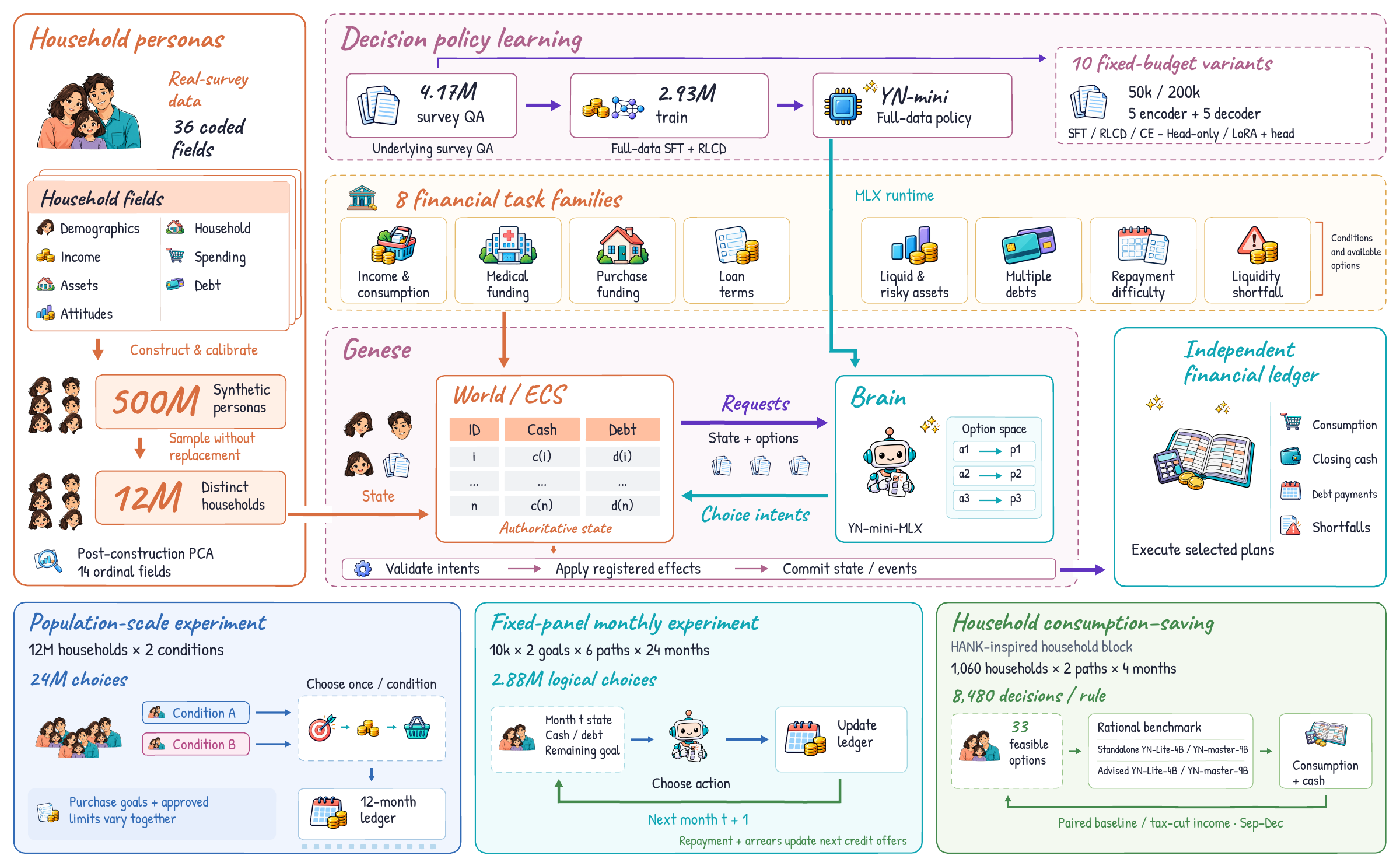}

\par\vspace{5pt}{\fontsize{8.5}{10.3}\selectfont\raggedright \refstepcounter{figure}\label{figure:main1}Figure \thefigure{} | Population-scale household decisions in S\textsuperscript{2}-MAS. The left column links real-survey data and 36 coded fields to 500M constructed personas and a 12M cohort sampled without replacement; PCA summarizes 14 ordinal fields. The upper row distinguishes the full-data YN-mini policy from ten fixed-budget encoder and decoder variants. Eight task families share the option-scoring interface. Genese World/ECS exchanges requests and choice intents with Brain; the application ledger executes selected plans and tracks consumption, cash, payments and shortfalls. The lower panels show the population-scale paired-choice workload, a repeated-decision sandbox (Supplementary Figure~\ref{figure:monthly-scenario}), and the 1,060-household consumption--saving comparison of rational, standalone and advised policies.\par}

\end{minipage}

\end{figure}

Language-model-based multi-agent simulation offers a way to study how heterogeneous populations respond to changing environments and interventions. At society scale, three challenges arise together: decisions must respond to individual circumstances, actions must produce consistent changes in resources and obligations, and inference and execution must remain efficient enough to repeat experiments across conditions. We introduce S\textsuperscript{2}-MAS, a society-scale multi-agent simulation framework integrating household populations, trainable decision policies and explicit state transitions (Figure~\ref{figure:main1}).

Survey-conditioned agents allow predictions to be evaluated against recorded human responses \href{https://www.cambridge.org/core/journals/political-analysis/article/abs/out-of-one-many-using-language-models-to-simulate-human-samples/035D7C8A55B237942FB6DBAD7CAA4E49}{[25]}, \href{https://arxiv.org/abs/2411.10109v3}{[26]}. MatrAIx constructs heterogeneous persona populations for simulated-user evaluation \href{https://arxiv.org/abs/2608.04205}{[1]}, while OneSim and OASIS support large-scale agent execution in social environments \href{https://arxiv.org/abs/2505.07581}{[16]}, \href{https://arxiv.org/abs/2411.11581}{[17]}. Economic experiments require these capabilities together: differences among households shape their decisions, and the resulting changes in resources and obligations shape what they can do next.

Household finance provides observed responses for decision learning and budget rules for calculating the consequences of selected plans. After an income loss, a household may draw on reserves to sustain consumption, reduce spending to preserve cash, or leave payments unmet. Buffer-stock saving, consumption commitments and heterogeneous-household models explain why resources and obligations matter for these adjustments \href{https://academic.oup.com/qje/article-abstract/112/1/1/1870884}{[22]}, \href{https://academic.oup.com/qje/article-abstract/122/2/831/1942152}{[23]}, \href{https://www.aeaweb.org/articles?id=10.1257/aer.20160042}{[31]}. Consumption, cash reserves and payment histories therefore reveal how the same financial condition affects households with different starting positions.

YN-mini and YN-Lite-4B implement encoder and decoder decision policies, respectively. YN-mini uses 2,925,263 examples for SFT + RLCD; YN-mini-MLX is its runtime used in the completed financial applications. Genese separates decision proposals from state updates, while independent application ledgers calculate consumption, cash, debt payments and shortfalls. Paired experiments retain the same households and accounting rules across conditions.

On the one-shot consumer financial-choice benchmark, YN-mini-MLX predicts recorded answers and coded respondent targets, improving on training-question majority by 2.78 accuracy points and 6.21 question-macro F1 points. Matched information controls measure the contribution of household profiles, and the common benchmark compares complete decision interfaces (Section 5.3). In a controlled 1,000-agent benchmark with the same model allocation, Genese completes accepted decisions and state writeback faster than OneSim and OASIS (Section 5.5).

YN-mini-MLX runs eight financial task families. The 500M resource supports cohort construction; the completed population application executes 24M one-shot choices for 12M distinct households and advances their independently calculated monthly ledgers. Jointly raising purchase goals and approved limits increases simulated consumption. Education, employment and household-size differences persist after standardization to a common age--income distribution, largely through transitions from deferral to purchase. A separate retrospective consumption experiment compares learned and fully rational household policies against observed annual spending.

\section*{2. Related work}

\subsection*{2.1. Survey-grounded populations and user simulation}

Survey-conditioned agents connect language modeling to empirical response distributions. Argyle et al. condition GPT-3 on sociodemographic backstories and compare generated subpopulation responses with survey samples \href{https://www.cambridge.org/core/journals/political-analysis/article/abs/out-of-one-many-using-language-models-to-simulate-human-samples/035D7C8A55B237942FB6DBAD7CAA4E49}{[25]}. Park et al. construct interview- and survey-grounded agents for 1,052 Americans and evaluate held-out responses against the represented individuals' test--retest consistency \href{https://arxiv.org/abs/2411.10109v3}{[26]}. YN-mini learns direct decisions from observed consumer survey responses. Question-majority and matched-profile controls measure the contribution of respondent information to held-out answer prediction. The application ledger calculates the financial consequences of selected plans.

MatrAIx combines dependency-aware population construction with biographies, reviews, surveys and self-reports \href{https://arxiv.org/abs/2608.04205}{[1]}. S\textsuperscript{2}-MAS uses a 36-field household representation constructed through processing and statistical calibration of real-survey data. PCA characterizes the economic structure of the resulting population.

MatrAIx compares AI persona ratings with human assessments \href{https://arxiv.org/abs/2608.04205}{[1]}. We use this evaluation approach for the processed-field cohort described in Supplementary Methods M7.

\subsection*{2.2. Direct decision models and financial supervision}

Direct decision models score named alternatives or return structured probabilities without generating a prose answer. Jev's official introduction names RLCD as its training approach and its API defines typed outputs \href{https://typesafe.ai/blog/introducing-system-one-models-and-jev}{[35]}, \href{https://docs.typesafe.ai/introduction}{[11]}. Kev-4B publishes option-classification CE for its LoRA/pointer adapter; Laya's open trainer combines noisy-logit proper-reward policy gradients with soft CE \href{https://raw.githubusercontent.com/jaredpalmer/kev/main/docs/model-cards/kev-4b.md}{[37]}, \href{https://github.com/NandhaKishorM/laya/blob/main/laya/train.py}{[36]}. Encoder interfaces include Laya and Von; decoder decision readouts include Kev, Clef and Nimble \href{https://github.com/LeonaDavinci/laya-system-one}{[2]}--\href{https://github.com/jaredpalmer/kev}{[4]}, \href{https://blog.cloudflare.com/clef-decision-models/}{[8]}, \href{https://github.com/bespokelabsai/nimble}{[9]}. S\textsuperscript{2}-MAS adapts Laya's existing option-scoring interface to recorded household answers and coded states. Supplementary Methods M5 compares available interfaces on the same held-out consumer questions.

Table S7 records the decision interfaces, backbones and readouts; Table S8 distinguishes financial adaptation from general decision training, instruction tuning and financial-language pretraining. Table S9 reports the populations and execution units of selected simulation studies.

Financial language adaptation spans large-corpus pretraining, instruction tuning and expert numerical QA. BloombergGPT uses 363B financial tokens plus 345B general tokens for pretraining \href{https://arxiv.org/abs/2303.17564}{[12]}. PIXIU's NeurIPS 2023 FIT release contains 128,640 instruction samples for financial NLP and prediction \href{https://proceedings.nips.cc/paper\_files/paper/2023/file/6a386d703b50f1cf1f61ab02a15967bb-Paper-Datasets\_and\_Benchmarks.pdf}{[13]}. FinGPT distributes task-specific instruction sets, including 76.8K sentiment, 82.2K headline and 17.1K financial-QA training rows \href{https://github.com/AI4Finance-Foundation/FinGPT}{[14]}. FinQA targets expert-written numerical reasoning over financial reports \href{https://aclanthology.org/2021.emnlp-main.300/}{[15]}. These resources measure document interpretation, linguistic tasks and market prediction.

Financial Laya checkpoints adapt the same direct-decision interface to narrower tasks. The public Laya-Finance card lists financial sentiment, topics, central-bank text, headlines and terminology \href{https://huggingface.co/hravi/laya-finance}{[27]}. A Banking77 adaptation trains typed decision heads on 7,994 customer-service messages while keeping its encoder frozen \href{https://huggingface.co/dimitrisdais/laya-banking77-typed-decisions}{[28]}. YN-mini updates its encoder and decision head with observed household and respondent survey targets. Supplementary Methods M5 evaluates the external Laya-Finance checkpoint on the same held-out consumer questions.

Laya-Fintiment uses 61,017 gradient-training examples for three-class financial sentiment \href{https://huggingface.co/shanaka95/laya-fintiment}{[32]}. The Ariadne-Laya finance-topic model trains a 1,049,600-parameter interface over a frozen 421M base for twenty news topics \href{https://huggingface.co/GoatHerder/Ariadne-Laya-FinanceTopics}{[33]}. Household survey supervision adds income, spending, wealth, debt, product participation and attitudes to the decision target, connecting the model's inputs to a consumer's economic circumstances.

The training corpus combines recorded financial answers with coded household-state, education and occupation targets. Its 2,925,263 training examples correspond to approximately 48 times Laya-Fintiment's 61,017 examples and 122 times Chinese-Jev Finance's 24,000 decisions \href{https://huggingface.co/shanaka95/laya-fintiment}{[32]}, \href{https://arxiv.org/abs/2609.36965}{[34]}.

\subsection*{2.3. Simulation execution and household financial analysis}

OneSim, OASIS, AgentSociety and Generative Agents study multi-agent activity at different population sizes and interaction depths \href{https://arxiv.org/abs/2505.07581}{[16]}--\href{https://arxiv.org/abs/2304.03442}{[19]}. Their published workloads use environment populations, events, calls or trials (Table S9). S\textsuperscript{2}-MAS completes a 12M-agent purchase-funding application.

Household-finance research studies consumption responses through the distribution of resources and obligations. Kaplan, Moll and Violante's heterogeneous-household framework links uninsured income risk and assets with different liquidity to heterogeneous consumption responses \href{https://www.aeaweb.org/articles?id=10.1257/aer.20160042}{[31]}. S\textsuperscript{2}-MAS represents income, spending, assets and debt jointly and evaluates how those states change the executable response to a shared financial condition.

\section*{3. From population resources to executable experiments}

An experiment in S\textsuperscript{2}-MAS specifies four inputs: a population, a policy adapter, a task and condition specification, and state-transition rules. The population supplies whole-agent attributes. The task converts current state and available information into a question with explicit alternatives. The adapter returns an option and its probabilities; the execution layer applies the registered consequence and records the resulting state. Ordinary agent, condition and task IDs connect these stages across partitions and retries.

A static episode evaluates a selected plan through its horizon. YN-Lite-4B and YN-mini provide two trainable policy routes within the same contract. Changing a policy or designated condition preserves the other inputs and the same-household pairing. Policy proposals and their feasible execution remain separately observable.

The financial task library instantiates this contract across eight families (Table~\ref{table:main1}). A medical-bill task changes the outflow to be funded; a loan-term task changes payment dates; an asset-sale task changes when proceeds enter cash. Each task defines its alternatives and action-to-ledger mapping while retaining the population and choice interfaces. Outputs link selected actions to monthly consumption, cash, debt and shortfalls.

The static task suite contains 45,473 cases per language, including 22,736 baseline/intervention pairs and one additional case. Task specifications separate model-visible information from execution-only realizations and provide 147,787 conditional option paths over twelve months. The completed suite uses YN-mini-MLX and the original Laya reference. The population-scale application fixes one T3 contrast.
\begin{table}[!htbp]
\begin{minipage}{\textwidth}
\fontsize{8.5}{10.3}\selectfont
\refstepcounter{table}\label{table:main1}Table \thetable{} | Financial task specifications and reported applications
\par\vspace{4pt}
\begin{tabularx}{\textwidth}{@{}>{\hsize=1.028571\hsize\linewidth=\hsize\raggedright\arraybackslash}X>{\hsize=0.990476\hsize\linewidth=\hsize\raggedright\arraybackslash}X>{\hsize=0.876190\hsize\linewidth=\hsize\raggedright\arraybackslash}X>{\hsize=1.104762\hsize\linewidth=\hsize\raggedright\arraybackslash}X@{}}
\toprule
Task / changed condition & Offered actions & Financial measures & Evidence reported here \\
\midrule
T1: income and consumption & Adjust discretionary spending & Consumption, cash, unmet needs & Paired static YN-mini/Laya execution (Appendix S11) \\
T2: medical expense funding & Use savings, borrow or combine both & Cash, debt, unmet needs & paired YN-mini/Laya execution (Appendix S11) \\
T3: purchase-funding need & Defer purchase or finance a partial/full amount & Consumption, cash, payments, debt & Paired bilingual execution and 12M application (§6.2, Tables S32--S33) \\
T4: loan product and term & Choose a short/long contract or defer purchase & Payments, cash, arrears & paired YN-mini/Laya execution (Appendix S11) \\
T5: liquid saving and risky assets & Allocate between liquid and risky holdings & Cash, assets, unfunded needs & paired YN-mini/Laya execution (Appendix S11) \\
T6: multiple-debt payments & Allocate a payment budget across debts & Debt balances, arrears, default & paired YN-mini/Laya execution (Appendix S11) \\
T7: repayment difficulty & Retain schedule, prepay or request extension & Payments, debt, extension outcome & paired YN-mini/Laya execution (Appendix S11) \\
T8: liquidity shortfall & Sell assets, refinance or combine both & Cash, assets, debt & paired YN-mini/Laya execution (Appendix S11) \\
\bottomrule
\end{tabularx}
\par\vspace{3pt}{\fontsize{8}{9.5}\selectfont All eight families have paired specifications, conditional option paths and completed YN-mini/original-Laya execution with Chinese variants included. The static task set is derived from 207 financial parent records; parameter variants are not independent households. The separate DeepSeek workflow study remains in Appendix S9.}
\end{minipage}
\end{table}

\noindent\begin{minipage}{\textwidth}
Consider a synthetic household assigned CNY 6,000 monthly income, CNY 3,000 opening cash, CNY 3,500 essential and CNY 1,800 usual adjustable spending, and no opening debt. A CNY 20,000 purchase goal offers deferral, half financing or full financing at 5\% nominal annual interest over 24 installments. Selecting full financing buys CNY 20,000 immediately and records the dated purchase contract. Month-0 cash is 3,000 + 6,000 + 20,000 - 20,000 - 3,500 - 1,800 = CNY 3,700, with repayment beginning in month 1.
\end{minipage}

\section*{4. Data, computable personas and simulation components}

\subsection*{4.1. Survey data and computable household personas}

Real-survey data support the household attributes, recorded responses and coded financial states used here. Cleaning, translation and statistical harmonization produce a common 36-field representation of demographics, household structure, income, expenditure, assets, debt and attitudes. Conditional construction preserves the dependence of financial attributes on household characteristics, and statistical calibration aligns selected population margins. Numeric codes support population operations, while decoded fields form model-visible personas. PCA summarizes joint economic structure after construction (Supplementary Methods M1--M3).

The reusable 10M consumer bank supports descriptive analysis of the constructed representation. The 12M execution cohort selects whole records without replacement from two 10M-row shards of the 500M resource, with the cohort fixed before inference.

The same preparation produces supervision for policy learning. The 4,172,421 underlying QA observations have aligned bilingual views, retaining wording, alternatives and monetary units. Target-related fields are omitted from the input, and all questions and language copies for a respondent group share a split. The grouped training partition contains 2,925,263 examples per language. YN-mini uses the full training partition; the matched downstream comparisons use nested 50k/200k training sets, including the selected 200k YN-Lite-4B route (Appendix S10).
\subsection*{4.2. Decision execution and financial accounting}

\begin{figure}[!htbp]

\begin{minipage}{\textwidth}

\centering

\includegraphics[width=\textwidth,height=.66\textheight,keepaspectratio]{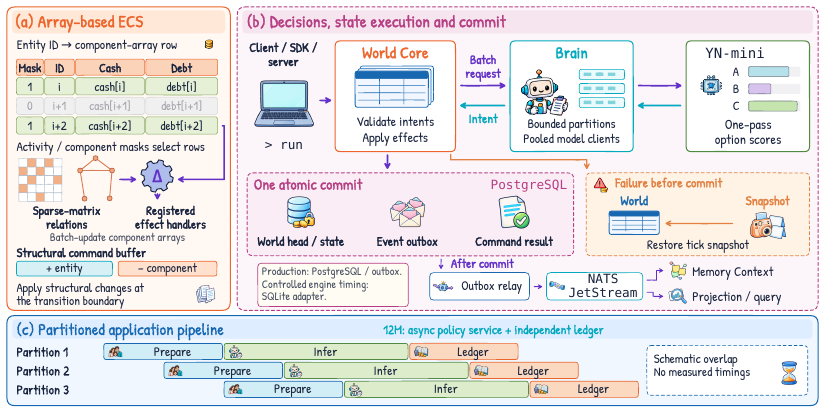}

\par\vspace{5pt}{\fontsize{8.5}{10.3}\selectfont\raggedright \refstepcounter{figure}\label{figure:main2}Figure \thefigure{} | ECS computation and the Genese state-transition boundary. a, Entity IDs index numeric component arrays; masks select active rows, registered systems update components, and a structural-command buffer controls structural changes. b, World Core requests decisions from Brain through bounded partitions and pooled model connections, then validates intents, resolves effects and commits authoritative world state, the event outbox and command result together in production PostgreSQL. The outbox relay publishes committed events to NATS JetStream; memory and projection consumers process them asynchronously. The shaded commit boundary covers database state and outbox records. c, Schematic overlap across workload partitions places input preparation, model inference and ledger calculation on distinct stages; the segment widths are illustrative.\par}

\end{minipage}

\end{figure}

Genese separates decision proposals from authoritative state updates (Figure~\ref{figure:main2}). World stores numeric state in entity--component--system arrays indexed by entity ID. Brain receives bounded decision partitions, reuses model connections and returns policy intents. World validates these intents, resolves registered effects and applies updates. Production PostgreSQL commits world state, the event outbox and command result together; a failed pre-commit transition restores the tick snapshot. An outbox relay publishes committed events asynchronously.

The household application evaluates financial consequences in a separate monthly ledger. Input preparation, decision inference and ledger calculation overlap across partitions, while household and condition IDs retain the pairing. Integer-cent accounting applies income, selected financing, consumption priorities and scheduled debt payments in a fixed order. The resulting paths record funded amounts and shortfalls.

Section 5.5 reports controlled decision-and-writeback benchmarks and end-to-end timing for the 12M application.

\subsection*{4.3. Learning and substituting decision policies}

\begin{figure}[!htb]
\begin{minipage}{\textwidth}
\centering
\includegraphics[width=\textwidth,height=.68\textheight,keepaspectratio]{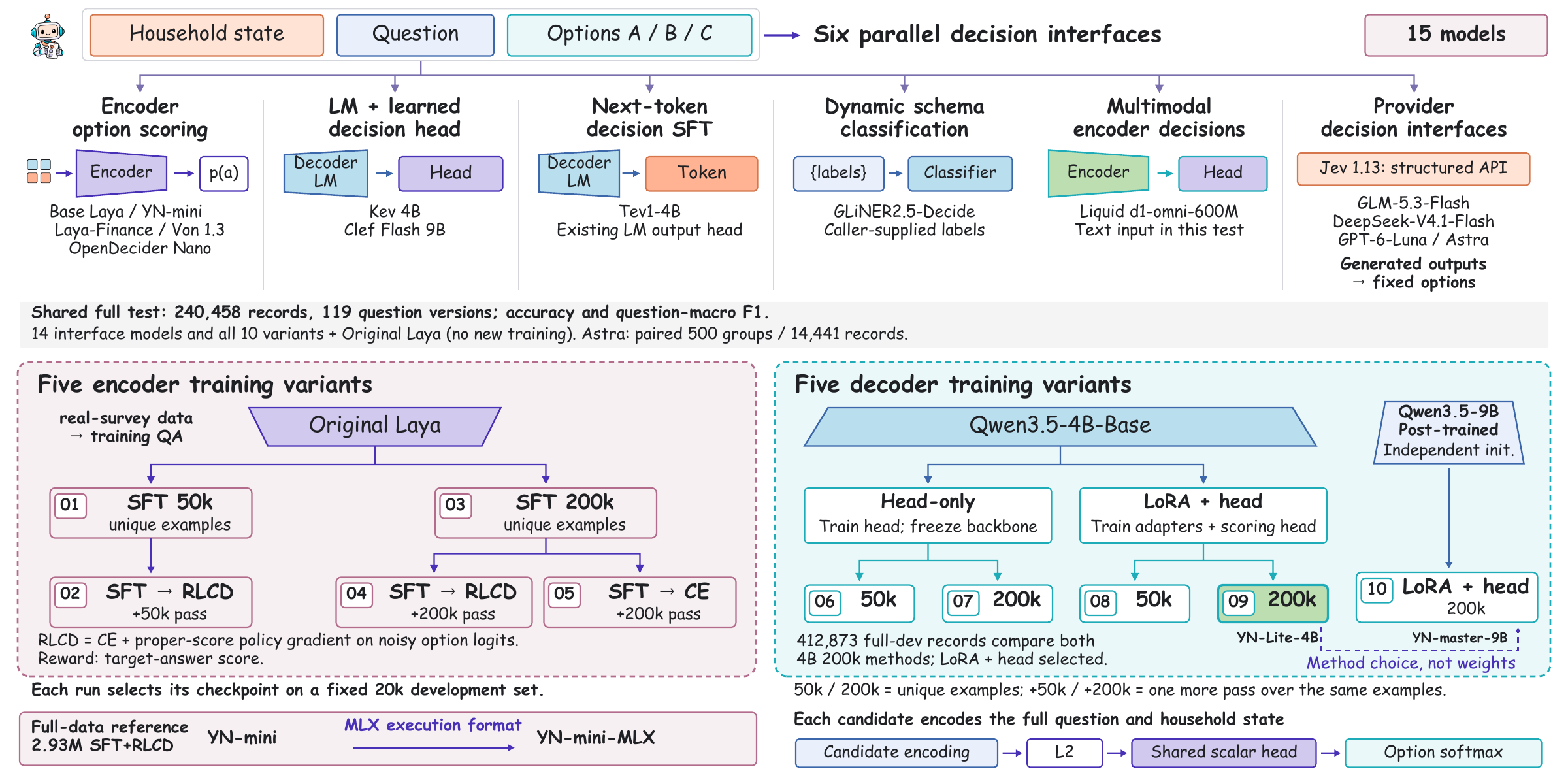}
\par\vspace{5pt}{\fontsize{8.5}{10.3}\selectfont\raggedright \refstepcounter{figure}\label{figure:main3}Figure \thefigure{} | Decision interfaces and training routes. The upper panel groups fifteen evaluated models into six architecture or interface routes; all receive text inputs. Fourteen use the complete test, while GPT-6-Astra uses a matched fixed-group sample (Table~\ref{table:main4}). The lower panel compares ten fixed-budget trained versions covering encoder adaptation, repeated CE training, low-rank adaptation and larger backbones. YN-Lite-4B is the selected Qwen3.5-4B-Base LoRA-plus-head checkpoint at the 200k budget. Qwen candidates are encoded separately, L2-normalized and scored by a shared scalar head; head-only training freezes the backbone. Native continuations start from the matching SFT development-best checkpoint, and native RLCD combines CE with a proper-score policy-gradient term over noisy logits. Full development data (412,873 records) compare the two 200k 4B methods; the dashed arrow applies the selected method to an independently initialized 9B backbone. Each run uses fixed 20k development data for checkpoint selection. All ten trained versions and the zero-new-training native reference share the held-out test. Full-data YN-mini is the principal encoder route and a separate system reference. CE, cross-entropy.\par}
\end{minipage}
\end{figure}

YN-Lite-4B and YN-mini learn choices from household states and task alternatives. YN-Lite-4B combines a decoder backbone with LoRA and a learned decision head; YN-mini adapts a compact encoder and option-marker head. Existing approaches include Jev's RLCD \href{https://typesafe.ai/blog/introducing-system-one-models-and-jev}{[35]}, Kev's option cross-entropy \href{https://raw.githubusercontent.com/jaredpalmer/kev/main/docs/model-cards/kev-4b.md}{[37]} and Laya's encoder-based SFT and RLCD \href{https://github.com/LeonaDavinci/laya-system-one}{[2]}, \href{https://github.com/NandhaKishorM/laya/blob/main/laya/train.py}{[36]}. Figure~\ref{figure:main3} groups the benchmarked interfaces into six routes. The fixed-budget study contains ten trained versions: five YN-mini encoder adaptations and five YN-Lite-4B or YN-master-9B head-only or LoRA adaptations, including YN-Lite-4B. Full-data YN-mini is reported separately.

YN-Lite-4B is the selected Qwen3.5-4B-Base LoRA-plus-decision-head checkpoint trained on 200,000 unique examples. Candidate contexts independently encode the state, question and alternatives with the evaluated candidate marked. The final nonpadding representation is L2-normalized and scored by a shared scalar head. LoRA updates the backbone adapters together with that head \href{https://arxiv.org/abs/2106.09685}{[38]}. The fixed 20k development set selects the checkpoint; the complete 412,873-example development partition compares the two 200k 4B methods before the corresponding Qwen3.5-9B run.

The YN family comprises the compact YN-mini encoder, YN-Lite-4B and YN-master-9B decoder policies; suffixes identify training variants and runtime formats. YN-mini is the full-data financial adaptation of Laya's 421M-parameter decision model, comprising a 395M ModernBERT-large backbone and an option-marker scoring head \href{https://github.com/LeonaDavinci/laya-system-one}{[2]}. Two SFT stages and native RLCD produce the YN-mini-3rounds policy. A further SFT and RLCD continuation broadens task coverage over 2.93M examples. Checkpoints are selected on development data, and probability temperatures are fitted on independent calibration data (Supplementary Methods M4--M6). YN-mini-MLX denotes execution of this full-data checkpoint through MLX; YN-mini-322M is the separate multilingual-base adaptation.

The controls retain nested 50k/200k YN-Lite-4B head-only and LoRA routes, the selected YN-master-9B LoRA route, and native Laya SFT on the same example sets. Additional native RLCD and CE-continuation controls start from the same SFT parent and share an added full-pass budget and development rule. Unique examples, repeated exposures, selected checkpoints and allocated GPU-hours are reported in Appendix S10.

All adapters preserve the state, question and semantic alternatives while translating the policy's scoring interface. The final 240,458-record test evaluates every fixed trained version and the original native reference after method and budget selection.
\FloatBarrier
\section*{5. Evaluating the framework components}

We evaluate household information, decision-policy training and substitution, and action execution. YN-Lite-4B and YN-mini provide the two principal training routes; matched controls compare adaptation choices, and execution benchmarks time accepted choices and state updates.

\subsection*{5.1. Population structure and cohort coverage}

PCA summarizes 14 standardized ordinal attributes in the constructed 10M bank. Among 7,227,206 complete records, PC1 explains 32.8904\% of variance and PC2 explains 11.5234\%. Income, spending, assets and debt load positively on the main axis, while age loads negatively. The 500M resource separately has a shard-mean PC1 explained variance of 32.62\%, with loading standard deviations at most 0.0003. Tables S17--S18 report the bank analysis, and M2--M3 give the descriptive protocol. The independent synthetic companion supplies a reusable population with assigned financial states: Figures S1 and S4 describe its 2M personas, including the income, liquidity and obligation dimensions that enter the scenario interface.

\begin{figure}[!htbp]

\begin{minipage}{\textwidth}

\centering

\includegraphics[width=.90\textwidth,trim=0pt 10pt 0pt 8pt,clip,keepaspectratio]{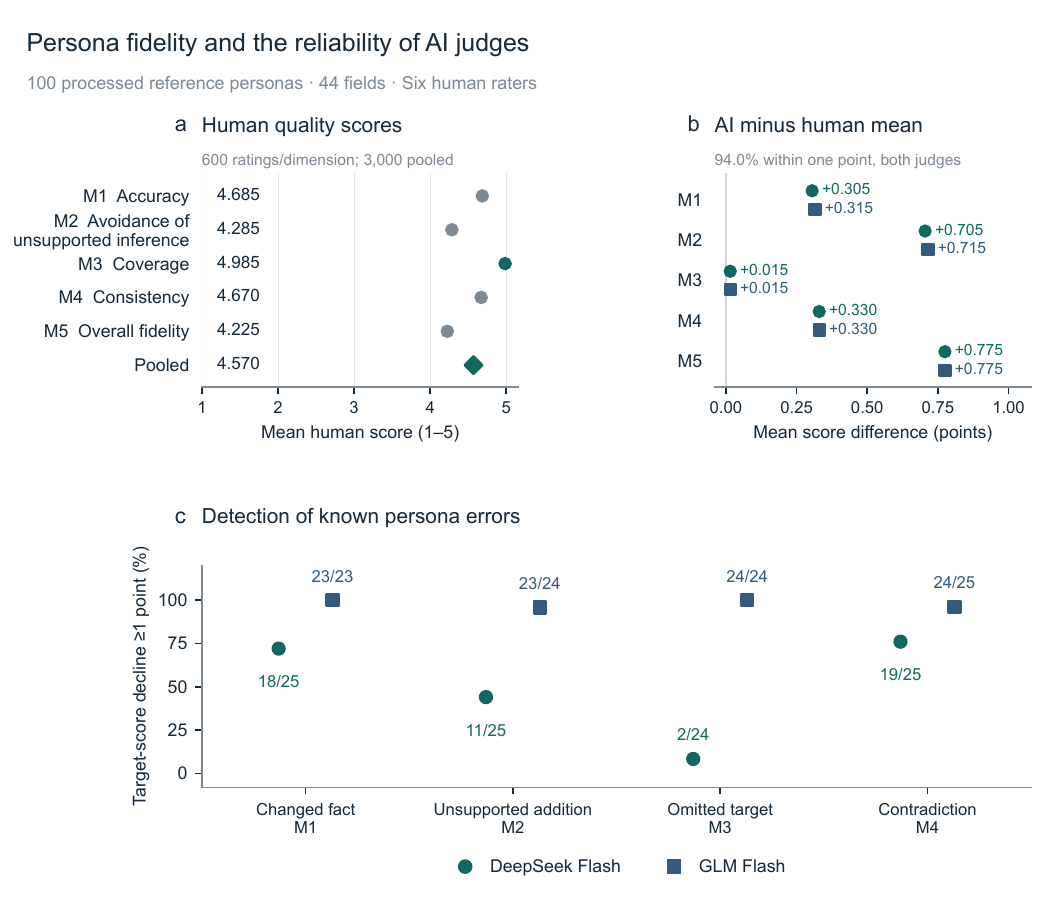}

\par\vspace{5pt}{\fontsize{8.5}{10.3}\selectfont\raggedright \refstepcounter{figure}\label{figure:main4}Figure \thefigure{} | Human persona fidelity, judge bias and known-error detection. a, Six raters score five dimensions for 100 processed survey-derived personas rendered from 44 processed fields, producing 600 ratings per dimension and 3,000 in total. Points show mean human scores on the five-point scale; pooled mean is 4.570. M2 measures avoidance of unsupported inference, so a higher score indicates fewer unsupported inferences. b, Mean AI score minus the six-rater mean for each dimension, based on 100 persona-level comparisons per judge. Both judges agree within one point on 470/500 comparisons (94.0\%), while giving higher M2 and M5 scores. c, The proportion of valid original–perturbed pairs with a target-dimension decline of at least one score point after a known error is injected. Labels give detected/valid pairs; valid denominators vary from 23 to 25 per judge and error type. DeepSeek detects 2/24 omitted targets, compared with 24/24 for GLM. Supplementary Figure S3 shows the full human-score distributions and Table S16 reports scoring thresholds and non-target specificity.\par}

\end{minipage}

\end{figure}

Rendered-persona evaluation finds high field coverage and uneven sensitivity to reconstruction errors (Figure~\ref{figure:main4}). Six human raters score 100 personas against processed reference fields, with a pooled mean of 4.570/5. DeepSeek and GLM agree within one point of the human mean on 94.0\% of persona-dimension comparisons, while giving higher scores for avoidance of unsupported inference and overall fidelity. Their omission sensitivity differs: GLM detects 24 of 24 valid omitted-target pairs, compared with 2 of 24 for DeepSeek. M7 describes the rendering, scoring and perturbation design.

\subsection*{5.2. Persona information and policy adaptation}

\begin{figure}[!htbp]

\begin{minipage}{\textwidth}

\centering

\includegraphics[width=.90\textwidth,trim=0pt 4pt 0pt 2pt,clip,keepaspectratio]{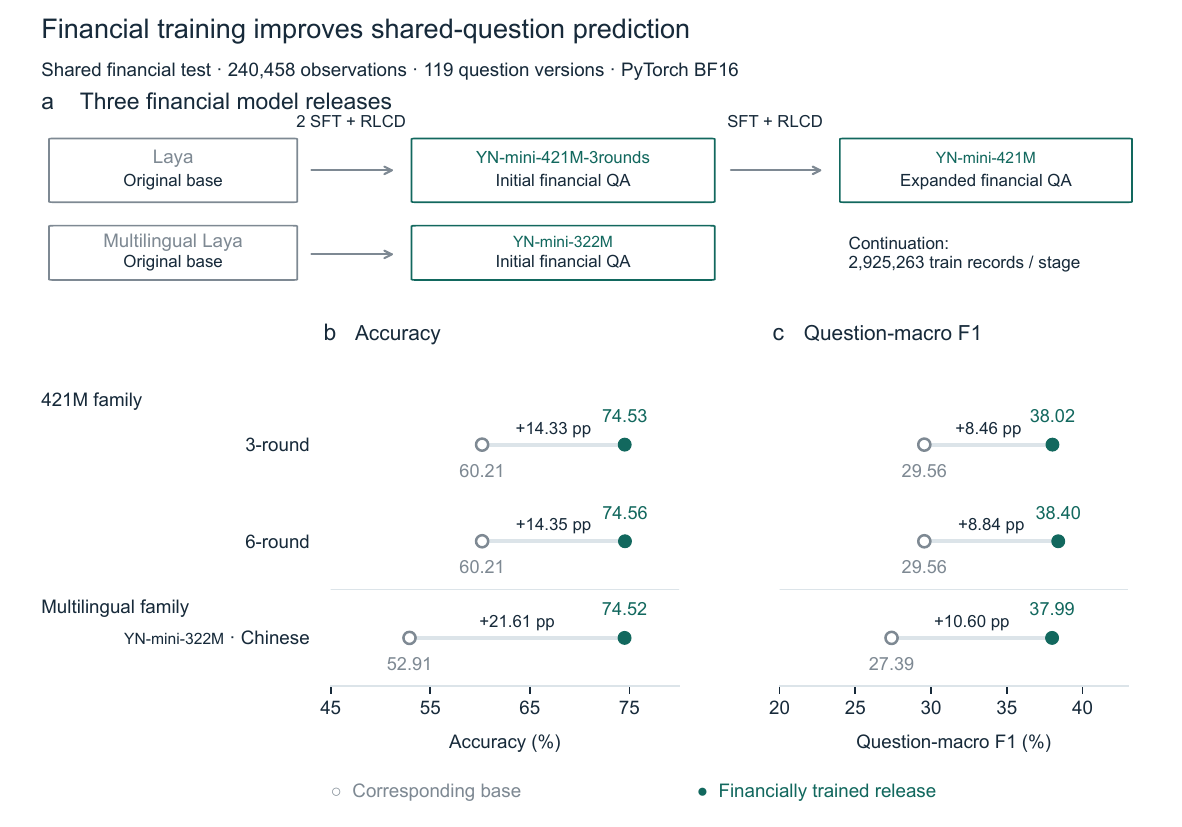}

\par\vspace{5pt}{\fontsize{8.5}{10.3}\selectfont\raggedright \refstepcounter{figure}\label{figure:main5}Figure \thefigure{} | Financial training improves shared-question prediction in the YN family. a, The YN-mini-3rounds policy uses financial-question supervision; full-data YN-mini adds expanded financial supervision through another SFT and RLCD stage. YN-mini-322M derives from multilingual Laya and remains an initial-supervision adaptation. b, Accuracy gains over the corresponding base are 14.33 and 14.35 percentage points for the native variants and 21.61 points for YN-mini-322M evaluated in Chinese. c, Question-macro F1 rises by 8.46, 8.84 and 10.60 points. Comparisons use 240,458 shared financial-question QA observations, 119 question versions and PyTorch BF16. Connectors pair adaptations with their corresponding base; labels show scores and percentage-point differences. Accuracy weights observations equally; F1 averages candidate-class F1 within version, then weights supported versions equally.\par}

\end{minipage}

\end{figure}

Financial supervision improves shared-question prediction in the YN family (Figure~\ref{figure:main5}). The three-round policy raises native BF16 accuracy from 60.2080\% to 74.5336\% and question-macro F1 from 29.5574\% to 38.0164\%. Full-data YN-mini maintains shared-question accuracy at 74.5586\% while extending supervised task coverage: accuracy on the expanded test rises from 56.3657\% to 67.2728\%. The auxiliary YN-mini-322M policy evaluated in Chinese gains 21.61 accuracy points and 10.60 F1 points over its multilingual base.

\begin{figure}[!htbp]

\begin{minipage}{\textwidth}

\centering

\includegraphics[width=\textwidth,trim=0pt 12pt 0pt 20pt,clip,keepaspectratio]{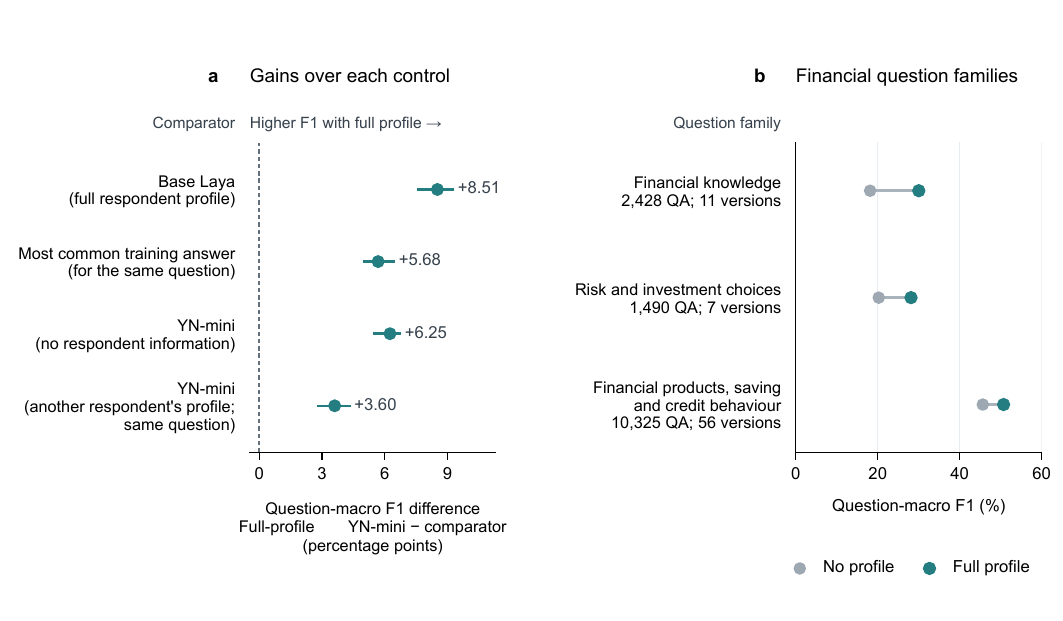}

\par\vspace{5pt}{\fontsize{8.5}{10.3}\selectfont\raggedright \refstepcounter{figure}\label{figure:main6}Figure \thefigure{} | Respondent information improves survey-answer prediction.
a, Question-macro F1 of full-profile YN-mini minus each labelled comparator, in percentage points; zero indicates equal scores and positive values favour full-profile YN-mini. Points show paired differences on 20,000 QA; lines show 95\% bootstrap intervals from 2,000 resamples of 6,752 household/respondent groups. The wrong-profile control supplies another respondent's information for the same question.
b, YN-mini scores with full profiles (teal) and no respondent information (grey), grouped by financial question family. Counts give QA and question versions. Full profiles improve F1 in all three families.\par}

\end{minipage}

\end{figure}

Respondent information also improves prediction when the trained policy and questions remain fixed (Figure~\ref{figure:main6}; Table~\ref{table:main3}). On the matched 20,000-QA panel, full profiles add 6.2475 question-macro F1 points over empty profiles (95\% interval 5.4958--6.7267) and 3.6047 points over another respondent's profile for the same question (2.7977--4.3155). Full profiles outperform both the question-only and training-majority controls, showing that household attributes contribute information beyond the question's answer distribution. The gain appears in financial knowledge, risk and investment attitudes, and saving and credit products.

The adaptation controls distinguish the training choices supported by the common interface (Tables 2 and 5). YN-Lite-4B is the selected 200k 4B LoRA-plus-head route; updating the backbone is more effective than training only its head. The evaluated YN-Lite-4B and YN-master-9B routes reach similar accuracy under the common downstream protocol. Native Laya controls show that supervised adaptation supplies most of the encoder's predictive gain. Additional RLCD raises accuracy slightly over SFT, whereas CE continuation yields higher question-macro F1 than RLCD. Table~\ref{table:main2} retains every fixed-budget version, and Appendix S10 reports development selection, repeated exposures, training cost and paired intervals.

YN-mini-SFT-200k uses 0.733 allocated H100 GPU-hours, versus 24.043 for YN-Lite-4B and 29.782 for YN-master-9B (Table S24). These downstream-stage totals include validation and saving.
\begin{table}[!htbp]
\begin{minipage}{\textwidth}
\fontsize{8.5}{10.3}\selectfont
\refstepcounter{table}\label{table:main2}Table \thetable{} | Common-test performance of every fixed-budget trained version
\par\vspace{4pt}
\begin{tabularx}{\textwidth}{@{}>{\hsize=1.65\hsize\linewidth=\hsize\raggedright\arraybackslash}X>{\hsize=.85\hsize\linewidth=\hsize}X>{\hsize=.9\hsize\linewidth=\hsize}X>{\hsize=.8\hsize\linewidth=\hsize}X>{\hsize=.8\hsize\linewidth=\hsize}X@{}}
\toprule
Model / adaptation & Unique training QA & Selected cumulative exposures & Accuracy (\%) & Question-macro F1 (\%) \\
\midrule
Original native Laya & 0 & 0 & 60.1972 & 29.5995 \\
YN-Lite-4B-head-50k & 50,000 & 100,000 & 67.3369 & 31.0166 \\
YN-Lite-4B-LORA-50k & 50,000 & 50,000 & 73.7289 & 36.6225 \\
YN-Lite-4B-head-200k & 200,000 & 200,000 & 67.2841 & 30.8995 \\
YN-Lite-4B & 200,000 & 200,000 & 74.1435 & 36.7343 \\
YN-master-9B & 200,000 & 200,000 & 74.0919 & 36.6622 \\
YN-mini-SFT-50k & 50,000 & 50,000 & 72.6942 & 34.8511 \\
YN-mini-SFT-200k & 200,000 & 200,000 & 73.8753 & 35.9949 \\
YN-mini-RLCD-50k & 50,000 & 100,000 & 72.9258 & 34.7841 \\
YN-mini-RLCD-200k & 200,000 & 400,000 & 73.9268 & 36.0503 \\
YN-mini-CE-200k & 200,000 & 250,688 & 73.8986 & 36.1732 \\
\bottomrule
\end{tabularx}
\par\vspace{3pt}{\fontsize{8}{9.5}\selectfont shared test: 240,458 records, 119 question versions and 8,275 respondent groups; invalid decisions count as incorrect. All ten trained versions use their development-best checkpoint. Exposures count downstream example presentations in this experiment, including the SFT parent for continued native models. Original Laya has zero new downstream training here, in addition to its upstream training. YN-mini's full-data training is reported separately. Native and Qwen inference use BF16; the full-data YN-mini system comparison in Table~\ref{table:main4} uses MLX FP16. The complete training costs and terminal exposures appear in Appendix S10.}
\end{minipage}
\end{table}

\begin{table}[!htbp]
\begin{minipage}{\textwidth}
\fontsize{8.5}{10.3}\selectfont
\refstepcounter{table}\label{table:main3}Table \thetable{} | Matched respondent-information controls
\par\vspace{4pt}
\begin{tabularx}{\textwidth}{@{}XXXX@{}}
\toprule
Model / information condition & QA & Weighted accuracy (\%) & Question-macro F1 (\%) \\
\midrule
Base Laya, full profile & 20000 & 60.14 & 29.26 \\
YN-mini, full profile & 20000 & 74.40 & 37.77 \\
YN-mini, no profile & 20000 & 71.19 & 31.53 \\
YN-mini, wrong same-question profile & 20000 & 69.56 & 34.17 \\
Training-question majority & 20000 & 71.71 & 32.09 \\
\bottomrule
\end{tabularx}
\par\vspace{3pt}{\fontsize{8}{9.5}\selectfont The same 20,000 held-out QA and 119 question versions enter every row. Accuracy uses full-test question frequencies. The majority baseline uses training answers only; wrong profiles are drawn within question version.}
\end{minipage}
\end{table}

\FloatBarrier
\subsection*{5.3. Comparing models on one-shot consumer choices}

Fourteen models and a question-majority baseline predict recorded answers and coded respondent targets on the same 240,458 one-shot consumer financial-choice records and 119 question versions (Table~\ref{table:main4}a). YN-mini-MLX achieves the highest accuracy, 74.55\%, with question-macro F1 of 38.37\%. Its accuracy exceeds GLM-5.3-Flash (68.88\%), Jev 1.13 (65.26\%), DeepSeek-V4.1-Flash (63.75\%) and GPT-6-Luna (62.10\%). Their question-macro F1 scores are 38.00\%, 37.48\%, 37.94\% and 35.74\%, respectively.

GPT-6-Astra is evaluated on a fixed 500-group sample containing 14,441 records and all 119 question versions. Restricting YN-mini-MLX to those same records gives 74.73\% accuracy and 37.30\% question-macro F1, compared with GPT-6-Astra's 69.34\% and 36.24\% (Table~\ref{table:main4}b). The paired accuracy gain is 5.39 percentage points (95\% interval 4.76--6.10), and the F1 gain is 1.07 points (0.37--1.85), using 1,000 joint resamples of the 500 complete groups.

The paired YN-mini-MLX--Jev accuracy gain is 9.2960 percentage points (95\% interval 9.0637--9.5174), and its F1 gain is 0.8953 points (0.5448--1.2834). Against GLM-5.3-Flash \href{https://docs.z.ai/guides/vlm/glm-5.3-flash}{[40]}, the paired accuracy gain is 5.6787 points (5.5090--5.8459), and the F1 gain is 0.3754 points (0.0398--0.7234). These intervals use 1,000 joint resamples of 8,275 household/respondent groups.

Clef Flash 9B has the highest F1, 38.5193\%; the paired YN-mini-MLX--Clef difference is -0.1447 percentage points (95\% interval -0.6208 to 0.4380). At 421M parameters, YN-mini reaches a similar measured F1 to the 9B comparator through this runtime, while scoring higher in accuracy. Relative to the question-majority baseline, its F1 increases by 6.2058 points.

\begin{table}[!htbp]
\begin{minipage}{\textwidth}
\fontsize{8}{9.7}\selectfont
\refstepcounter{table}\label{table:main4}Table \thetable{} | One-shot consumer financial-choice prediction benchmark
\par\vspace{4pt}
a, Complete test: 240,458 records; 119 question versions; 8,275 household/respondent groups
\par\vspace{3pt}
\begin{tabularx}{\textwidth}{@{}>{\hsize=1.45\hsize\linewidth=\hsize\raggedright\arraybackslash}X>{\hsize=.65\hsize\linewidth=\hsize\raggedright\arraybackslash}X>{\hsize=1.00\hsize\linewidth=\hsize\raggedright\arraybackslash}X>{\hsize=1.25\hsize\linewidth=\hsize\raggedright\arraybackslash}X>{\hsize=.75\hsize\linewidth=\hsize\raggedright\arraybackslash}X>{\hsize=.90\hsize\linewidth=\hsize\raggedright\arraybackslash}X@{}}
\toprule
Model & Model size & Backend & QA / valid & Accuracy (\%) & Question-macro F1 (\%) \\
\midrule
YN-mini-MLX & 421M & MLX CUDA FP16 & 240,458 / 240,458 & 74.5544 & 38.3746 \\
Question-majority baseline & — & CPU; training labels & 240,458 / 240,458 & 71.7710 & 32.1688 \\
GLM-5.3-Flash & 320B & API; provider precision & 240,458 / 240,360 & 68.8756 & 37.9991 \\
Jev 1.13 & Not disclosed & API; provider precision & 240,458 / 240,458 & 65.2584 & 37.4793 \\
Kev 4B & 4B & CUDA BF16 & 240,458 / 240,458 & 64.9120 & 36.0435 \\
DeepSeek-V4.1-Flash & 552B + 196B & Provider API & 240,458 / 240,400 & 63.7546 & 37.9400 \\
Clef Flash 9B & 9B & CUDA BF16 & 240,458 / 240,458 & 63.6490 & 38.5193 \\
GPT-6-Luna & Not disclosed & Provider API & 240,458 / 240,458 & 62.1048 & 35.7412 \\
Base Laya & 421M & MLX CUDA FP16 & 240,458 / 240,458 & 60.1831 & 29.5760 \\
Laya-Finance (hravi) & 421M & CUDA FP32 & 240,458 / 240,458 & 59.8317 & 28.7871 \\
Tev1 4B & 4B & CUDA BF16 & 240,458 / 240,458 & 58.8909 & 35.4105 \\
Liquid d1-omni-600M & 587M & CUDA FP16 & 240,458 / 240,458 & 58.4227 & 28.2825 \\
OpenDecider Nano & \textasciitilde{}400M & Ettin / CUDA FP32 & 240,458 / 240,458 & 57.3572 & 28.3874 \\
Von 1.3 & 395M & CUDA FP32 & 240,458 / 240,458 & 42.4873 & 26.8372 \\
GLiNER2.5-Decide & 486M & CUDA FP32 & 240,458 / 240,458 & 30.8815 & 19.6371 \\
\bottomrule
\end{tabularx}
\par\vspace{7pt}
b, Matched fixed sample: 14,441 records; 119 question versions; 500 groups
\par\vspace{3pt}
\begin{tabularx}{\textwidth}{@{}>{\hsize=1.45\hsize\linewidth=\hsize\raggedright\arraybackslash}X>{\hsize=.65\hsize\linewidth=\hsize\raggedright\arraybackslash}X>{\hsize=1.00\hsize\linewidth=\hsize\raggedright\arraybackslash}X>{\hsize=1.25\hsize\linewidth=\hsize\raggedright\arraybackslash}X>{\hsize=.75\hsize\linewidth=\hsize\raggedright\arraybackslash}X>{\hsize=.90\hsize\linewidth=\hsize\raggedright\arraybackslash}X@{}}
\toprule
Model & Model size & Backend & QA / valid & Accuracy (\%) & Question-macro F1 (\%) \\
\midrule
YN-mini-MLX & 421M & MLX CUDA FP16 & 14,441 / 14,441 & 74.7317 & 37.3026 \\
GPT-6-Astra & Not disclosed & Provider API & 14,441 / 14,441 & 69.3442 & 36.2366 \\
\bottomrule
\end{tabularx}
\par\vspace{3pt}{\fontsize{7.7}{9.3}\selectfont Panel a uses the complete 240,458-record test across 119 question versions. Panel b uses the same fixed 500-group sample of 14,441 records for YN-mini-MLX and GPT-6-Astra, covering all 119 versions. QA / valid gives total records and legal decisions; invalid responses count as incorrect. GLM-5.3-Flash has 98 invalid responses and DeepSeek-V4.1-Flash has 58; the remaining rows have none. GLM reports 320B total and 18B active parameters \href{https://docs.z.ai/guides/vlm/glm-5.3-flash}{[40]}. DeepSeek reports 552B backbone plus 196B Engram parameters, with 8B active during prefill and 16B during decode \href{https://huggingface.co/deepseek-ai/DeepSeek-V4.1-Flash/blob/main/DeepSeek_V41_Tech_Report.pdf}{[41]}. GLiNER\textquotesingle{}s complete checkpoint has 486M parameters; its card uses the 340M label \href{https://huggingface.co/fastino/GLiNER2.5-Decide}{[6]}. Kev, Clef and Tev sizes are their published backbone classes; other local sizes follow released weights or model cards. These are model sizes, not trainable-adapter counts. The majority baseline uses training labels. YN-mini and base Laya use MLX CUDA FP16 here; Table~\ref{table:main2} reports the controlled BF16 adaptations. Supplementary Methods M5 describes adapters and input settings.}
\end{minipage}
\end{table}

\FloatBarrier
The system comparison in Table~\ref{table:main4} captures differences across decision interfaces; matched controls in Table~\ref{table:main5} isolate the effects of adaptation within each backbone.
\begin{table}[!htbp]
\begin{minipage}{\textwidth}
\fontsize{8.2}{10.0}\selectfont
\refstepcounter{table}\label{table:main5}Table \thetable{} | What the component controls and policy comparisons establish
\par\vspace{4pt}
\begin{tabularx}{\textwidth}{@{}>{\hsize=1.02\hsize\linewidth=\hsize\raggedright\arraybackslash}X>{\hsize=1.18\hsize\linewidth=\hsize\raggedright\arraybackslash}X>{\hsize=.80\hsize\linewidth=\hsize\raggedright\arraybackslash}X@{}}
\toprule
Tested component and contrast & Observed result & Contribution to the framework \\
\midrule
Respondent information: full versus empty profile, same YN-mini policy and questions; 20,000 QA & Full profiles add 6.2475 points of question-macro F1 (95\% interval 5.4958--6.7267) & Agent attributes contribute predictive information beyond the question alone \\
\addlinespace[3pt]
Compact domain adaptation: original native Laya versus SFT on 200k unique QA; same 421M architecture, BF16 & Accuracy rises from 60.1972\% to 73.8753\% & Survey supervision supplies most of the native policy's downstream gain \\
\addlinespace[3pt]
Continuation objective: same SFT200 parent and added 200k-example pass budget; development-selected RLCD and CE checkpoints & RLCD raises accuracy from SFT's 73.8753\% to 73.9268\%. Against CE continuation, RLCD's F1 is 0.1229 points lower (95\% interval -0.2073 to -0.0330) & The continuation objective changes the metric balance after SFT; CE has higher F1 in this comparison \\
\addlinespace[3pt]
Representation adaptation: YN-Lite-4B-head-200k versus YN-Lite-4B on the same 200k QA & Accuracy is 67.2841\% with head-only training and 74.1435\% with LoRA & Updating the representation adds value beyond the frozen-backbone route \\
\addlinespace[3pt]
Backbone substitution: YN-Lite-4B and YN-master-9B, same 200k-example one-pass LoRA protocol; different upstream histories & Accuracy is 74.1435\% and 74.0919\%; 9B minus 4B is -0.0516 points (-0.0981 to -0.0071) & These two adaptation routes reach similar accuracy on the common test \\
\addlinespace[3pt]
External policy substitution: full-data YN-mini versus complete decision-model systems on common records and alternatives & YN-mini exceeds Jev by 9.2960 accuracy points and 0.8953 F1 points. Clef has the highest F1: 38.5193\%, versus YN-mini's 38.3746\%; their paired F1 difference includes zero & A common interface exposes accuracy/F1 tradeoffs across decision systems with different readouts \\
\bottomrule
\end{tabularx}
\par\vspace{3pt}{\fontsize{8}{9.5}\selectfont F1 denotes question-macro F1; differences are percentage points. The information control uses 6,752 respondent/household groups. The remaining rows use the common test of 240,458 records, 119 question versions and 8,275 respondent/household groups. Matched adaptations use BF16; the full-data YN-mini system comparison uses MLX FP16. Equal continuation pass budgets permit development to select different cumulative exposures: 400,000 for RLCD and 250,688 for CE. Tables 2 and 4 retain all version scores; Appendix S10 gives the training work and paired comparisons. External systems differ in upstream training and readout, so their row evaluates policy substitution rather than an architecture ablation.}
\end{minipage}
\end{table}

\FloatBarrier
\subsection*{5.4. MLX deployment paths}

YN-mini-MLX deploys the full-data encoder policy in MLX CUDA FP16 for the population applications (Table S5a).

The separate single-request deployment benchmark evaluates YN-mini-322M in PyTorch and YN-mini-322M-MLX through MLX (Table S5b). On that measured workload, YN-mini-322M-MLX delivers 2.42 times the same checkpoint's PyTorch shared-input-token throughput and 22.72 times Jev API's measured throughput. These rates count shared input tokens processed by a candidate-scoring request, rather than generated output tokens; the 322M timing workload is separate from the 421M population application and the 4B/9B household experiment. The three fixed Chinese inputs and single-request timing protocol are described in Supplementary Methods M9.
\FloatBarrier
\subsection*{5.5. Scaling execution and persistent state}

\begin{figure}[!htbp]

\begin{minipage}{\textwidth}

\centering

\includegraphics[width=\textwidth,height=.60\textheight,keepaspectratio]{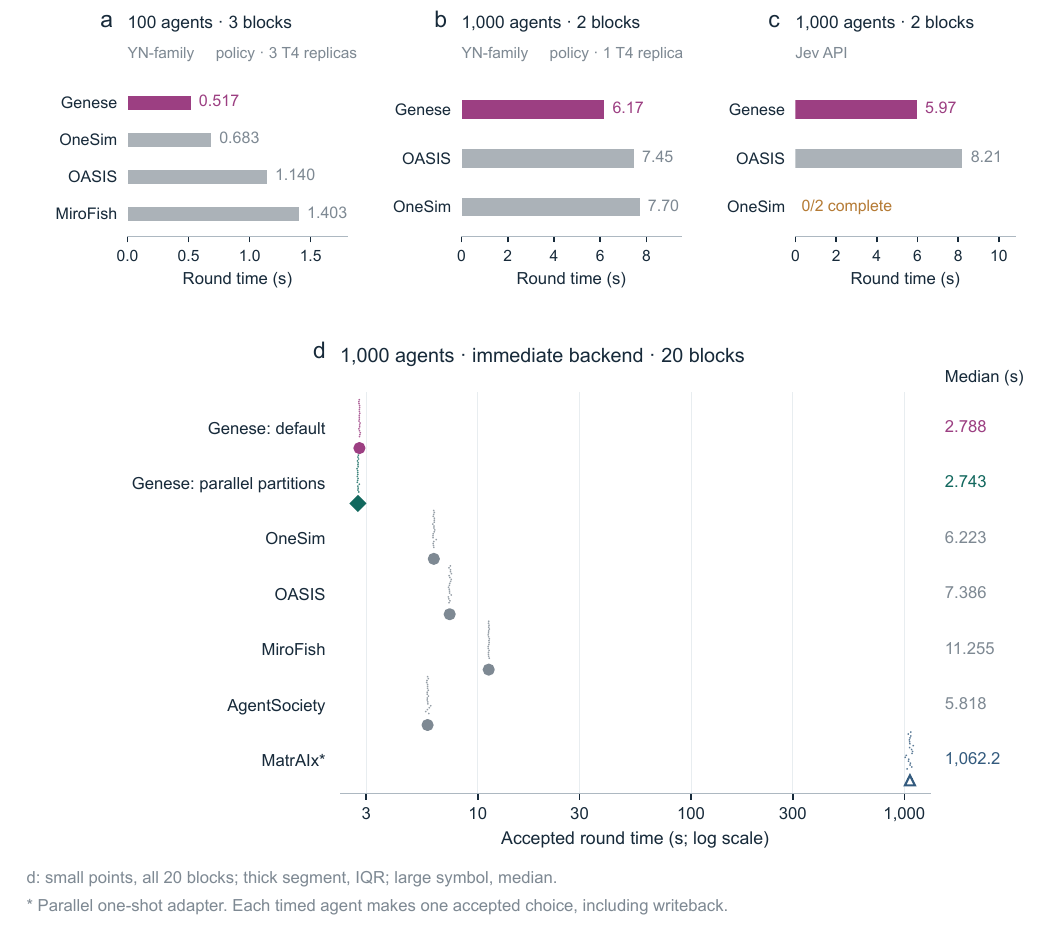}

\par\vspace{5pt}{\fontsize{8.5}{10.3}\selectfont\raggedright \refstepcounter{figure}\label{figure:main7}Figure \thefigure{} | Agent-round times under fixed upstream allocations. Each timed agent makes one accepted model choice, including answer writeback; initialization is separate. a--b, 100 agents with three T4 YN-mini replicas and 1,000 agents with one replica, with medians over three and two randomized blocks, respectively. c, 1,000 agents with Jev API, two blocks; OneSim completes 0/2 attempts. d, Immediate-response execution at 1,000 agents, twenty randomized interleaved blocks per configuration after one warm-up. Small points show rounds, large symbols medians and thick segments inclusive interquartile ranges. Genese default and parallel partitions have distinct symbols and colors; an asterisk and open triangle mark the parallel MatrAIx one-shot adapter. Supplementary Methods M9 gives the common host, allocation and persistence settings.\par}

\end{minipage}

\end{figure}

Fixed-upstream comparisons hold the task and model allocation common, timing one accepted choice and writeback per agent. At 1,000 agents with one T4 replica, Genese takes 6.17 seconds, versus 7.45 for OASIS and 7.70 for OneSim. Twenty randomized interleaved immediate-response blocks isolate execution overhead: Genese's medians are 2.788 seconds with default partitions and 2.743 with parallel partitions. All 280 rounds across the two population sizes accept and write back every requested choice (Figure~\ref{figure:main7}; Table~\ref{table:main6}). These local adapters use a memory-backed SQLite WAL state--event store; production uses PostgreSQL.

The 12M purchase-funding run tests the preparation--inference--ledger pipeline at population scale using YN-mini-MLX.

The 12M-household application executes 24M one-shot choices and 288M monthly ledger states through the asynchronous HTTP policy--ledger runner. A total of 14 GPUs (eight H100 and six A100) complete the common end-to-end interval in 3 h 59 min 17 s, sustaining 1,671.66 choices/s (Figure~\ref{figure:main8}a). The runner overlaps input preparation, direct option scoring, ledgers and compressed writes across 2,400 work units, using service batches of at most sixteen requests. Supplementary Methods M9 defines the measured interval, token accounting and completed-result recovery.

\begin{table}[!htbp]
\begin{minipage}{\textwidth}
\fontsize{8.5}{10.3}\selectfont
\refstepcounter{table}\label{table:main6}Table \thetable{} | Fixed benchmark matrices and completed execution
\par\vspace{4pt}
\begin{tabularx}{\textwidth}{@{}XXXX@{}}
\toprule
Measurement & Configuration / scope & Genese result & Other results \\
\midrule
1,000 agents, immediate backend & 2\ensuremath{\times}Xeon Platinum 8268; 96 threads; 20 interleaved blocks & Default 2.788 s; partitions 2.743 s & AgentSociety 5.818; OneSim 6.223; OASIS 7.386; MiroFish 11.255; MatrAIx 1,062.153 s \\
100 agents, immediate backend & 2\ensuremath{\times}Xeon Platinum 8268; 96 threads; 20 interleaved blocks & Default 0.362 s; partitions 0.360 s & AgentSociety 1.063; OneSim 0.521; OASIS 0.931; MiroFish 1.163; MatrAIx 96.529 s \\
100 agents, YN-mini fl4 & 3 T4 model replicas; 3 blocks & 0.517 s & OneSim 0.683; OASIS 1.140; MiroFish 1.403 s \\
100 agents, YN-mini fl5 & Separate adapter matrix; 3 blocks & 0.509 s & MatrAIx uses this separate reference (Supplementary Figure S5d) \\
1,000 agents, YN-mini & 1 T4 model replica; 2 blocks & 6.17 s & OASIS 7.45; OneSim 7.70 s \\
1,000 agents, Jev & 2 blocks & 5.97 s & OASIS 8.21 s; OneSim 0/2 completed \\
\bottomrule
\end{tabularx}
\par\vspace{3pt}{\fontsize{8}{9.5}\selectfont In the agent-round benchmarks, each agent makes one choice; round time includes answer writeback and excludes initialization. Immediate-backend entries use all twenty accepted block times per configuration. Framework CPU50 and proxy CPUs60–63 have nonexclusive affinity during concurrent training, with memory-backed comparison storage. MatrAIx uses a parallel one-shot adapter.}
\end{minipage}
\end{table}
\FloatBarrier

\section*{6. Applications across financial tasks and horizons}

\subsection*{6.1. Application groups}

We evaluate YN-mini-MLX in an eight-task suite and the 12M purchase-funding experiment (Table~\ref{table:main7}). A separate two-million-household fixed-plan credit application supplies supporting consumption and cash results in Appendix S12.

In that sixteen-month task, a 2,000-household comparison holds households, the available menu and the ledger fixed. YN-mini selects fully funded plans for 87.00\% of households, compared with 79.30\% for original Laya and 88.05\% for the fixed restrained-consumption plan C0. Relative to C0, YN-mini executes CNY 1,146.60 more consumption and retains CNY 1,115.34 less cash. Information and menu controls separate changes in what agents are told from changes in the available plans (Appendix S12).

\begin{center}
\begin{minipage}{\textwidth}
\fontsize{8.3}{10.1}\selectfont
\refstepcounter{table}\label{table:main7}Table \thetable{} | Applications connecting task reuse and population scale
\par\vspace{4pt}
\begin{tabularx}{\textwidth}{@{}XXXX@{}}
\toprule
Application & Population and conditions & Choices and state horizon & Framework capability \\
\midrule
Eight one-shot financial families & 207 financial parents; YN-mini and original Laya, with Chinese variants & 181,892 one-shot choices; twelve-month ledgers & Reuse one decision interface across eight mechanisms \\
12M purchase-funding simulation & 12M distinct synthetic households; goals and tied approved limits of CNY 20k/30k & 24M one-shot choices; 288M monthly ledger states & Execute the same task and accounting rules at population scale \\
\bottomrule
\end{tabularx}
\par\vspace{3pt}{\fontsize{8}{9.5}\selectfont One-shot selections govern the subsequent fixed ledger path. Tables S32--S33 give full 12M results and measurement scopes.}
\end{minipage}
\end{center}

\subsection*{6.2. Reusing one-shot task interfaces at scale}

YN-mini-MLX and original Laya completed 181,892 one-shot choices across the eight-task suite in English and Chinese: 181,888 choices in complete condition pairs and four supplemental T1 choices. Each task contains 2,842 pairs from its eligible financial parents within a 207-parent pool. Table S27 specifies the income, medical-bill, purchase, loan-rate, return, payment-budget and asset-value changes, together with their action consequences. Task-specific parent counts define the response intervals. In the exploratory T3 grid, raising the purchase goal and tied approved limit by 50\% raised YN-mini-MLX's consumption by CNY 17,148.49 (95\% interval 14,818.58--19,601.23), versus 1,194.03 with Chinese inputs (638.59--1,825.40). These intervals resampled the 207 parents. Both input languages use the same English-base checkpoints with identical financial specifications. The input-language difference accompanied changes in payments, debt and unmet routine needs (Appendix S11).

\begin{figure}[!htbp]
\begin{minipage}{\textwidth}
\centering
\includegraphics[width=\textwidth]{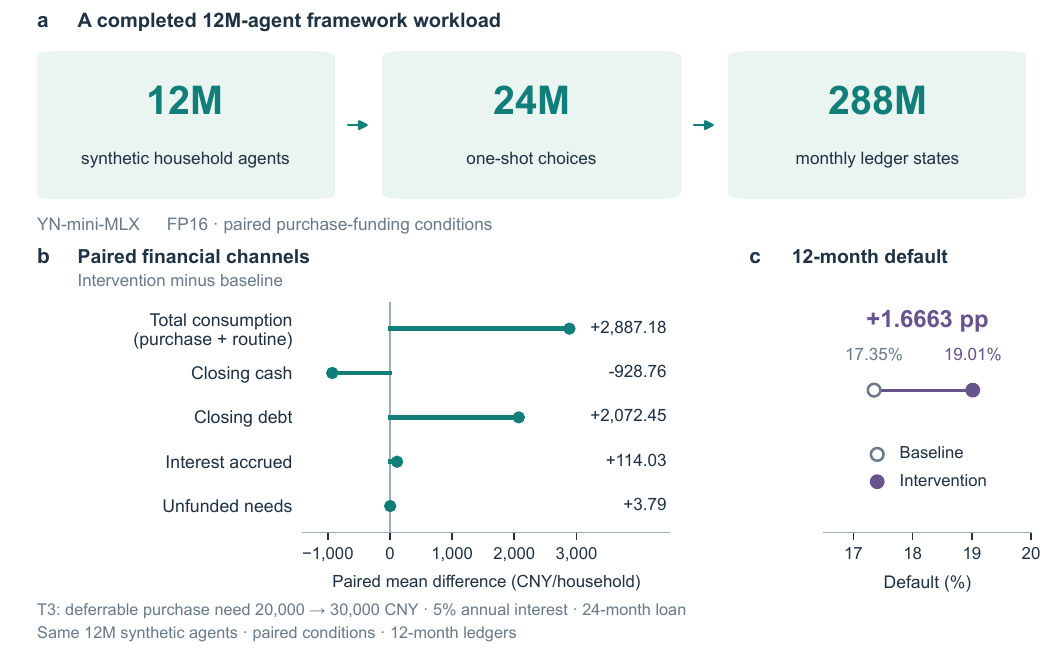}
\par\vspace{5pt}{\fontsize{8.5}{10.3}\selectfont\raggedright \refstepcounter{figure}\label{figure:main8}Figure \thefigure{} | Population-scale execution and paired financial outcomes. a, YN-mini-MLX FP16 executes two frozen T3 conditions for 12M synthetic household agents, producing 24M one-shot choices and 288M monthly ledger states. Each selected action is executed through twelve accounting months. b, Full-panel paired mean changes, with total consumption comprising purchase and routine spending, when the deferrable purchase goal and tied approved limit rise from CNY 20,000 to 30,000, with an assigned 5\% annual rate and 24-month loan. c, Simulated default over the twelve-month ledger horizon rises from 17.35\% to 19.01\%. Differences are intervention minus baseline for the same synthetic agents. Tables S32--S33 give complete financial means and measurement definitions.\par}
\end{minipage}
\end{figure}

The purchase-funding application uses one fixed T3 comparison for 12M distinct synthetic household agents, with English inputs and frozen YN-mini-MLX: the full-data English-base YN-mini checkpoint in MLX CUDA FP16. Each household faced a deferrable purchase goal and tied approved limit of CNY 20,000 or 30,000, with alternatives to defer, finance half, or finance the full purchase at an assigned 5\% annual rate over 24 months. The bounded HTTP decision service returned one selection per condition, which the independent accounting path carried through a twelve-month ledger (Figure~\ref{figure:main8}a; Table S33).

Executed purchase spending rose by CNY 2,890.97 per household; a CNY 3.79 reduction in routine consumption yielded a CNY 2,887.18 increase in total twelve-month consumption. Closing cash fell by CNY 928.76 and closing debt rose by CNY 2,072.45; actual debt payments and accrued interest also increased. Unfunded routine needs rose by CNY 3.79, and simulated default rose by 1.6663 percentage points (Figure~\ref{figure:main8}b--c). The ledger followed the first twelve months of the twenty-four-month financing contract (Tables S32--S33).

\FloatBarrier
\subsection*{6.3. Consumption heterogeneity and its financial channels}

Households whose decision makers are aged 25--34 show a consumption increment of CNY 9,161.69, compared with 12.88 at ages 55--64 and zero at ages 65 and above. Reweighting each age group to the full panel's income distribution preserves the contrast: the 25--34 mean becomes 9,181.62 and the age-65+ mean remains zero. Across annual-income groups, the mean increment rises from CNY 1,059.67 below 20,000 to 6,052.72 at 500,000 or above. Within ages 25--34, the income pattern is nonmonotonic, linking the marginal gradient to the joint household profile (Tables S38--S40). All 2,364,307 households with decision makers aged 65+ deferred the purchase in both conditions. At ages 55--64, 99.96\% also deferred in both; 859 switched from deferral to the full purchase, 36 kept the full-purchase option and two switched to deferral, producing the CNY 12.88 mean increment.

\begin{figure}[!htbp]
\begin{minipage}{\textwidth}
\centering
\includegraphics[width=\textwidth]{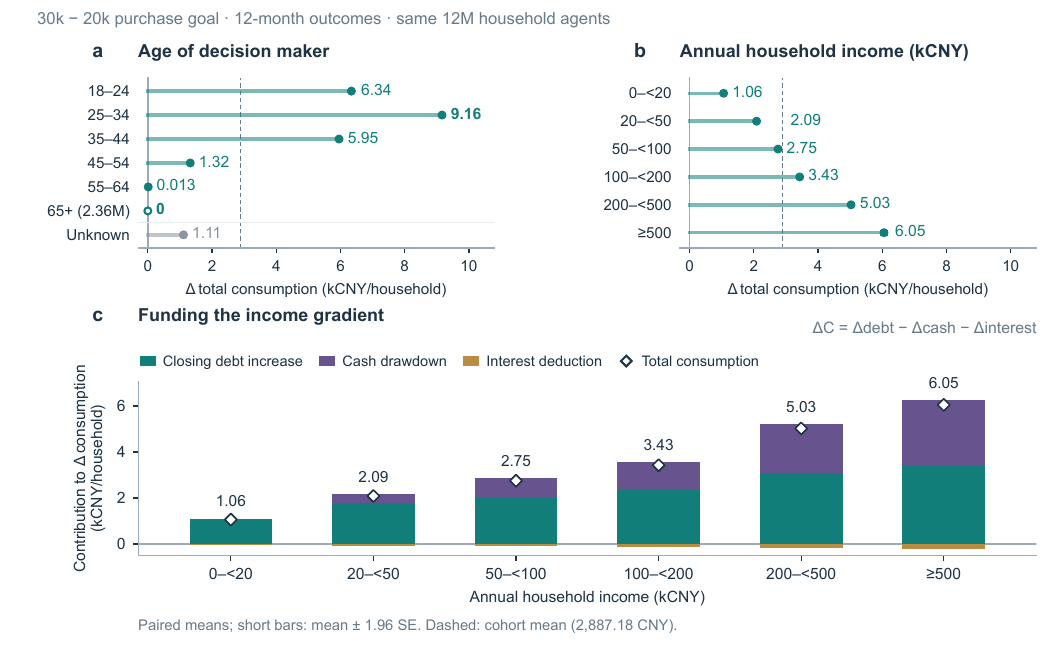}
\par\vspace{5pt}{\fontsize{8.5}{10.3}\selectfont\raggedright \refstepcounter{figure}\label{figure:main9}Figure \thefigure{} | Household profiles shape the consumption response and its financial channels. a, Paired mean consumption differences by decision-maker age. b, The same contrast by annual household income. Differences are CNY 30k minus 20k purchase goal and tied approved limit over twelve accounting months. All 12M households contribute to each partition; unknown age remains a separate group. Dashed lines mark the full-panel mean; short bars show mean $\pm1.96$ synthetic-row standard-error references. c, For each income bracket, additional closing debt plus cash drawdown minus additional accrued interest equals the additional consumption. Amounts are thousand CNY per household; closing debt includes unpaid interest. Table S38 gives actual debt payments. Tables S38--S42 and source aggregates give group sizes, financial-stock partitions, standardized contrasts and variation decompositions.\par}
\end{minipage}
\end{figure}

The ledger connects these selected purchases to subsequent resources and payments (Figure~\ref{figure:main9}c). The lowest income bracket increases purchase spending by CNY 1,061.72 and reduces routine spending by 2.04; the highest bracket increases purchases by 6,052.72 with routine spending unchanged. Their additional actual debt payments are 60.45 and 2,877.92, respectively. T3 purchases are fully financed at origination; subsequent cash use and repayment determine the debt carried through the horizon. With income and opening resources fixed within each pair, $\Delta C=\Delta D-\Delta K-\Delta I$ links additional consumption to additional closing debt, cash drawdown and accrued interest.

Education, employment and household size also distinguish purchase responses under the same age--income mixture. The consumption increment is CNY 398.12 for decision makers with primary education or below and 3,820.99 for the college-educated group; it is 1,406.09 for those not working and 3,437.93 for the employed. One-person households show an increment of 1,666.17, compared with 3,797.16 for three-person households. Each group contains all 42 age--income cells (Table S44).

\begin{samepage}
The choice transitions explain most of these differences. Of the CNY 3,427.57 difference in purchase response between the college-educated and primary-or-below groups, 3,183.17 comes from switching from deferral to a purchase and 234.46 from expanding the amount under the same half- or full-purchase choice. The routine-spending difference is $-4.70$, giving a consumption-response difference of 3,422.87. This difference accompanies 1,004.78 more cash drawdown, 1,009.48 more actual debt payments and 2,555.05 more closing debt. Employment and household-size contrasts follow the same dominant deferral-to-purchase channel. Risk preference and housing tenure add only 0.13 and 0.07 percentage points, respectively, to the consumption variation explained by age--income groups (Table S43).
\par
\end{samepage}

Financial assets distinguish consumption size from the way households carry the obligation. The raw consumption increment is CNY 2,580.88 for financial assets below 10,000 and 3,054.60 for assets of 1--2 million. Reweighting both asset groups to the same age--income distribution reverses that ordering to 3,014.26 and 2,361.25. The higher-asset group nevertheless makes more additional actual payments (1,137.87 versus 905.61), uses more cash (1,136.28 versus 905.61) and retains less additional closing debt (1,311.52 versus 2,229.04). Higher financial-asset stock therefore accompanies a larger cash-and-payment contribution to a smaller consumption increment under the common profile mixture (Table S41).

Opening debt reveals a different adjustment. Under the same age--income mixture, households without old debt increase consumption by CNY 2,864.61 and actual payments by 1,063.72. Households with CNY 2--5 million of opening debt increase consumption by 1,811.13 and actual payments by 44.31; their cash drawdown is 43.97 and closing-debt increment 1,848.23. The high-debt group's additional consumption is carried predominantly as debt through the twelve-month window. Adding debt stock to age--income groups increases the between-group share of consumption-response variation by 1.40 percentage points and payment-response variation by 2.15 points (Tables S41--S42).

\FloatBarrier
\clearpage
\input{household_experiment}
\FloatBarrier
\clearpage
\section*{8. Discussion and future work}

The two applications address complementary questions: how purchase responses vary across a large population, and how well learned policies predict household consumption. In S\textsuperscript{2}-MAS, the shared state and action interface lets different policies be compared using the same household profiles and accounting rules.

\subsection*{Purchase choices and financial channels}

The purchase-funding results separate the decision to buy from how households carry the resulting obligation. Age and income organize purchase responses; education, employment and household size retain differences under a common age--income mixture, mainly through transitions from deferral to purchase. Assets and existing debt distinguish cash use and repayment: standardization reverses the consumption-increment ordering of the illustrated asset groups, while the high-debt group carries much of its added consumption as closing debt.

The learned policy and purchase menu produce an extreme age boundary: all 2,364,307 households with decision makers aged 65 or above defer in both conditions. They comprise 19.70\% of the panel and contribute zero to its consumption increment, even after income standardization. This concentration identifies a policy behavior to test against observed lifecycle demand before using the simulated age gradient for demand forecasting.

\subsection*{Trainable policies and framework extension}

The adaptation controls show that LoRA improves on head-only training and that supervised adaptation supplies most of native Laya's gain; the evaluated 4B and 9B routes reach similar accuracy. Input language also affects the resulting financial behavior: the authored T3 grid produces consumption contrasts of CNY 17,148 with English inputs and 1,194 with Chinese inputs under the same English-base checkpoint. This sensitivity makes input formulation part of the household-policy specification (Appendix S11).

Extending the ledger to asset repricing, employment transitions and transfers would allow the same household policies to be studied under changes in both income and wealth.
\section*{9. Conclusion}

S\textsuperscript{2}-MAS combines household personas, trained decision policies and financial accounting to simulate responses across heterogeneous populations. In the 12M-household application, purchase transitions explain much of the consumption variation, while cash use, repayment and closing debt identify how households fund those purchases. The 1,060-household consumption--saving study connects learned decisions to observed annual consumption, group errors and information sensitivity. Together, these applications support learned decision rules as another route to large-scale heterogeneous-household modeling.
\section*{Data and code availability}

YN-Lite-4B and YN-mini are the principal decision policies; YN-mini-MLX provides the full YN-mini checkpoint in MLX format. Model weights and cards are available at \href{https://huggingface.co/ya9919/sharp-4b}{YN-Lite-4B}, \href{https://huggingface.co/ya9919/sharp-421m}{YN-mini} and \href{https://huggingface.co/ya9919/sharp-421m-mlx}{YN-mini-MLX}.

Scientific reproduction code and reported non-identifying aggregates accompany the \href{https://github.com/KataDavidXD/s2-mas}{project repository}. The \href{https://github.com/KataDavidXD/s2-mas/blob/main/resources/MODEL_CATALOG.md}{model catalog} identifies the ten trained release versions and additional checkpoints and runtime formats. Project model releases use noncommercial research terms prohibiting redistribution and reverse engineering. Survey-derived decision QA, study-cohort records and individual trajectories remain under restricted access.

A \href{https://github.com/KataDavidXD/s2-mas/tree/main/resources/public_companion}{companion resource} contains 2M independently constructed financial household personas and 32,000 unlabelled task inputs from 2,000 households, eight authored task families and two conditions. Its resource entry provides generation code, schema, component terms and the data-archive entry. A separate \href{https://github.com/KataDavidXD/s2-mas/tree/main/resources/wikidata}{CC0 Wikidata component} contains 38 entities and 66 relations. These companion records are separate from the populations used in model training and reported simulation. The public scientific code provides aggregate figure reproduction, decision-model training and evaluation interfaces, and independent companion generation. The GENESE/MAS engine implementation remains private.

\FloatBarrier
\begingroup
\fontsize{9}{11}\selectfont
\hypersetup{urlcolor=ink,linkcolor=ink}

\endgroup

\clearpage\small\raggedbottom

\section*{Supplementary Methods M1. Financial observations and split construction}

Inputs and evaluations use English by default; Chinese and bilingual settings are labelled explicitly.

The prepared corpus contains 4,172,421 QA observations with aligned Chinese and English views. Repeated questions and both languages are assigned together within an operational household/respondent group. Target-related attributes are excluded before prediction, and groups are partitioned across training, development, calibration and test.

Supervision combines recorded questionnaire answers and targets formed from observed financial, occupation and education categories. Cleaning preserves question wording, complete alternatives, monetary units and the distinction between missing values and zero. Target-related fields are removed from each input as specified below.

\begin{center}
\begin{minipage}{\textwidth}
\fontsize{8.5}{10.3}\selectfont
Table S1 | Grouped bilingual supervision
\par\vspace{4pt}
\begin{tabularx}{\textwidth}{@{}XX@{}}
\toprule
Partition & QA observations per language \\
\midrule
Training & 2,925,263 \\
Development & 412,873 \\
Calibration & 415,688 \\
Test & 418,597 \\
Total & 4,172,421 \\
\bottomrule
\end{tabularx}
\par\vspace{3pt}{\fontsize{8}{9.5}\selectfont bilingual views represent the same underlying observations.}
\end{minipage}
\end{center}

Translation uses Grok 4.7 and GPT-6 Luna across preparation, with glossary review and static translations. Amounts, rates, units, negation and option meanings remain aligned. Input profiles contain 27 harmonized dimensions; the population bank uses 36 coded fields, while persona-text evaluation uses 44 processed reference fields.

Income tasks remove income inputs; spending tasks remove total and component expenditure; debt tasks remove amount and type. Housing, occupation and education tasks similarly exclude target-related fields. Missing, skipped, refused and undefined targets are excluded, while questionnaire-defined do-not-know alternatives remain eligible. Development chooses checkpoints; a separate calibration partition fits probability temperatures. The financial task rules define conditions and execution independently of these observed-answer targets.

\section*{Supplementary Methods M2. Population construction and economic structure}

\subsection*{Constructed household records}

The population resource contains 500 million constructed household financial decision-maker records in 50 Parquet shards of 10 million rows. Real-survey data inform a common household representation. Financial attributes are sampled jointly conditional on age, household income and total spending, while eligibility constraints maintain coherent category combinations. Unknown categories remain explicit.

The 500M expansion calibrates household size in five bins from one person to five or more, together with selected debt, homeownership and financial-asset margins; rural debt retains its sampled distribution. It also refines upper monetary categories. The 10M consumer bank uses its specified 36-field representation and five income, spending and asset brackets. Population operations use numeric codes; the policy receives decoded text.

Supplementary Methods M3 defines the descriptive bank and the 12M execution cohort.

\subsection*{Economic structure and cohort preservation}

PCA evaluates standardized ordinal attributes in the completed constructed records. In the 10M bank, 7,227,206 rows have all 14 fields known. The analysis forms their correlation matrix, extracts its eigenvectors and orients PC1 toward positive total spending. PC1 explains 32.8904\% of variance and PC2 11.5234\%.

The separate 500M resource has approximately 4.626 million complete records per shard for the same analysis. Its shard-mean PC1 explains 32.62\% of variance, with across-shard loading standard deviations at most 0.0003 (Table S15). The independent companion contains 2M separately constructed synthetic personas with assigned household and financial states. Household size and cash-to-income multiples use discrete uniform draws with the settings below. Figure S1 summarizes the assigned income and debt distributions. Figure S4 applies PCA to household size and seven financial variables after log transformation of CNY amounts and standardization. All 2M records are complete for these variables; PC1 and PC2 explain 45.6450\% and 18.0043\% of variance. The companion's schema and generation code accompany the resource entry.

\begin{center}
\begin{minipage}{.91\textwidth}
\centering
{\small\textbf{Assigned generation settings for the independent synthetic companion}\par}
\vspace{4pt}
{\small
\begin{tabularx}{\textwidth}{@{}lXc@{}}
\toprule
Variable & Assigned values & Probability per value \\
\midrule
Household members & 1, 2, 3, 4, 5 & $1/5$ \\
Opening cash / monthly income & 0, 1, 2, 3, 4, 5, 6 & $1/7$ \\
\bottomrule
\end{tabularx}}
\end{minipage}
\end{center}

\begin{center}

\begin{minipage}{\textwidth}

\centering

\includegraphics[width=\textwidth,height=.66\textheight,keepaspectratio]{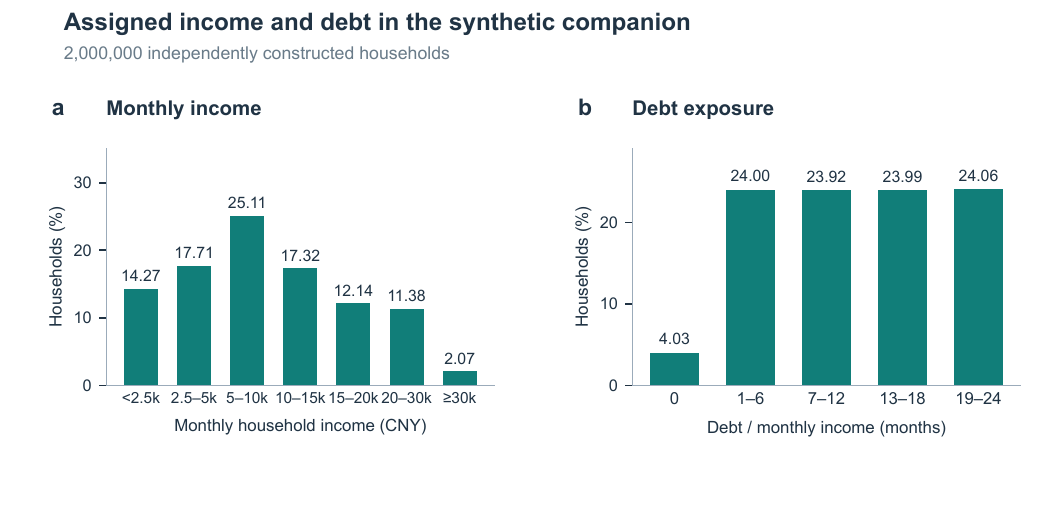}

\par\vspace{5pt}{\fontsize{8.5}{10.3}\selectfont\raggedright Supplementary Figure S1 | Assigned income and debt distributions in the independent synthetic companion. All 2,000,000 constructed households contribute to each distribution. a, Monthly household income in CNY; k denotes thousands, and finite income intervals include their lower endpoint and exclude their upper endpoint. b, Opening debt principal divided by monthly income, grouped by the assigned integer income multiples. Heights and labels give the percentage of households. The companion provides reusable scenario inputs, separate from the 12M purchase-funding execution cohort.\par}

\end{minipage}

\end{center}

The 500M resource supports population construction; the 10M bank supports descriptive analysis of the representation; the separate 12M cohort executes the purchase-funding application.

\section*{Supplementary Methods M3. Experimental-bank sampling and descriptive checks}

The 10M consumer bank contains constructed, survey-derived records. Its delivered selection retains the first 10,000,000 distinct eligible rows in shard order from the documented 100M parent; eligible records have known financial assets, debt amount and type, investment activity and trust.

The 14-field PCA uses complete cases from the delivered 10M bank; loading and completeness summaries appear in Tables S17--S18.

For the 12M purchase-funding application, two staged shards contain 20M distinct constructed records from the 500M resource. Seed 20261018, finite-pool inclusion caps and randomized cell-count rounding select whole rows without replacement before inference; uniform row draws operate within the selected cells. The 12M panel retains each selected profile, including 2,818,405 households with assigned baseline deficits and 543,327 with debt above CNY 500,000. The 35,000-household pilot is a nested subset whose completed choices are retained in the full run. Equal household weights define the reported fixed-panel means.

\section*{Supplementary Methods M4. Full prediction and respondent-information results}

YN-mini receives a household description, question and named alternatives through Laya’s existing option-scoring interface. We evaluate 3-round, 6-round and bilingual variants against their corresponding Laya bases. Laya combines a 395M ModernBERT-large backbone and a decision head for 421M total parameters; the multilingual base uses mmBERT with 322M \href{https://github.com/LeonaDavinci/laya-system-one}{[2]}. Native PyTorch and MLX implementations support training and deployment.

The serialized input places the question and candidate descriptions before the respondent state, with a marker for each candidate. The encoder and choice head score those markers; inference returns the semantic option with the highest calibrated probability. SFT and native Laya RLCD adapt the model to observed answers or coded household targets. The reused RLCD objective adds a normalized-advantage policy-gradient term over Gaussian-perturbed option logits to soft cross-entropy; the perturbations are scored against target distributions by proper rewards \href{https://github.com/NandhaKishorM/laya/blob/main/laya/train.py}{[36]}.

The primary comparison uses 240,458 shared financial-question test records and 119 question versions in PyTorch BF16. Accuracy pools correct choices across records. Question-macro F1 averages candidate-class F1 within each question version and then weights versions equally; classes with zero denominators receive zero (Figure~\ref{figure:main5}).

Figure~\ref{figure:main5} reports the shared-question native and bilingual predictions. On the expanded-supervision test, native question-macro F1 rises from 19.2395\% for YN-mini-3rounds to 27.8355\% for YN-mini. Appendix S3 reports the shared- and expanded-test results in MLX.

A training-question majority baseline predicts each question's most frequent observed training answer. On all 240,458 shared test records, it obtains 71.7710\% accuracy and 32.1688\% question-macro F1. YN-mini in MLX obtains 74.5544\% and 38.3746\%, gains of 2.7834 and 6.2058 percentage points.

Matched information controls use 20,000 held-out QA drawn proportionally within all 119 question versions, with at least one record per version and seed 20261014. Empty-profile requests remove only the respondent state. Wrong-profile requests use another held-out respondent's profile for the same question; questions, candidate meanings and gold answers remain fixed. Accuracy weights question versions by their full-test frequencies; F1 retains equal weighting of versions. Table~\ref{table:main3} reports all five conditions.

Intervals in Figure~\ref{figure:main6} use 2,000 paired bootstrap draws over the 6,752 household/respondent groups in this panel. The corresponding accuracy gains for full versus empty and wrong profiles are 3.2058 points (95\% interval 2.7336–3.6138) and 4.8406 points (4.3626–5.3266).

The registry defines three financial families: financial knowledge, risk and investment attitudes, and financial-product, saving and credit behavior. Full-profile versus empty-profile F1 is 30.0124\% versus 18.1357\% for financial knowledge, 28.1147\% versus 20.2383\% for risk and investment, and 50.6979\% versus 45.6257\% for saving and credit products (Figure~\ref{figure:main6}). The panel contains 14,243 QA across 74 versions in these families. Table S22 reports the remaining registry domains.

On the complete shared test, YN-mini exceeds the training-majority F1 in 80 of 119 question versions, falls below it in two and ties in 37. The equal-version gain is 6.2058 percentage points (95\% interval 5.8711–6.4951); weighting those same version-level F1 scores by test frequency gives 7.4109 points (7.2111–7.6073). The two versions with fewer than 30 responses contain 22 QA and contribute -0.2489 points to the overall equal-version gain. Financial-knowledge, risk-attitude and saving/credit families each improve over majority (Table S2).

\begin{center}
\begin{minipage}{\textwidth}
\fontsize{8.5}{10.3}\selectfont
Table S2 | Complete financial-family gains over training majority
\par\vspace{4pt}
\begin{tabularx}{\textwidth}{@{}>{\hsize=1.500000\hsize\linewidth=\hsize\raggedright\arraybackslash}X>{\hsize=0.600000\hsize\linewidth=\hsize\raggedright\arraybackslash}X>{\hsize=0.750000\hsize\linewidth=\hsize\raggedright\arraybackslash}X>{\hsize=0.800000\hsize\linewidth=\hsize\raggedright\arraybackslash}X>{\hsize=1.350000\hsize\linewidth=\hsize\raggedright\arraybackslash}X@{}}
\toprule
Financial family & Test QA & Question versions & Macro-F1 gain, pp & 95\% confidence interval \\
\midrule
Financial knowledge & 29,198 & 11 & 12.40 & 11.51 to 13.28 \\
Financial products, saving and credit & 124,129 & 56 & 4.79 & 4.25 to 5.32 \\
Risk and investment attitudes & 17,923 & 7 & 7.79 & 7.36 to 8.18 \\
\bottomrule
\end{tabularx}
\par\vspace{3pt}{\fontsize{8}{9.5}\selectfont All 119 versions remain in the primary shared 240,458-QA comparison. Intervals use 1,000 joint resamples of 8,275 linked household/respondent groups. Equal-version and sample-weighted results accompany the complete version-level data.}
\end{minipage}
\end{center}

\section*{Supplementary Methods M5. One-shot consumer financial-choice benchmark}

All evaluated models receive the same financial-test questions, respondent states and semantic alternatives. Model-specific adapters map returned choices to the original candidate IDs. Table~\ref{table:main4}a reports fourteen models and the question-majority baseline on 240,458 records from 8,275 household/respondent groups. Table~\ref{table:main4}b compares GPT-6-Astra and YN-mini-MLX on the same fixed 500-group sample of 14,441 records; both panels cover all 119 question versions. The YN-mini-MLX predictions in panel b are restricted to the exact records evaluated by GPT-6-Astra. Invalid responses count as incorrect: GLM-5.3-Flash has 98 and DeepSeek-V4.1-Flash has 58. Question-macro F1 averages class-macro F1 over the original alternatives within each question version, then gives the 119 versions equal weight. Reported full-test paired intervals use 1,000 whole-group bootstrap resamples of the 8,275 groups, retaining all records from each sampled group. Table~\ref{table:main2} gives the controlled BF16 adaptations, and Figure~\ref{figure:main5} gives the native full-data predictions.

The 119-version registry in Table S22 was fixed for the initial shared-model comparison and retained for the API extension. Every eligible held-out record in that registry enters the complete test. Targets comprise recorded questionnaire answers and coded respondent attributes; the authored financial applications execute selected plans through their specified action menus and accounting rules.

GPT-6-Luna and GPT-6-Astra were evaluated from 20 to 22 October 2026 with requested IDs \texttt{gpt-6-luna} and \texttt{gpt-6-astra}, low reasoning effort and a 4,096-token completion budget. DeepSeek-V4.1-Flash-1017 used a 4,096-token output budget with thinking disabled. These three text-output adapters serialize complete state and question objects as compact JSON, preserve candidate order without truncation, and request one original candidate ID as \texttt{\{"choice":"ID"\}}. After removing an enclosing code fence, the parser accepts only a one-key JSON object containing a legal candidate. The first successful provider response is scored; transport failures retry the same payload, and invalid answers receive no repair call.

The Astra sample was drawn before the new model outputs using seed 20261020: 500 household/respondent groups are sampled at random, retaining every record in each selected group. It contains all 119 versions. YN-mini-MLX is restricted to exactly those records. Paired intervals jointly resample the 500 complete groups 1,000 times with seed 20261018 and use the 2.5th and 97.5th percentiles; the question-macro calculation retains all original candidate classes and equal version weights.

The public Laya-Finance checkpoint contains 421,293,830 parameters, rounded to 421M in Table~\ref{table:main4} \href{https://huggingface.co/hravi/laya-finance}{[27]}.

Jev uses \texttt{jev-1.13.0} typed-choice outputs; GLM uses \texttt{glm-5.3-flash} text outputs. Both map to original candidates.

QA caps are 512 input and 256 head tokens for YN-mini/base MLX, 512/192 for Laya-Finance, and 2,048 input tokens for OpenDecider. Decoder adapters preserve their released longer windows. CPU replay of the evaluated MLX input preparation finds no profile truncation on the full test or any matched information-control condition; full-profile inputs average 159.75 tokens and reach at most 230. The supplementary training procedure specifies the separate complete-context budgets used for financial deployment.

YN-mini has the highest answer accuracy in the complete comparison. Clef Flash 9B obtains 38.5193\% question-macro F1 versus YN-mini's 38.3746\%. The paired YN-mini-minus-Clef difference is -0.1447 F1 points (95\% interval -0.6208 to 0.4380), alongside a 10.9054-point accuracy gain (10.6636–11.1576). The external Laya-Finance checkpoint obtains 59.8317\% accuracy and 28.7871\% F1; YN-mini's paired gains are 14.7227 points (14.5391–14.9281) and 9.5874 points (9.2879–9.8604).

\section*{Supplementary Methods M6. Training procedure and continuation variability}

The legacy 6-round identifier denotes the full-data release. Its optimization recipe comprises three SFT stages and two RLCD stages, following the unadapted base. The sequence is original-base \ensuremath{\rightarrow} sft-1 \ensuremath{\rightarrow} sft-2 \ensuremath{\rightarrow} rlcd-1 \ensuremath{\rightarrow} sft-3 \ensuremath{\rightarrow} rlcd-2. Two SFT stages and one RLCD stage on financial questions produce the 3-round model from 1,721,623 training observations per language. The continuation for YN-mini adds expanded financial supervision through one SFT epoch and one native RLCD epoch over all 2,925,263 training examples, using BF16 on six A100 GPUs. The continuation updates encoder and choice-head parameters. YN-mini-322M uses two SFT stages and one RLCD stage on initial-supervision material. Household or respondent groups and their language copies stay together within each split.

Two independent full continuation replicas repeat the full-data continuation from the same completed 3-round initialization, each with one SFT pass and one RLCD pass over 2,925,263 training QA. All three runs use stage-specific calibration temperatures and evaluate the same 119-version, 240,458-QA financial-question test with RLCD fixed as the final stage.

Across the three native BF16 RLCD runs, accuracy averaged 74.5647\% with sample standard deviation 0.0067 percentage points and range 74.5586–74.5719\%. Question-macro F1 averaged 38.3277\% with sample standard deviation 0.1254 points and range 38.1829–38.4016\% (Table S3). These repeats quantify continuation variability conditional on the common initialization, survey preparation and splits; the respondent bootstrap quantifies sample variation for a fixed checkpoint. The direct decision comparison in Table~\ref{table:main4} evaluates the 6-round rlcd-2 checkpoint in MLX CUDA FP16 on the same shared test.

On the expanded 418,597-record test, YN-mini obtains 67.2726\% accuracy and 27.8955\% F1 in MLX, and 67.2728\% and 27.8355\% natively. The paired MLX shared-QA gain over base is 14.3713 percentage points, with a stratified household/respondent bootstrap 95\% interval of 14.1686–14.5719. Supplementary Figure S2 provides deployment and expanded-scope results.

\begin{center}
\begin{minipage}{\textwidth}
\fontsize{8.5}{10.3}\selectfont
Table S3 | Continuation for YN-mini: seed variability
\par\vspace{4pt}
\begin{tabularx}{\textwidth}{@{}XXXX@{}}
\toprule
Training run & SFT / RLCD seed & SFT Accuracy / F1 (\%) & RLCD Accuracy / F1 (\%) \\
\midrule
Reference & 202610120/202610121 & 74.5315 / 37.8775 & 74.5586 / 38.3986 \\
Replica 1 & 202610140/202610141 & 74.5519 / 37.9777 & 74.5635 / 38.4016 \\
Replica 2 & 202610150/202610151 & 74.5739 / 37.8331 & 74.5719 / 38.1829 \\
\bottomrule
\end{tabularx}
\par\vspace{3pt}{\fontsize{8}{9.5}\selectfont Both stages evaluate the identical 240,458 shared QA and 119 versions. Accuracy weights QA records; F1 is the equally weighted mean of per-version candidate-macro F1.}
\end{minipage}
\end{center}

\section*{Supplementary Methods M7. Persona rendering and judge evaluation}

Six human raters evaluate 100 processed survey-derived personas against the same 44 processed fields. Five dimensions assess accuracy (M1), avoidance of unsupported inference (M2), coverage (M3), internal consistency (M4) and overall fidelity (M5). The 3,000 scores average 4.570/5; 87.8\% are 4 or 5. Coverage averages 4.985, with 591 of 600 ratings equal to 5. Nine predefined targets cover income, consumption, debt, budget status, risk preference, financial assets, insurance and income source (Figure~\ref{figure:main4}; Table S4).

Accuracy and consistency average 4.685 and 4.670; avoidance of unsupported inference and overall fidelity average 4.285 and 4.225. DeepSeek Flash and GLM Flash each score within one point of the six-rater mean on 94.0\% of 500 persona-dimension comparisons. Relative to the human mean, AI scores are higher for avoidance of unsupported inference (M2, +0.705 and +0.715 points) and overall fidelity (M5, +0.775 points for both judges). Reevaluation with DeepSeek V4 Pro and GLM 5.3 gives 94.3\% on 98 valid personas and 94.0\% on 100, respectively. Supplementary Figure S3 gives the full human-score distributions.

Perturbed personas test changed facts, unsupported additions, missing targets and contradictions. GLM lowers the intended score on 95.8–100\% of valid pairs. DeepSeek's response ranges from 8.3\% for omissions to 76.0\% for contradictions. For omitted targets, GLM lowers the intended score in all 24 valid pairs, whereas DeepSeek does so in 2 of 24. Scoring proximity and known-error sensitivity therefore measure different properties of these judges (Appendix Table S16). These evaluations assess reconstruction of processed survey fields.

Appendix S9 reports a separate DeepSeek financial-workflow study, comparing single-pass choice, self-check and multi-agent execution on constraint and goal adherence.

\begin{center}
\begin{minipage}{\textwidth}
\fontsize{8.5}{10.3}\selectfont
Table S4 | Construction-quality and judge comparison design
\par\vspace{4pt}
\begin{tabularx}{\textwidth}{@{}XXXXX@{}}
\toprule
Evaluation & Reference & Raters / judges & Scoring units & Main result \\
\midrule
Human construction quality & 100 processed survey-derived personas; 44 processed fields & Six human raters & 3,000 dimension scores & 4.570 / 5 \\
Coverage, M3 & Nine predefined financial targets & Six human raters & 600 scores & 4.985 / 5; 591 scores of 5 \\
Human–AI proximity & AI score vs six-rater mean & DeepSeek / GLM & 500 comparisons per judge & 94.0\% / 94.0\% within one \\
Pro human–AI proximity & Same processed-field cohort & DeepSeek V4 Pro / GLM 5.3 & 98 / 100 valid personas & 94.3\% / 94.0\% within one \\
\bottomrule
\end{tabularx}
\par\vspace{3pt}{\fontsize{8}{9.5}\selectfont Within-one agreement compares each AI score with the six-human mean for the same persona and dimension. Known-error pairs and judge-specific valid denominators appear in Table S16.}
\end{minipage}
\end{center}

\section*{Supplementary Methods M8. State ownership and execution paths}

The shared contract passes a household state, a question and named alternatives to a policy adapter, which returns an option ID and available probabilities. The execution layer retains the state to which that choice applies. Figure~\ref{figure:main2} describes the Genese state-transition architecture.

In the controlled Genese benchmarks, World owns authoritative state and Brain requests bounded decision partitions through pooled model connections. World validates the returned intents and applies registered effects. Numeric components use indexed NumPy arrays, Boolean activity masks and registered systems; a structural-command buffer defers entity changes to the transition boundary. The timed local adapter stores state, events and command results in a SQLite WAL transaction on memory-backed storage.

The static-task runners call the policy service through bounded asynchronous requests and advance the assigned financial ledger. The existing MAS scheduler overlaps input preparation, inference and accounting across independent episodes, with accepted responses indexed by ordinary task IDs.

Full Genese inference uses the separately supplied runtime, model service and deployment configuration. The independent synthetic fixture exercises local decision recording and accounting. The 12M deployment follows the asynchronous HTTP service and independent-ledger path. The production PostgreSQL/outbox transaction remains the reference architecture described in Section 4.2; local benchmark measurements use the SQLite adapter.

\section*{Supplementary Methods M9. Execution benchmarks and measured scopes}

We time fixed-input framework benchmarks and the complete 12M one-shot workload separately.

\subsection*{Complete 12M execution and trace accounting}

The completed purchase-funding application comprises 2,400 disjoint work units of 5,000 households. Ordinary unit IDs cover the complete allocation, with 1,548 completed units on the eight-H100 pool and 852 on the six-A100 pool. Each household completes the two frozen T3 conditions once. The same YN-mini checkpoint, MLX CUDA FP16, tokenizer and semantic candidates serve both pools; service batches contain at most sixteen requests, with sixty-four bounded in-flight requests per endpoint. Preparing the next work unit overlaps current inference, CPU accounting and compressed trace writes.

The common end-to-end interval starts with preparation of the first nested-pilot inputs and ends when the final work unit completes. It includes input transfer waits, subsequent preparation, tokenization, policy calls, twelve-month ledgers, trace persistence and recovery waits. Its 14,356.9775 seconds cover 24M choices and 12M paired households, giving 1,671.661 choices/s and 835.831 households/s. Actual nonpadding candidate input totals 12,719,522,472 tokens, or 885,947.092 tokens/s over the same interval. The discriminative service reports zero autoregressive output tokens. Table S33 separates this interval from batch-processing and summed forward-call wall measurements.

The private compressed traces total 67,664,771,925 bytes. They retain full model inputs and state, semantic alternatives, selected options and available probabilities, ordinary household/condition/task/attempt IDs, model and service usage, preparation/inference/save timestamps, and each twelve-month cash, debt and payment ledger. All 24M requests have recorded responses and complete-input indicators; HTTP and response-validation errors are zero. Fourteen process/status finalization retries affect eleven already complete work units. Their saved summaries are adopted with the existing choices, preserving one forward result per logical request.

Final aggregation checks complete condition counts, unique work-unit coverage and disjoint pool ownership. Per-household differences retain both conditions. Table S32 reports exact full-panel financial means and defines the supplementary row-variation reference.

Fixed-input benchmarks hold the agent task and upstream model allocation constant; immediate-response backends measure dispatch, storage and adapter overhead.

Each timed agent produces one model choice in one round. Round time includes accepted-answer writeback; initialization is measured separately. The immediate-response matrices comprise twenty randomized interleaved measured blocks per configuration at 100 and 1,000 agents, after one warm-up. All 280 measured rounds accept and write back every requested choice.

With YN-mini and three T4 model replicas, the 100-agent Genese round takes 0.517 seconds, compared with 0.683 for OneSim, 1.140 for OASIS and 1.403 for MiroFish. At 1,000 agents on one T4 replica, Genese takes 6.17 seconds, OASIS 7.45 and OneSim 7.70 (Figure~\ref{figure:main7}a–b). The completed 1,000-agent immediate-backend matrix gives Genese default a median of 2.788 seconds and parallel partitions 2.743. The corresponding medians are 5.818 for AgentSociety, 6.223 for OneSim, 7.386 for OASIS, 11.255 for MiroFish and 1,062.153 for the parallel MatrAIx questionnaire adapter (Figure~\ref{figure:main7}d). The 1,000-agent Jev API matrix reports 5.97 seconds for Genese and 8.21 for OASIS; OneSim completes neither of its two attempted runs (Figure~\ref{figure:main7}c).

At 100 agents, Genese default and parallel partitions take 0.362 and 0.360 seconds (Supplementary Figure S6). The host has two Intel Xeon Platinum 8268 processors and 96 logical threads. Framework CPU50 and proxy CPUs60–63 have nonexclusive affinities during concurrent training; the isolated comparison directory is memory-backed. Initialization is timed separately: Genese default and parallel partitions have 1,000-agent medians of 0.400 and 0.405 seconds. Observed one-minute host load spans 18.87–66.69 in the 100-agent matrix and 4.19–31.89 at 1,000 agents. The immediate-response Genese arms use the local Brain–World adapter and a SQLite WAL commit store on tmpfs, with state, outbox and result committed in one transaction. The production PostgreSQL design is described in Section 4.2.

Tables S12--S14 compare Genese's array-based state, batched decision requests and transactional state--event design with native action constraints, saved-state continuation and request-level controls in the other frameworks. These measurements cover the local SQLite adapter and application ledgers; they exclude production crash recovery and outbox delivery.

\subsection*{Single-request model deployment benchmark}

The 11 October 2026 deployment benchmark uses three fixed Chinese financial-choice cases: income loss, a payment plan and loan-product choice. YN-mini-322M is the released multilingual-base adaptation, evaluated in local PyTorch and as YN-mini-322M-MLX in MLX CUDA on one NVIDIA A100 PCIe 40GB. The third local arm is its original multilingual Laya base. The hosted arm requests Jev and records the returned version jev-1.13.0. Every arm receives three warm-ups and twenty timed repetitions per case, with batch size and concurrency one. The local models use FP16; Jev's server hardware and precision are provider-managed.

The timing interval covers input preparation, tokenization, synchronized prediction and response formatting. Jev includes its HTTPS round trip. The client does not cache responses; provider prompt caching is uncontrolled. The shared mmBERT tokenizer counts compact, sorted JSON requests without special tokens outside the timed interval. The three cases contain 218, 393 and 398 shared tokens, producing 20,180 tokens across sixty measured calls per arm. Input-token throughput is total shared tokens divided by total measured time; choice throughput is sixty divided by that time.

For income loss, payment planning and product choice respectively, YN-mini-322M-MLX reaches 26,177, 37,506 and 46,840 shared input tokens/s; Jev reaches 1,051, 1,888 and 1,944. Pooled throughputs are 36,955.47 and 1,626.63 tokens/s, a ratio of 22.72. The same-checkpoint PyTorch arm reaches 15,247.34 tokens/s, giving an MLX/PyTorch ratio of 2.42373. The MLX and PyTorch financial models agree on all 69 warm-up and timed choices, with maximum probability difference approximately 0.0006. Source aggregates contain per-case summaries, individual elapsed times and the common-token protocol.

\begin{center}
\begin{minipage}{\textwidth}
\fontsize{8.5}{10.3}\selectfont
Table S5 | Financial prediction and deployment measurements
\par\vspace{4pt}
\textbf{a, Common question benchmark}
\par\vspace{3pt}
\begin{tabularx}{\textwidth}{@{}>{\hsize=1.50\hsize\linewidth=\hsize\raggedright\arraybackslash}X>{\hsize=.75\hsize\linewidth=\hsize\raggedright\arraybackslash}X>{\hsize=.75\hsize\linewidth=\hsize\raggedright\arraybackslash}X@{}}
\toprule
Model and runtime & Accuracy (\%) & Question-macro F1 (\%) \\
\midrule
YN-mini-MLX & 74.5544 & 38.3746 \\
Jev 1.13 · API & 65.2584 & 37.4793 \\
\bottomrule
\end{tabularx}
\par\vspace{6pt}
\textbf{b, Single-request deployment}
\par\vspace{3pt}
\begin{tabularx}{\textwidth}{@{}>{\hsize=1.72\hsize\linewidth=\hsize\raggedright\arraybackslash}X>{\hsize=.80\hsize\linewidth=\hsize\raggedright\arraybackslash}X>{\hsize=.68\hsize\linewidth=\hsize\raggedright\arraybackslash}X>{\hsize=.80\hsize\linewidth=\hsize\raggedright\arraybackslash}X@{}}
\toprule
Model and runtime & Mean latency (ms) & Choices/s & Input tokens/s \\
\midrule
YN-mini-322M · PyTorch A100 & 22.06 & 45.33 & 15,247 \\
YN-mini-322M-MLX · MLX CUDA A100 & 9.10 & 109.88 & 36,955 \\
Jev 1.13 · API & 206.77 & 4.84 & 1,627 \\
\bottomrule
\end{tabularx}
\par\vspace{3pt}{\fontsize{8}{9.5}\selectfont a, The 6-round release on 240,458 held-out QA records and 119 question versions. b, Measured 11 October 2026 using the YN-mini-322M checkpoint, three fixed Chinese inputs and sixty timed calls per arm. Local FP16 inference uses one NVIDIA A100 PCIe 40GB, batch size and concurrency one. Timings include input preparation and synchronized prediction; Jev includes the API round trip and service response. Loading is excluded. Shared input tokens use the mmBERT tokenizer on identical serialized requests. Supplementary Methods M9 gives the complete protocol.}
\end{minipage}
\end{center}

\section*{Supplementary Methods M10. Authored tasks, purchase funding and financial accounting}

The eight task families cover income and consumption adjustment, medical funding, purchase funding, product terms, asset allocation, multiple-debt payments, restructuring and asset sale or refinancing. Each language contains 45,473 decision inputs: 22,736 complete baseline/intervention pairs and one additional T1 intervention input. Each pair retains the household while changing a designated condition. The suite uses financial-anchor income, spending, deposits, loan principal and scheduled payments, with adjustable spending assigned as 10--30\% of usual spending. Table S27 defines the sampled conditions and the action-to-ledger mapping.

Buffer-stock saving motivates joint measurement of consumption and cash \href{https://academic.oup.com/qje/article-abstract/112/1/1/1870884}{[22]}; consumption commitments motivate harder-to-adjust and adjustable spending \href{https://academic.oup.com/qje/article-abstract/122/2/831/1942152}{[23]}. Household liquidity and income risk motivate comparing the response across financial states \href{https://www.aeaweb.org/articles?id=10.1257/aer.20160042}{[31]}. The 12M purchase-funding application assigns the following monetary amounts and contracts to constructed household records.

Monetary brackets use their midpoints; an open upper bracket uses 1.25 times its lower bound. Annual income and consumption are divided by twelve. Missing annual income is assigned CNY 72,000, consumption 80\% of income, assets one income month and debt CNY 50,000. Opening liquid cash is assigned 39.1\% of financial assets and essential spending 70\% of total spending. Mixed old debt is split equally between housing and other debt. Housing contracts use 5.42\% annual interest and 240 remaining months; other debt uses 7.68\% and 36 months, with rates referenced to March 2022 newly issued household lending \href{https://app.www.gov.cn/govdata/gov/202205/07/484673/article.html}{[24]}. These assignments are common within household pairs and retain initial budget deficits.

Opening debt in the 12M application is assigned from the selected 500M record's debt-principal category. The no-debt category assigns zero; finite positive brackets assign their midpoint; the open bracket above CNY 5 million assigns CNY 6.25 million; and an unknown category assigns CNY 50,000. Debt type selects the housing or other-debt contract, with mixed housing debt split equally between the two. The two conditions share these opening amounts, scheduled payments and remaining terms.

Each household faces a deferrable purchase goal of CNY 20,000 or 30,000 and a tied approved borrowing limit equal to that goal. Both conditions permit financing. Options defer the purchase, finance half, or finance its full amount. Borrowed funds purchase the good at disbursement. The assigned contract has a 5\% annual nominal rate, 24 monthly installments, no fee and its first payment in month 1. Both arms share the profile, opening cash, existing obligations and income; the goal and tied approved amount change jointly. The twelve-month ledger begins on 1 November 2026.

Income arrives before current-month spending. Routine essential and selected adjustable consumption precede dated bank payments; oldest dues are paid first and interest before principal. Integer-cent cash remains nonnegative, unmet selected spending remains a gap, and unpaid obligations stay in arrears. Default records bank dues of at least CNY 1 remaining over ninety calendar days overdue, retaining an event after cure. Exposure at first default includes outstanding principal and accrued unpaid interest. A 24-month loan can retain debt at the end of the twelve-month observation window. Table S32 reports mean total, purchase and routine consumption, cash, payments, interest, debt and risk.

\subsection*{Consumption heterogeneity from the completed 12M panel}

Crossing decision-maker age, annual household-income bracket, financial-asset-stock bracket and debt-principal-stock bracket defines 4,509 occupied analysis cells. For metric $Y$, the group contrast is $\Delta\bar Y_g=(1/n_g)\sum_{i\in g}(Y_{i,30k}-Y_{i,20k})$. Marginal partitions sum to all 12M households and retain unknown categories. Groups use the constructed profile fields supplied to the completed application. Tables S38--S39 report all occupied marginal groups.

For age comparisons, direct income standardization uses $\Delta\bar C_a^{\mathrm{std}}=\sum_{j=1}^{6}(n_j/N)\Delta\bar C_{a,j}$, where $n_j/N$ is the full-panel share in income bracket $j$. Every age group has observations in all six brackets. The standardized contrast uses a common income mixture and retains each group's other profile composition. Table S40 gives the raw and standardized means. Source aggregates include age-by-income and financial-stock cross-tabulations. Fixed-panel means are primary; optional mean $\pm1.96$ synthetic-row standard-error references use the saved paired sums and squared sums.

With $B$ the cumulative new disbursement, $P$ actual debt payments, $K$ closing cash, $D$ closing debt including unpaid interest, and $I$ accrued interest, the paired cash and debt changes satisfy $\Delta K=\Delta B-\Delta C-\Delta P$ and $\Delta D=\Delta B+\Delta I-\Delta P$. Thus $\Delta C=\Delta D-\Delta K-\Delta I$. Each purchase is fully financed at origination, so its contrast is $\Delta B=\Delta P+\Delta D-\Delta I$; the routine-spending contrast is minus the change in unfunded routine needs. Integer-cent sums satisfy these identities in every occupied cell.

\subsection*{Financial stocks under a common household composition}

For a financial-stock group $g$, direct standardization uses $\Delta\bar Y_g^{\mathrm{std}}=\sum_{a,j}w_{aj}\Delta\bar Y_{g,a,j}$, where $w_{aj}=n_{aj}/N$ is the full panel's joint age--income share. The four illustrative stock groups in Table S41 each contain all 42 age--income cells. Source aggregates give raw and standardized means for each stock category, with full-mixture estimates for categories covering all cells. Financial assets denote the initial stock bracket used in the common cash assignment; debt denotes opening principal. Standardized cash, debt and interest means retain the same funding identity as the raw contrasts.

For paired household response $d_i=Y_{i,30k}-Y_{i,20k}$ and grouping $G$, the between-group variation share is $R_G^2=\sum_g n_g(\bar d_g-\bar d)^2/\sum_i(d_i-\bar d)^2$. Saved paired sums and squared sums supply both terms. Table S42 first reports each marginal partition, then nested partitions that add financial assets or debt to the 42 age--income groups. The increment is the nested share minus the age--income share, in percentage points. Each partition includes all 12M households and unknown categories.

\subsection*{Extended profile groups and purchase-choice decomposition}

We reaggregated the completed result events for all 2,400 work units into 18 additional profile partitions: household size; child and older-member indicators; employment; main income source; housing tenure; debt type; risk preference; financial literacy; investment activity; education; urban/rural residence; region; gender; marital status; income--expenditure status; general trust; and annual household spending. Ordinary unit and household IDs preserve the same-household pairing. Each partition retains all 12M households, including unknown categories. For each partition, we retained paired sums and squared sums for consumption, executed initial purchases, routine spending, closing cash, closing debt, actual payments, accrued interest and unfunded needs, together with the nine baseline-to-intervention choice counts.

The same joint age--income weights standardize each profile category. Table S44 reports full-mixture means for categories with all 42 cells. Categories missing any of the 42 reference cells have blank standardized entries in Table S44. Pairwise shared-support comparisons use the intersection of the compared categories' occupied cells, with full-panel weights renormalized within that set. The older-member comparison uses 36 cells, with reference population 9,635,693 (80.30\% of the panel); the no-older-member category has no age-65+ decision makers. Table S43 extends the nested variation calculation to all 18 attributes. These group comparisons retain the other characteristics within each profile category.

Options A, B and C defer the purchase, purchase half the goal and purchase the full goal. The baseline-to-intervention transition contributions to purchases are CNY 0, 15,000, 30,000, $-10,000$, 5,000, 20,000, $-20,000$, $-5,000$ and 10,000 for A$\to$A, A$\to$B, A$\to$C, B$\to$A, B$\to$B, B$\to$C, C$\to$A, C$\to$B and C$\to$C, respectively. Weighting these amounts by group transition frequencies exactly reconstructs the purchase response. A$\to$B and A$\to$C form the deferral-to-purchase channel; B$\to$B and C$\to$C capture amount increases under an unchanged partial- or full-purchase choice. Adding the routine-consumption response gives the total consumption response. Table S43 applies this decomposition after the same age--income standardization.

\section*{Supplementary Resources}

Table S6 describes the recorded resources, aggregate evidence and reproduction interfaces. Published model weights use the project model terms; household records, decision QA and individual trajectories remain restricted. The public aggregates and training interfaces support the reported analyses without distributing those records.

\begin{center}
\begin{minipage}{\textwidth}
\fontsize{8.5}{10.3}\selectfont
Table S6 | Recorded resources and reproduction interfaces
\par\vspace{4pt}
\begin{tabularx}{\textwidth}{@{}XXX@{}}
\toprule
Component & Recorded scope & Interface or evidence \\
\midrule
Survey-derived QA & 4,172,421 underlying observations; bilingual views & Grouped partitions, target exclusion and aggregate evaluation \\
Training partition & 2,925,263 examples per language & Restricted observed-answer and coded-state supervision \\
Constructed bank & 10M rows; 36 coded fields & Restricted survey-derived household representation \\
Financial tasks & 8 rule families; 45,473 executed cases per language & Authored task rules are separable from restricted household-instantiated cases \\
Decision adaptations & Ten trained versions; 50k/200k budgets & Defined candidate interface, exposure counts and aggregate scores \\
Common-question benchmark & 240,458 records; 119 versions & Accuracy, candidate-class macro-F1 and grouped paired intervals \\
Population application & 24M choices / 12M households / 12 months & Defined ledgers, complete aggregate outcomes and restricted trajectories \\
Synthetic companion & 2M independently constructed financial household personas; 32,000 unlabelled inputs from 2,000 households, 8 families and 2 conditions & Generation code, schema, component terms and companion archive \\
Wikidata companion & 38 entities and 66 parent/type relations & Separate CC0 knowledge component \\
Synthetic task-example subset & 20 generated households; 8 families; 320 inputs & Input-only illustration under CC BY 4.0, without model labels or ledgers \\
\bottomrule
\end{tabularx}
\end{minipage}
\end{center}

\section*{Supplementary tables: related-system comparisons}

\begin{center}
\begin{minipage}{\textwidth}
\fontsize{8.5}{10.3}\selectfont
Table S7 | Representative direct decision interfaces, October 2026
\par\vspace{4pt}
\begin{tabularx}{\textwidth}{@{}XXX@{}}
\toprule
Model & Backbone / decision-model size & Decision readout \\
\midrule
Laya & ModernBERT-large / 421M & Bidirectional option scoring \\
Von 1.3 & ModernBERT-large / 395M & Option-marker head \\
OpenDecider Nano & Ettin encoder / \textasciitilde{}400M & Encoder scoring \\
GLiNER2.5-Decide & DeBERTa-v3-large / 486M checkpoint & Schema label scoring \\
d1-omni-600M & LFM2.5 encoder; multimodal / 587M & Shared-trunk option probabilities \\
Kev & Qwen3.5 / Qwen3.8; 0.8B–27B & Pointer decision head \\
Clef / Clef-Flash & Qwen3.8-27B / Qwen3.5-9B & Decision head; option probabilities \\
Nimble & Qwen3.5 / 9B & Candidate-token logits \\
Tev1 & Qwen3.5 / 4B & Answer-letter output \\
Jev & Hosted service; undisclosed size & Structured decision API \\
\bottomrule
\end{tabularx}
\par\vspace{3pt}{\fontsize{8}{9.5}\selectfont Parameter entries follow the named model cards and released checkpoints. GLiNER2.5-Decide has 486,444,053 parameters in the complete checkpoint; its card labels the model 340M. Encoder classification and decoder decision readouts share a structured-output objective.}
\end{minipage}
\end{center}

\begin{center}
\begin{minipage}{\textwidth}
\fontsize{8.5}{10.3}\selectfont
Table S8 | Financial decision post-training and related training resources
\par\vspace{4pt}
\begin{tabularx}{\textwidth}{@{}XXXX@{}}
\toprule
Training family & Resource / model & Disclosed training scale & Supervision and objective \\
\midrule
Financial decision model & YN-mini & 2,925,263 training examples & Observed survey answers and coded socioeconomic states; encoder + head SFT and RLCD \\
Financial decision model & Laya-Fintiment & 61,017 examples & Financial sentiment; three-way choice; RLCD \\
Financial decision model & Chinese-Jev Finance & 24,000 decisions & Financial knowledge, language tasks and rule-based choice/noul/score; full-parameter RLCD \\
Financial decision model & Laya, Banking77 & 7,994 messages & Banking service intents; frozen encoder and trained typed heads \\
Financial decision model & Laya-Finance (hravi) & Not disclosed & Financial sentiment, topics, headlines and terminology \\
Financial decision model & Ariadne FinanceTopics & Not disclosed & Financial-news topics; trained interface over frozen Laya \\
General decision model & Chinese-Jev General & 10,000,000 decisions & General Chinese choice/noul/score adaptation; separate from financial specialization \\
General decision model & Von 1.3 & \textasciitilde{}290K initial items & Operational decisions; additional continuation \\
General decision model & Nimble, original-2676 & 2,676 train; 324 held out & Curated typed decisions; answer-token fine-tuning \\
General decision model & Tev1, 4B new-v1 & 37,840 train; 4,568 validation & Decision examples; answer-letter fine-tuning \\
Financial generative & PIXIU / FIT & 128,640 instruction samples & Financial NLP and prediction; FinMA instruction tuning \\
Financial generative & FinGPT task sets & 76.8K sentiment; 82.2K headline; 17.1K QA train & Task-specific financial instruction tuning \\
Financial generative & BloombergGPT & 363B financial + 345B general tokens & Mixed-domain language pretraining \\
\bottomrule
\end{tabularx}
\par\vspace{3pt}{\fontsize{8}{9.5}\selectfont YN-mini has 4,172,421 underlying corpus observations, including 2,925,263 training examples per language. The financial decision-model comparison uses disclosed domain post-training decisions; general decision training, instruction samples and pretraining tokens are reported in their original units. Bilingual translations represent the same underlying records.}
\end{minipage}
\end{center}

\begin{center}
\begin{minipage}{\textwidth}
\fontsize{8.5}{10.3}\selectfont
Table S9 | Populations in selected completed simulation studies
\par\vspace{4pt}
\begin{tabularx}{\textwidth}{@{}XXX@{}}
\toprule
Framework & Simulated population & Cited study / cohort \\
\midrule
S\textsuperscript{2}-MAS & 12,000,000 synthetic households & Completed T3 application; 24M paired-condition choices and 12 ledger months \\
OASIS [17] & 1,000,000 users & Social-media population; inference follows user activation \\
OneSim [16] & Approximately 100,000 agents & 2025 v1 cultural-dissemination study; three rounds \\
AgentSociety [18] & 30,000 agents & Published LLM-agent scaling test; ten rounds \\
MatrAIx [1] & Approximately 1,000 personas & Per study and model in Survey, Chatbot and Web evaluations \\
Generative Agents [19] & 25 agents & Smallville deployment; two game days \\
MiroFish [20] & Not specified & Primary project documentation gives no exact completed-run population \\
TwinMarket [21] & 1,000 agents & Selected scaling study; social participation with daily trading \\
\bottomrule
\end{tabularx}
\par\vspace{3pt}{\fontsize{8}{9.5}\selectfont Counts refer to the cited study populations or model-specific cohorts. Persona resources are described in Section 4.1; MatrAIx's 8.3B records and S\textsuperscript{2}-MAS's 500M records are resource sizes.}
\end{minipage}
\end{center}

\begin{center}
\begin{minipage}{\textwidth}
\fontsize{8.5}{10.3}\selectfont
Table S10 | Roles of population resources and executed cohorts
\par\vspace{4pt}
\begin{tabularx}{\textwidth}{@{}XXX@{}}
\toprule
Quantity & Financial research role & Experimental role or literature comparison \\
\midrule
500M persona resource & Reusable constructed household states & Population resource for selecting experimental cohorts \\
10M consumer bank & Reusable income–spending–wealth–debt population & JPMorgan purchasing-power study: account inflows covering 10M U.S. households [29] \\
12M executed panel & T3 purchase choices with twelve-month state transitions & A separate constructed cohort selected from 20M staged resource rows \\
\bottomrule
\end{tabularx}
\par\vspace{3pt}{\fontsize{8}{9.5}\selectfont S\textsuperscript{2}-MAS materializes survey-derived household states and executes two financial choices per household.}
\end{minipage}
\end{center}

\section*{Appendix S1. Bilingual supervision}

Bilingual views express the same underlying financial observations. Table S1 reports the four grouped partitions per language; the common model comparison uses the fixed 240,458-record, 119-version subset described in M4--M5.

\section*{Appendix S2. Financial units and execution assignments}

Recorded income and consumption categories describe annual CNY flows; assets and debt describe balance stocks. Applications assign representative monetary amounts and convert annual flows to monthly inputs before ledger execution. Cash, payments and unpaid obligations use integer cents. Both conditions use the same assignments; M10 gives the accounting rules.

\section*{Appendix S3. MLX and expanded-scope model results}

\begin{center}

\begin{minipage}{\textwidth}

\centering

\includegraphics[width=\textwidth,height=.66\textheight,keepaspectratio]{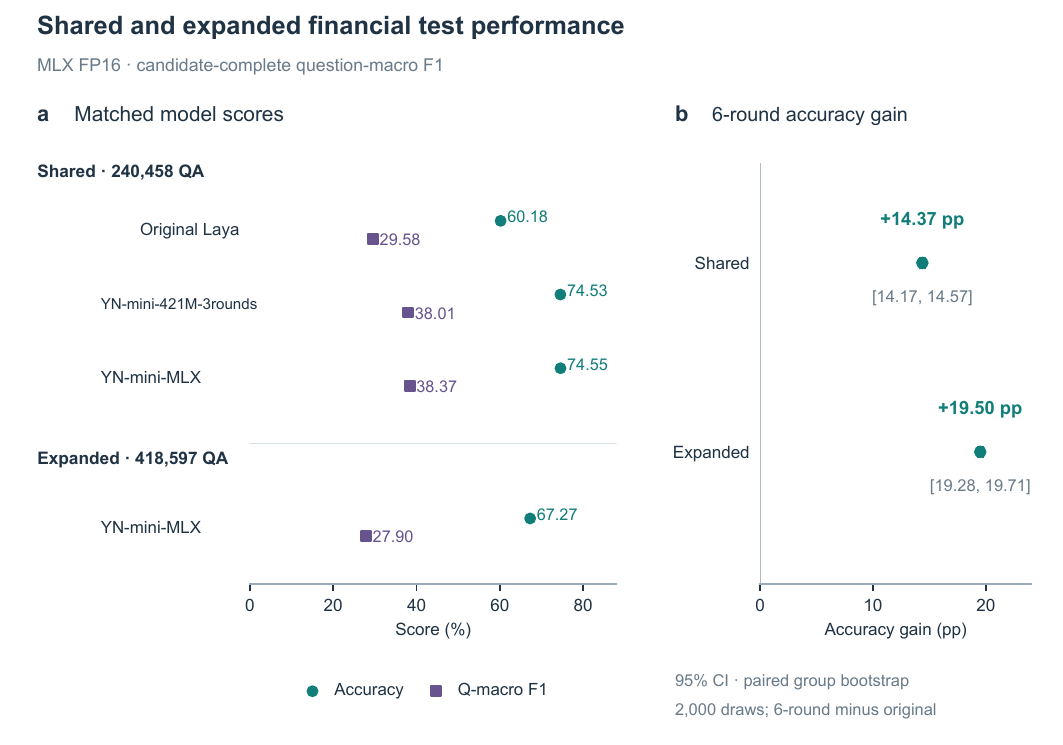}

\par\vspace{5pt}{\fontsize{8.5}{10.3}\selectfont\raggedright Supplementary Figure S2 | MLX performance on shared and expanded financial questions. a, Accuracy and question macro F1 for Original Laya, YN-mini-3rounds and YN-mini on 240,458 shared QA records (119 question versions). The expanded test contains 418,597 records and 862 question versions; its 6-round scores are shown separately. Accuracy weights QA records equally; F1 averages over the complete candidate classes and then equally over question versions. b, Paired accuracy gains of YN-mini over Original Laya for the shared and expanded tests. Error bars show stratified household/respondent-group bootstrap 95\% intervals from 2,000 draws. The shared-test gain is 14.3713 percentage points (14.1686--14.5719).\par}

\end{minipage}

\end{center}

\section*{Appendix S4. Human-score distributions and judge shifts}

\begin{center}

\begin{minipage}{\textwidth}

\centering

\includegraphics[width=\textwidth,height=.66\textheight,keepaspectratio]{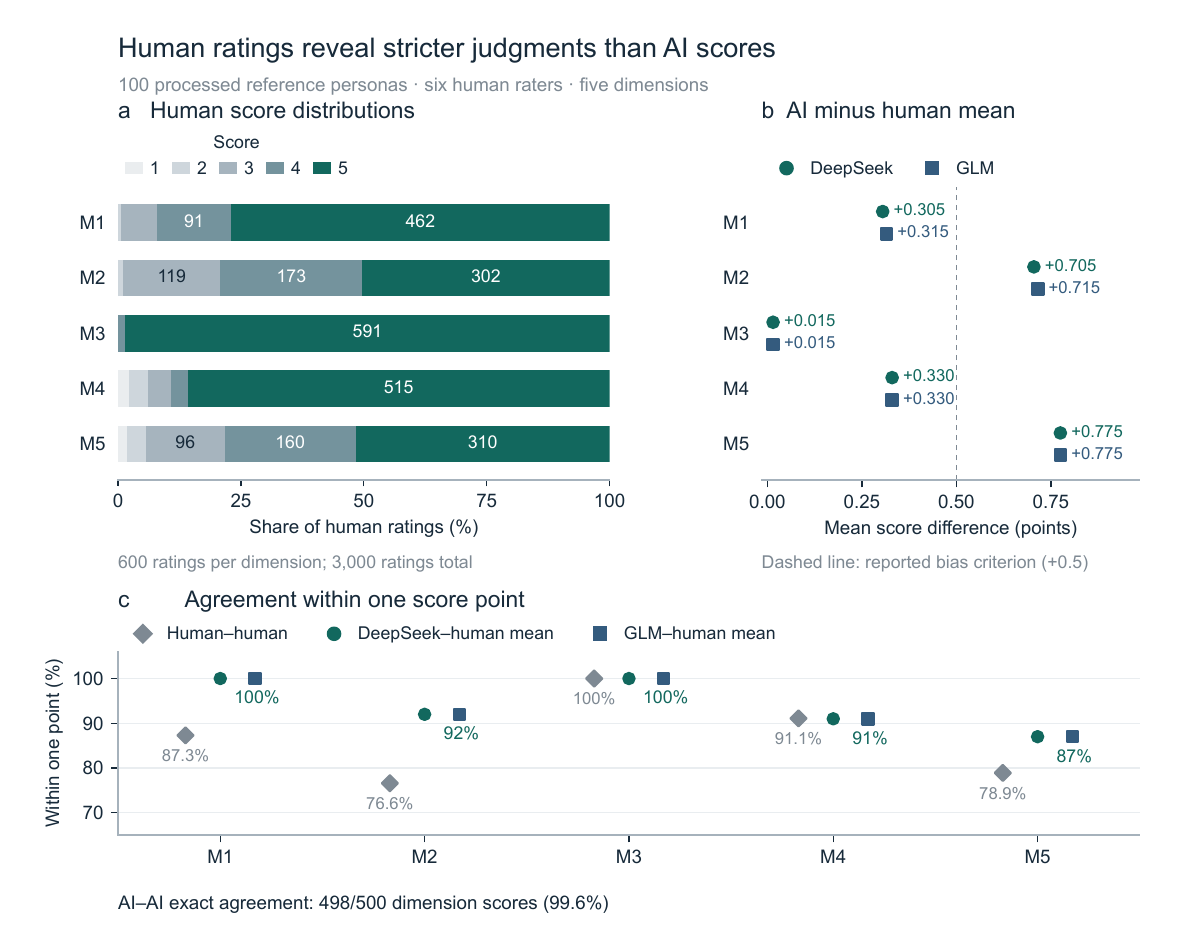}

\par\vspace{5pt}{\fontsize{8.5}{10.3}\selectfont\raggedright Supplementary Figure S3 | Human ratings identify the dimensions on which AI judges are more permissive. a, Rating distributions for 100 fixed processed survey-derived personas, each evaluated by six human raters on five dimensions (600 ratings per dimension; 3,000 total). Numbers inside bars are rating counts. M1 denotes accuracy; M2, absence of over-inference; M3, coverage of nine predefined important fields; M4, internal consistency; and M5, independent overall fidelity. M3 receives 591/600 maximum scores, whereas M2 and M5 have broader distributions. b, DeepSeek (\texttt{deepseek-flash}) and GLM (\texttt{glm-5.3-flash}) mean scores minus human means; the dashed line marks the reported +0.5-point bias criterion. M2 and M5 exceed that criterion. c, Reported within-one-point agreement; each AI is compared with the six-rater mean for each persona (100 comparisons per dimension). AI–AI exact agreement is 498/500 scores. High within-one agreement coexists with systematically higher AI scores.\par}

\end{minipage}

\end{center}

\section*{Appendix S5. Joint structure of the synthetic companion}

\begin{center}

\begin{minipage}{\textwidth}

\centering

\includegraphics[width=\textwidth,height=.66\textheight,keepaspectratio]{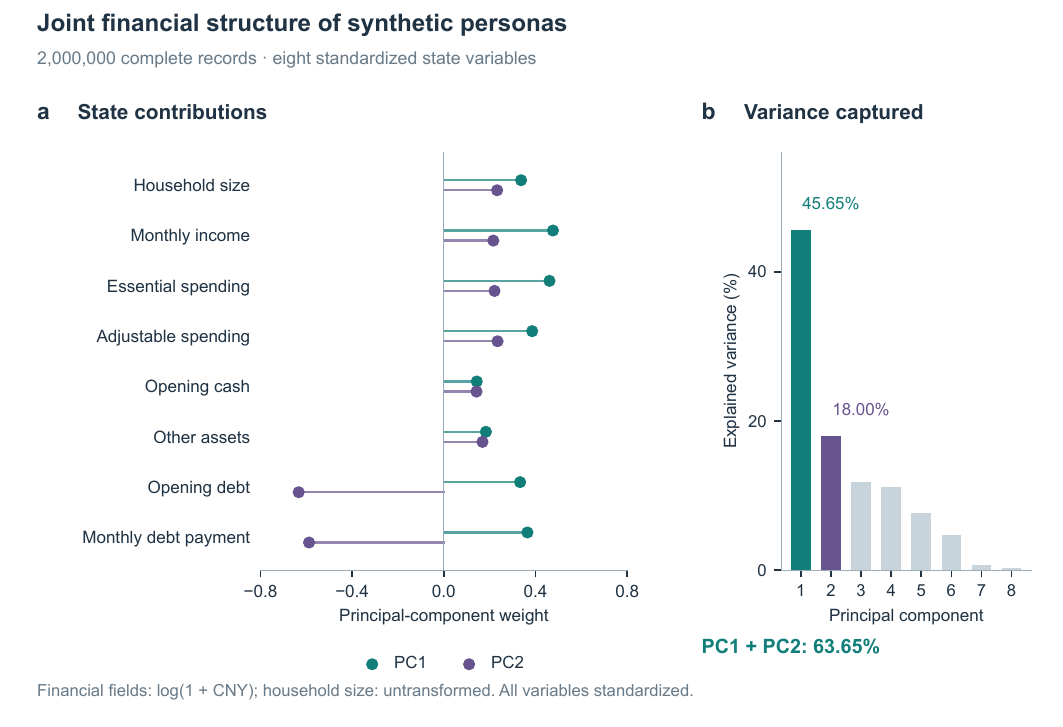}

\par\vspace{5pt}{\fontsize{8.5}{10.3}\selectfont\raggedright Supplementary Figure S4 | Principal-component structure of the independent synthetic persona companion. The analysis includes all 2,000,000 households and eight complete state variables: household size, monthly income, essential and adjustable spending, opening cash, other assets, opening debt and monthly debt payment. The seven financial variables use log(1 + CNY); household size is untransformed, and all eight variables are standardized. a, Unit-length eigenvector weights for PC1 and PC2, with each sign oriented toward positive income. b, Variance explained by each component. PC1 accounts for 45.6450\% and PC2 for 18.0043\%, together 63.6493\%.\par}

\end{minipage}

\end{center}

\section*{Appendix S6. Initialization, resources and matrix settings}

\begin{center}

\begin{minipage}{\textwidth}

\centering

\includegraphics[width=\textwidth,height=.66\textheight,keepaspectratio]{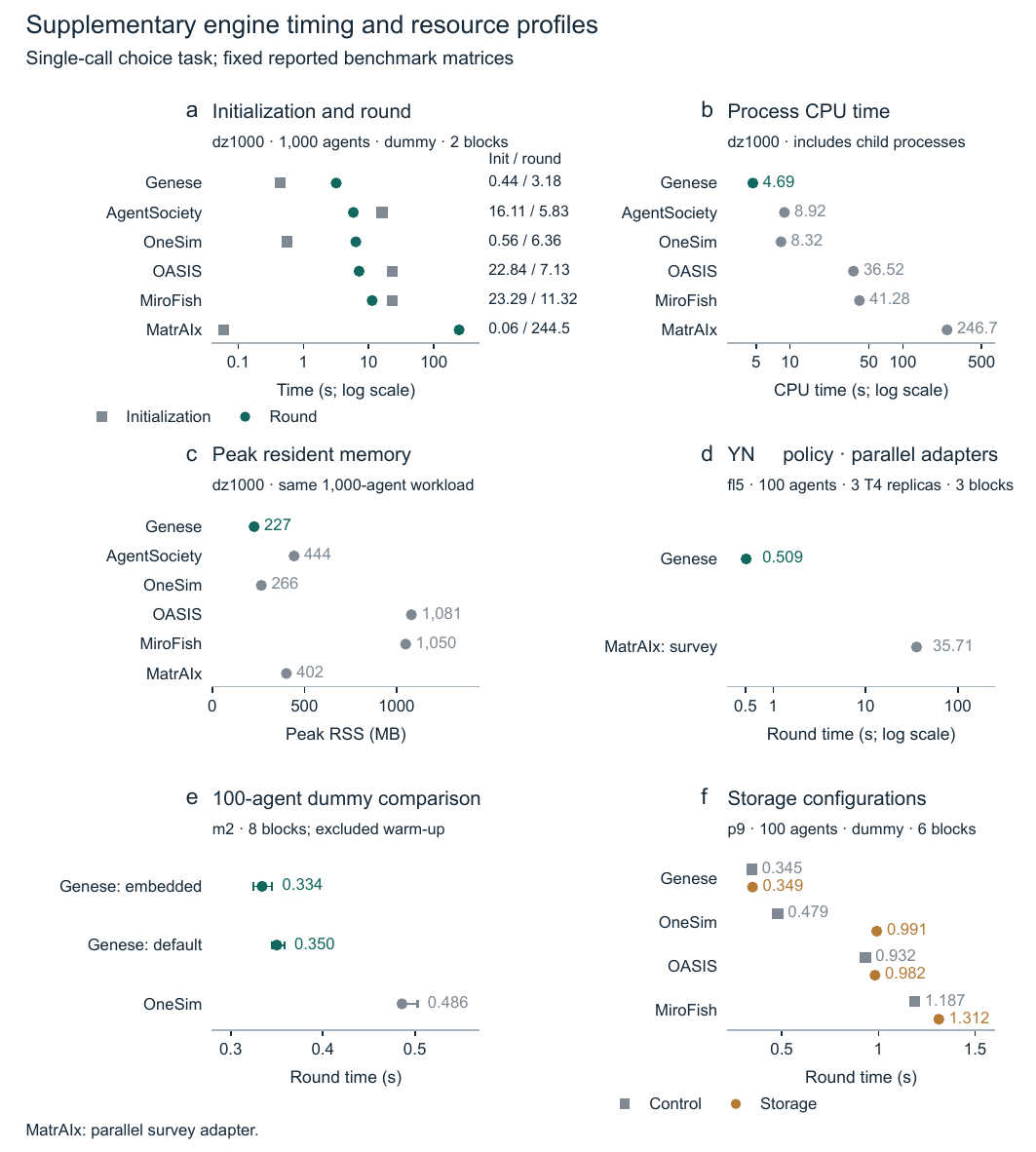}

\par\vspace{5pt}{\fontsize{8.5}{10.3}\selectfont\raggedright Supplementary Figure S5 | Initialization, resources and configuration-specific engine costs. a–c, The dz1000 matrix compares six single-call framework configurations at 1,000 agents per round on dual Xeon Platinum 8268 (96 threads): initialization and round time, process CPU time including child processes, and peak RSS. Genese reports 4.69 s CPU time. d, The separate fl5 YN-mini matrix uses 100 agents and three T4 model replicas; Genese's reference is 0.509 s. e, The m2 dummy matrix contrasts embedded and default Genese with OneSim. f, p9 contrasts framework-specific storage controls: Genese disk versus RAM-disk SQLite FULL + WAL; OneSim PostgreSQL versus memory; OASIS and MiroFish SQLite FULL versus OFF. Points show reported medians. Only panel e shows the three reported 95\% intervals from eight randomized blocks; the other panels show point summaries. Randomized blocks number 2, 3, 8 and 6 for dz1000, fl5, m2 and p9, respectively. Round timing excludes initialization. MatrAIx uses a parallel single-call adapter.\par}

\end{minipage}

\end{center}

Indexed action and partition lookups replace quadratic scans with linear work. Shared action descriptors are decoded once per request. Connection pools scale with concurrency; parallel partitions retain bounded queues. State, events and command results use one durable transaction. The optimization study reduces 1,000-agent rounds from 7.37 to 3.10 seconds and 5,000-agent rounds from 90.14 to 16.66 seconds. At 100 agents, durable and in-memory paths take 0.349 and 0.345 seconds, respectively (Table S11).

\begin{center}
\begin{minipage}{\textwidth}
\fontsize{8.5}{10.3}\selectfont
Table S11 | Execution-path optimization measurements
\par\vspace{4pt}
\begin{tabularx}{\textwidth}{@{}XXXX@{}}
\toprule
Path & Previous implementation & Optimized implementation & Effect \\
\midrule
Action lookup and synchronization & Repeated linear searches; O(N\textsuperscript{2}) total work & Indexed lookup; O(N) work & Lower overhead as the choice batch grows \\
Operation-state queries & Revalidate all agents for each query & Read the relevant shard and state & Avoid repeated cohort-wide scans \\
Decision-request construction & Copy whole requests; decode repeated descriptors per agent & Reduce copying; decode identical descriptors once & Less allocation and serialization \\
Snapshot handling & Copy matrix rows in Python & Compact matrix representation with C-level copying & Lower snapshot cost \\
Model connections & Shared client limited to 100 connections & Scale the pool with concurrency; shard pools above 100 & Higher utilization of model capacity \\
Partition decisions & Process same-session partitions sequentially & Optional parallel partitions with bounded dispatch & Overlap inference without overrunning the service \\
Durable writeback & Multiple independent write operations & One atomic transaction per round & 100-agent storage control: 0.349 s vs 0.345 s in memory-backed storage \\
\bottomrule
\end{tabularx}
\end{minipage}
\end{center}

\begin{center}
\begin{minipage}{\textwidth}
\fontsize{8.5}{10.3}\selectfont
Table S12 | Native execution paths and tested adapters
\par\vspace{4pt}
\begin{tabularx}{\textwidth}{@{}XXXX@{}}
\toprule
Framework & Native execution organization & State persistence & Tested choice adapter \\
\midrule
Genese & Partitioned World/Brain requests; bounded queues and pooled model connections & Production PostgreSQL: world state, event outbox and command result in one transaction & Accepted option actions and World writeback \\
OneSim & Event-driven agents; distributed workers and reusable model/connection resources & PostgreSQL environment and agent states when the data module is enabled & One decision event per sampled agent \\
OASIS & Asynchronous agents share a platform Channel & Platform SQLite; action-level commits & Question-to-platform action mapping \\
MiroFish & Simulation subprocess; concurrent OASIS platform coroutines & Per-platform SQLite and action logs & Native OASIS simulation path \\
AgentSociety v2 & Ray worker batches and local adaptive LLM admission & Per-step workspace checkpoints and replay JSONL & Single-worker Ray choice task \\
MatrAIx & Concurrent independent Harbor trials and resumable jobs & Trial results and completed-job progress & Added threaded questionnaire trial \\
\bottomrule
\end{tabularx}
\par\vspace{3pt}{\fontsize{8}{9.5}\selectfont Genese uses PostgreSQL for production state transitions.}
\end{minipage}
\end{center}

Capability labels describe the inspected native execution paths. Yes means the framework implements the defined mechanism, including through a configurable native module. Partial means only part of that function is implemented, and N/D means the mechanism is not documented. Array-based ECS requires indexed numeric component arrays. Synchronized state persistence requires a database transaction or explicit synchronized writes protecting saved state. OneSim's enabled PostgreSQL path qualifies. State saving and restart support are evaluated separately from synchronization and whole-world atomicity.

Native action-rule enforcement requires programmatic execution constraints on business state, permissions or spatial movement. OASIS and MiroFish enforce platform business rules; AgentSociety enforces permissions and workspace boundaries. Event registration or answer-format checks alone receive Partial. Policy-intent/world-effect separation requires the engine to resolve a proposed intent before applying authoritative effects. Atomic state–event commit couples world state, outbox and command result. The combined isolation/rollback row receives Partial for AgentSociety's independent agent tasks and MatrAIx's independent trials, which isolate failures without rolling back the complete world step.

Recovery/resume requires a native restart path from saved state or completed-task progress: AgentSociety checkpoints and MatrAIx job progress qualify. Execution/connection pooling includes framework-managed reusable workers, connections or model instances; OneSim's connection pools and AgentSociety's Ray workers qualify. Task admission/flow control requires waiting for execution capacity or reducing admission under overload. OASIS and MiroFish gate LLM calls, AgentSociety uses local adaptive admission, and MatrAIx gates trials. Decision/trial concurrency control requires a configured or adaptive execution allowance. OneSim's event queues and distributed coordination receive Partial because the inspected decision path launches tasks without such an allowance. Local admission does not establish a globally bounded waiting queue.

The comparison uses versioned native sources: \href{https://github.com/RUC-GSAI/YuLan-OneSim/tree/9829d72}{OneSim 9829d72}, \href{https://github.com/camel-ai/oasis/tree/dded86a}{OASIS dded86a}, \href{https://github.com/666ghj/MiroFish/tree/7657031}{MiroFish 7657031} with OASIS 0.2.5, \href{https://github.com/tsinghua-fib-lab/agentsociety/tree/8cc5bb9}{AgentSociety 8cc5bb9}, \href{https://github.com/MatrAIx-ai/MatrAIx-Persona-8B/tree/3633d8d}{MatrAIx 3633d8d}.

\begin{center}
\begin{minipage}{\textwidth}
\fontsize{8.5}{10.3}\selectfont
Table S13 | Native execution capabilities
\par\vspace{4pt}
\begin{tabularx}{\textwidth}{@{}>{\hsize=2.200000\hsize\linewidth=\hsize\raggedright\arraybackslash}X>{\hsize=0.640000\hsize\linewidth=\hsize\raggedright\arraybackslash}X>{\hsize=0.800000\hsize\linewidth=\hsize\raggedright\arraybackslash}X>{\hsize=0.800000\hsize\linewidth=\hsize\raggedright\arraybackslash}X>{\hsize=0.800000\hsize\linewidth=\hsize\raggedright\arraybackslash}X>{\hsize=0.960000\hsize\linewidth=\hsize\raggedright\arraybackslash}X>{\hsize=0.800000\hsize\linewidth=\hsize\raggedright\arraybackslash}X@{}}
\toprule
Capability & Genese & OneSim & OASIS & MiroFish & AgentSociety & MatrAIx \\
\midrule
Array-based ECS storage & Yes & N/D & N/D & N/D & N/D & N/D \\
Synchronized state persistence & Yes & Yes & No & No & No & No \\
Native action-rule enforcement & Yes & Partial & Yes & Yes & Yes & Partial \\
Policy-intent / world-effect separation & Yes & No & No & No & No & No \\
Atomic state–event commit & Yes & No & No & No & No & No \\
World-step isolation / rollback & Yes & No & No & No & Partial & Partial \\
Recovery / resume & Yes & No & No & No & Yes & Yes \\
Execution / connection pooling & Yes & Yes & No & No & Yes & No \\
Task admission / flow control & Yes & Partial & Yes & Yes & Yes & Yes \\
Decision / trial concurrency control & Yes & Partial & Yes & Yes & Yes & Yes \\
\bottomrule
\end{tabularx}
\par\vspace{3pt}{\fontsize{8}{9.5}\selectfont Action-rule scope: OASIS/MiroFish platform rules and AgentSociety permissions/workspace. Resume scope: checkpoints (AgentSociety) and jobs (MatrAIx). Admission covers native LLM or trial entrypoints. OneSim retains Partial for event/interface checks and decision-flow control; MatrAIx retains Partial for questionnaire-output checks. AgentSociety and MatrAIx isolate tasks but do not roll back a complete world step.}
\end{minipage}
\end{center}

\begin{center}
\begin{minipage}{\textwidth}
\fontsize{8.5}{10.3}\selectfont
Table S14 | Resource pooling and native persistence
\par\vspace{4pt}
\begin{tabularx}{\textwidth}{@{}>{\hsize=0.570000\hsize\linewidth=\hsize\raggedright\arraybackslash}X>{\hsize=0.330000\hsize\linewidth=\hsize\raggedright\arraybackslash}X>{\hsize=2.100000\hsize\linewidth=\hsize\raggedright\arraybackslash}X@{}}
\toprule
Framework & Pooling & State saving and synchronization \\
\midrule
Genese production & Yes & PostgreSQL; one transaction for world head/state, event outbox and command result \\
OneSim & Yes & PostgreSQL state saving when the data module is enabled; reusable connection pools \\
OASIS & No & SQLite action commits; native synchronous=OFF \\
MiroFish & No & OASIS SQLite action commits and simulation logs; native synchronous=OFF \\
AgentSociety & Yes & Ray execution workers; workspace checkpoints and replay JSONL; atomic file replacement without explicit fsync \\
MatrAIx & No & Trial results and job-progress files; restart continues remaining trials; ordinary file writes \\
\bottomrule
\end{tabularx}
\par\vspace{3pt}{\fontsize{8}{9.5}\selectfont Pooling includes framework-managed execution or connection resources. SQLite entries describe action-level saving with the inspected native synchronization setting. AgentSociety checkpoints use temporary-file replacement without an explicit fsync barrier. Saved-state continuation, synchronized persistence and atomic world-state transactions are separate mechanisms. Benchmark-added thread pools are excluded from native pooling.}
\end{minipage}
\end{center}

\section*{Appendix S7. Stability of the population economic axis}

\begin{center}
\begin{minipage}{\textwidth}
\fontsize{8.5}{10.3}\selectfont
Table S15 | Economic-axis stability across 50 population shards
\par\vspace{4pt}
\begin{tabularx}{\textwidth}{@{}XX@{}}
\toprule
Quantity & 500M expansion \\
\midrule
Complete records in the 14-field PCA & Approximately 4.626M per shard \\
PC1 variance explained & 32.62\% (shard mean) \\
Largest across-shard loading SD & At most 0.0003 \\
\bottomrule
\end{tabularx}
\par\vspace{3pt}{\fontsize{8}{9.5}\selectfont Standardized ordinal attributes and a positive-expenditure PC1 sign are used within each resource. Values summarize 50 constructed-population shards.}
\end{minipage}
\end{center}

\section*{Appendix S8. Randomized immediate-backend execution at 100 agents}

\begin{center}

\begin{minipage}{\textwidth}

\centering

\includegraphics[width=\textwidth,height=.66\textheight,keepaspectratio]{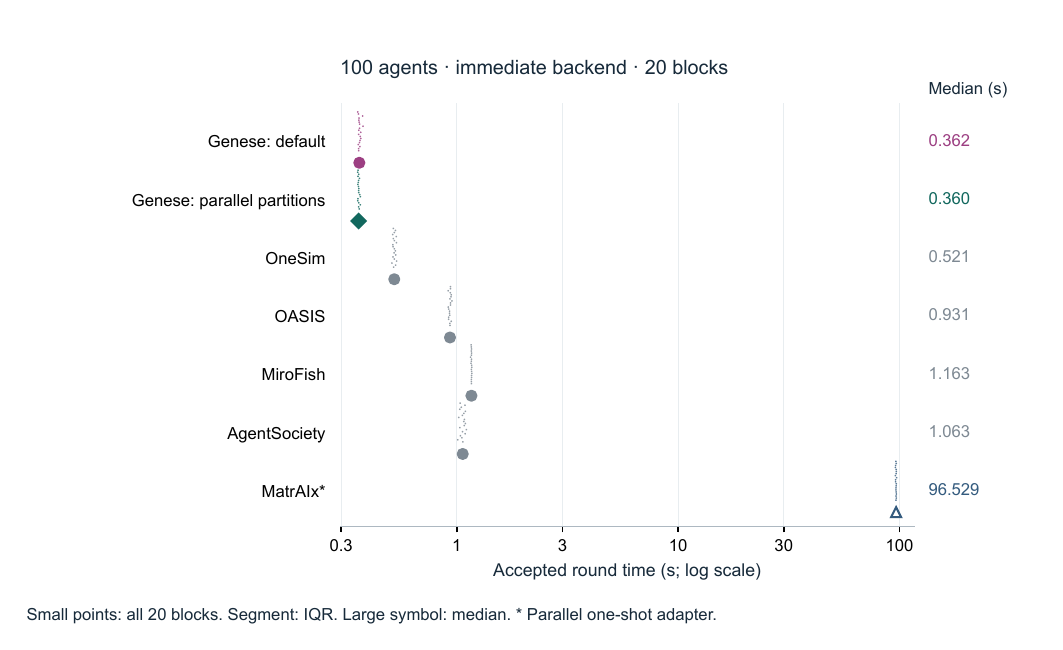}

\par\vspace{5pt}{\fontsize{8.5}{10.3}\selectfont\raggedright Supplementary Figure S6 | Complete 100-agent immediate-backend runtime distribution. Seven configured execution arms each contribute twenty randomized interleaved measured blocks after one warm-up. Every round accepts and writes back all 100 choices. Small points show individual times with vertical offsets by block number; large symbols mark medians and thick segments the inclusive 25th–75th percentiles. Configurations follow protocol order. Genese default and parallel partitions are distinct; an asterisk and open triangle identify the parallel MatrAIx one-shot adapter. The host has two Intel Xeon Platinum 8268 processors and 96 logical threads. Framework CPU50 and proxy CPUs60–63 have nonexclusive affinities during concurrent training; the isolated comparison directory is memory-backed. Initialization is measured separately, with medians of 0.157 seconds for Genese default and 0.158 seconds for parallel partitions. Source Data includes all 280 measured cells across both agent counts, fourteen warm-ups, initialization, process time, throughput and observed host load.\par}

\end{minipage}

\end{center}

\section*{Supplementary table: known-error judge sensitivity}

\begin{center}
\begin{minipage}{\textwidth}
\fontsize{8.5}{10.3}\selectfont
Table S16 | Responses to injected persona errors
\par\vspace{4pt}
\begin{tabularx}{\textwidth}{@{}XXXXXX@{}}
\toprule
Error / target & Judge & Valid n & Score ≤3 & Paired decline ≥1 & Non-target specificity \\
\midrule
Changed fact / M1 & DeepSeek Flash & 25 & 64\% & 72\% & 98.7\% \\
Changed fact / M1 & GLM Flash & 23 & 91.3\% & 100\% & 89.9\% \\
Unsupported addition / M2 & DeepSeek Flash & 25 & 0\% & 44\% & 100\% \\
Unsupported addition / M2 & GLM Flash & 24 & 79.2\% & 95.8\% & 97.2\% \\
Omitted target / M3 & DeepSeek Flash & 24 & 0\% & 8.3\% & 100\% \\
Omitted target / M3 & GLM Flash & 24 & 100\% & 100\% & 100\% \\
Contradiction / M4 & DeepSeek Flash & 25 & 44\% & 76\% & 100\% \\
Contradiction / M4 & GLM Flash & 25 & 96\% & 96\% & 58\% \\
\bottomrule
\end{tabularx}
\par\vspace{3pt}{\fontsize{8}{9.5}\selectfont Each of 100 personas receives one known error, with 25 per type; the 44-field reference is unchanged. Rates use each judge’s valid responses. Specificity is the proportion of non-target dimension scores remaining ≥4, under the protocol’s exclusions.}
\end{minipage}
\end{center}

\section*{Supplementary analysis: actual consumer-bank structure}

\begin{center}
\begin{minipage}{\textwidth}
\fontsize{8.5}{10.3}\selectfont
Table S17 | Ordinal PC1 loadings in the actual 10M bank
\par\vspace{4pt}
\begin{tabularx}{\textwidth}{@{}XX@{}}
\toprule
Field & 10M bank \\
\midrule
household\_size & 0.037 \\
age\_group & -0.156 \\
education & 0.204 \\
household\_income\_bracket & 0.345 \\
spending.total & 0.429 \\
spending.food & 0.341 \\
spending.clothing & 0.297 \\
spending.housing & 0.298 \\
spending.transport\_comm & 0.315 \\
spending.education & 0.185 \\
spending.health & 0.182 \\
spending.other & 0.335 \\
financial\_asset\_bracket & 0.170 \\
debt\_amount\_bracket & 0.168 \\
\bottomrule
\end{tabularx}
\par\vspace{3pt}{\fontsize{8}{9.5}\selectfont The bank is standardized on the same 14 ordinal attributes. Unknown values are jointly excluded; PC1 sign is oriented toward positive total spending. Complete records: 7,227,206. PC1 explains 32.8904\% and PC2 11.5234\% of variance.}
\end{minipage}
\end{center}

\begin{center}
\begin{minipage}{\textwidth}
\fontsize{8.5}{10.3}\selectfont
Table S18 | Representation completeness and consistency
\par\vspace{4pt}
\begin{tabularx}{\textwidth}{@{}XXX@{}}
\toprule
Property & Records & Share of 10M bank \\
\midrule
Known income, assets and debt brackets & 10,000,000 & 100\% \\
Known total-spending bracket & 9,499,649 & 94.9965\% \\
One-person household with a child & 0 & 0\% \\
Decision maker aged 65+ with explicit no-elder flag & 0 & 0\% \\
Debt amount/type none mismatch (known fields) & 0 & 0\% \\
Personal-income bracket above household-income bracket & 82,864 & 0.8286\% \\
\bottomrule
\end{tabularx}
\par\vspace{3pt}{\fontsize{8}{9.5}\selectfont Consistency counts use the consumer bank. Personal-income comparisons require both income fields to be known. Household financial assignments use household income.}
\end{minipage}
\end{center}

The delivered consumer-bank reader selects the first 10,000,000 distinct rows in shard order from its documented 100M parent. Selection requires known financial-asset bracket, debt amount, debt type, investment activity and general trust. The separate 500M expansion applies additional calibration and is evaluated as a population resource.

\section*{Supplementary analysis: training trajectory and prediction controls}

\subsection*{Training procedure and input lengths}

SFT minimizes cross-entropy against the observed answer or coded target. Native Laya RLCD adds a policy-gradient term over four zero-mean Gaussian logit perturbations to this cross-entropy \href{https://github.com/NandhaKishorM/laya/blob/main/laya/train.py}{[36]}. Proper scoring rules motivate its probability-based rewards \href{https://doi.org/10.1198/016214506000001437}{[39]}. Choice examples receive a clipped log-score and spherical-score reward, with spherical weight 0.75. Each example's sample rewards are centered, their advantages are standardized, and those advantages weight the perturbed logits' Gaussian log-density. Noise decreases linearly from 0.4 to 0.1. The ranked-probability reward is used for Score-typed targets; the survey examples use Choice targets.

Training uses AdamW, weight decay 0.01, gradient-norm clipping at 1.0, 5\% linear warm-up and cosine decay to 5\% of the initial learning rate. Separate encoder and head rates, batch sizes and stage seeds appear in Table S19 and Table S21. Option order is shuffled during training. Checkpoint temperatures are fitted on the independent calibration split. Development question-macro negative log-likelihood records checkpoint selection before test evaluation; the released 6-round model is final rlcd-2.

The QA training encoding uses 512 total tokens, a 256-token question/options allocation and up to 48 tokens per option. One-shot financial deployment uses a 2,048-token complete input, a 1,024-token question/options allocation and up to 192 tokens per option so that monetary paths and repayment terms remain in the model input. The original option markers, weights, probability calibration and decoder are retained. The 12M purchase-funding application processes a mean 529.98 actual nonpadding candidate input tokens per choice.

Two independent full continuation replicas repeated the 6-round recipe from the same completed 3-round initialization. Each processed 2,925,263 training QA in one SFT pass and one RLCD pass, using six cards, effective batch 768 and 3,809 optimizer updates per stage. Stage-specific temperatures were fitted on the fixed calibration split. All three runs evaluated the same 119-version, 240,458-QA financial-question test with RLCD fixed as the final stage.

\begin{center}
\begin{minipage}{\textwidth}
\fontsize{8.5}{10.3}\selectfont
Table S19 | Continuation for YN-mini: training configuration
\par\vspace{4pt}
\begin{tabularx}{\textwidth}{@{}XXX@{}}
\toprule
Parameter & SFT continuation & Native RLCD continuation \\
\midrule
Starting checkpoint & 3-round rlcd-1 & 6-round sft-3 \\
Training examples / epoch & 2,925,263 & 2,925,263 \\
Epochs & 1 & 1 \\
Precision / hardware & BF16 / 6 \ensuremath{\times} A100 PCIe 40GB & BF16 / 6 \ensuremath{\times} A100 PCIe 40GB \\
Micro-batch / accumulation & 32 / 4 & 32 / 4 \\
Effective batch & 768 & 768 \\
Encoder / head learning rate & 1.25 \ensuremath{\times} 10\textsuperscript{-}\textsuperscript{5} / 5 \ensuremath{\times} 10\textsuperscript{-}\textsuperscript{5} & 5 \ensuremath{\times} 10\textsuperscript{-}\textsuperscript{6} / 2 \ensuremath{\times} 10\textsuperscript{-}\textsuperscript{5} \\
Input / head length & 512 / 256 & 512 / 256 \\
RLCD samples; sigma; weights & — & 4; 0.4 \ensuremath{\rightarrow} 0.1; w\_sph 0.75, w\_rps 1.0 \\
\bottomrule
\end{tabularx}
\par\vspace{3pt}{\fontsize{8}{9.5}\selectfont Training sequence: original-base \ensuremath{\rightarrow} sft-1 \ensuremath{\rightarrow} sft-2 \ensuremath{\rightarrow} rlcd-1 \ensuremath{\rightarrow} sft-3 \ensuremath{\rightarrow} rlcd-2. 3-round and bilingual training use 1,721,623 training observations per language.}
\end{minipage}
\end{center}

\begin{center}
\begin{minipage}{\textwidth}
\fontsize{8.5}{10.3}\selectfont
Table S20 | 6-round training trajectory on expanded QA
\par\vspace{4pt}
\begin{tabularx}{\textwidth}{@{}XXX@{}}
\toprule
Checkpoint & Accuracy (\%) & Question-macro F1 (\%) \\
\midrule
Original Laya & 47.8324 & 18.1810 \\
SFT1 & 44.1382 & 18.2350 \\
SFT2 & 57.4760 & 19.3338 \\
RLCD1 / 3-round release & 56.3657 & 19.2395 \\
SFT3 / continuation & 67.2152 & 27.6787 \\
RLCD2 / 6-round release & 67.2728 & 27.8355 \\
\bottomrule
\end{tabularx}
\par\vspace{3pt}{\fontsize{8}{9.5}\selectfont Native PyTorch; 418,597 expanded QA. Each checkpoint has independently fitted calibration temperatures. The SFT continuation stage contributes most of the gain from YN-mini-3rounds to YN-mini; final RLCD adds 0.0576 accuracy and 0.1568 F1 points along this training trajectory.}
\end{minipage}
\end{center}

\begin{center}
\begin{minipage}{\textwidth}
\fontsize{8.5}{10.3}\selectfont
Table S21 | training-stage configuration
\par\vspace{4pt}
\begin{tabularx}{\textwidth}{@{}XXXXX@{}}
\toprule
Stage & Unique training QA & Replay QA & Effective batch & Encoder / head LR \\
\midrule
SFT1; 21 versions & 307,807 & 0 & 192 & 2.5e-5 / 1.0e-4 \\
SFT2; 119 versions & 1,721,623 & 43,935 & 768 & 1.25e-5 / 5.0e-5 \\
RLCD1 / 3-round release & 1,721,623 & 43,935 & 768 & 5.0e-6 / 2.0e-5 \\
SFT3 / continuation & 2,925,263 & 0 & 768 & 1.25e-5 / 5.0e-5 \\
RLCD2 / 6-round release & 2,925,263 & 0 & 768 & 5.0e-6 / 2.0e-5 \\
\bottomrule
\end{tabularx}
\par\vspace{3pt}{\fontsize{8}{9.5}\selectfont All stages use one pass through the unique training examples. Initial-supervision stages replay supported questions with 200–3,999 training examples up to 4,000 per question. Expanded-supervision stages have no replay. SFT1 uses microbatch 16 and accumulation 2 across six GPUs; later stages use 32 and 4. Continuation SFT/RLCD seeds: 202610120/202610121. AdamW weight decay 0.01; clipping 1.0; 5\% warm-up, cosine decay to 5\% of initial LR.}
\end{minipage}
\end{center}

\begin{center}
\begin{minipage}{\textwidth}
\fontsize{8.5}{10.3}\selectfont
Table S22 | Matched-control F1 by question-registry domain
\par\vspace{4pt}
\begin{tabularx}{\textwidth}{@{}XXXXXXX@{}}
\toprule
Domain & QA & Versions & Base & No profile & Wrong profile & Full YN-mini \\
\midrule
risk and investment attitudes & 1490 & 7 & 20.88 & 20.24 & 22.69 & 28.11 \\
financial product saving and credit behavior & 10325 & 56 & 44.34 & 45.63 & 46.90 & 50.70 \\
well being & 642 & 4 & 9.81 & 13.38 & 16.68 & 17.61 \\
income change & 149 & 3 & 16.89 & 16.32 & 19.02 & 20.31 \\
pension trust & 826 & 4 & 7.23 & 11.39 & 16.13 & 20.18 \\
financial knowledge & 2428 & 11 & 7.62 & 18.14 & 22.36 & 30.01 \\
financial education & 530 & 2 & 53.95 & 48.14 & 50.67 & 52.93 \\
retirement preference & 517 & 2 & 32.18 & 27.12 & 31.46 & 36.33 \\
financial attention & 928 & 4 & 7.69 & 12.23 & 16.69 & 22.91 \\
retirement planning & 528 & 2 & 37.99 & 37.99 & 39.41 & 38.81 \\
retirement responsibility & 591 & 3 & 8.52 & 12.14 & 24.12 & 28.54 \\
social trust & 195 & 2 & 13.73 & 12.73 & 13.70 & 15.31 \\
bank service satisfaction & 284 & 2 & 14.56 & 12.93 & 15.16 & 13.41 \\
social fairness & 67 & 1 & 4.86 & 12.93 & 12.93 & 12.93 \\
insurance participation & 265 & 4 & 20.07 & 30.30 & 30.33 & 34.92 \\
perceived economic status & 66 & 1 & 20.65 & 13.60 & 13.93 & 18.77 \\
wage structure & 15 & 1 & 6.25 & 7.95 & 7.95 & 8.33 \\
cash flow & 15 & 1 & 23.19 & 21.21 & 42.74 & 33.17 \\
work schedule & 74 & 5 & 18.43 & 23.96 & 28.20 & 29.16 \\
work satisfaction & 15 & 1 & 25.91 & 13.91 & 13.91 & 20.26 \\
income satisfaction & 21 & 1 & 5.56 & 6.41 & 8.03 & 11.41 \\
liquidity & 15 & 1 & 20.10 & 10.53 & 19.58 & 13.81 \\
work compensation & 14 & 1 & 10.00 & 16.67 & 25.14 & 21.90 \\
\bottomrule
\end{tabularx}
\par\vspace{3pt}{\fontsize{8}{9.5}\selectfont Question-macro F1 (\%), on matched selected QA. Results are aggregated by domain. The three financial families are defined in M4.}
\end{minipage}
\end{center}

\section*{Appendix S9. Auxiliary financial workflow evaluation}

A separate controlled behavioral study contains 1,120 decisions with DeepSeek Flash as actor. Twenty personas drawn from a 1,000-persona parent cohort complete four financial tasks under two conditions. Single-pass choice (SP), candidate generation with self-check (SC), and multi-agent planning, constraint checking and execution use one, two and three model calls per decision. All 480 workflow decisions return legal options (Table S23).

Strict adherence requires all hard ledger constraints and the assigned persona goal. Across 160 decisions per workflow, adherence is 67.5\% for SP, 81.9\% for SC and 86.9\% for multi-agent workflow. Hard-constraint violations are 23.1\%, 14.4\% and 6.9\%. Multi-agent workflow improves adherence over SP by 19.4 percentage points, with a persona-clustered exploratory 95\% bootstrap interval of 13.1–25.0; the multi-agent workflow–SC difference is 5.0 points, with an interval of -1.9 to 11.9 (Supplementary Figure S7a).

Borrowing-amount adherence rises from 22.5\% with SP to 77.5\% with SC and 90.0\% with multi-agent workflow. Consumption adjustment rises from 57.5\% to 65.0\% and 80.0\%. Risk allocation stays near 95\%; debt-payment allocation is 95.0\%, 87.5\% and 82.5\%. Additional decision steps are most useful for borrowing and consumption in this task set (Supplementary Figure S7c).

A second study fixes ledger and options while changing one goal at a time in 20 constructed personas. Four axes, two polarities, two workflows and two repeats produce 640 decisions and 160 pairs per workflow. Multi-agent workflow changes choices in 96.2\% of pairs, compared with 77.5\% for SC. Both ends satisfy their goals in 66.2\% and 58.1\% of pairs. Persona-clustered exploratory intervals are 15.0–22.5 points for the 18.8-point response difference and 1.2–15.0 for the 8.1-point attainment difference (Supplementary Figure S7b).

On the borrowing-versus-purchase axis, multi-agent workflow changes choices in 87.5\% of pairs and attains both targets in 60.0\%; SC reaches 10.0\% for both metrics. On the loss-cap axis, both workflows change every choice, while two-end attainment is 10.0\% for multi-agent workflow and 22.5\% for SC. Goal sensitivity and successful goal attainment capture different aspects of persona-conditioned financial behavior.

\begin{center}

\begin{minipage}{\textwidth}

\centering

\includegraphics[width=\textwidth,height=.66\textheight,keepaspectratio]{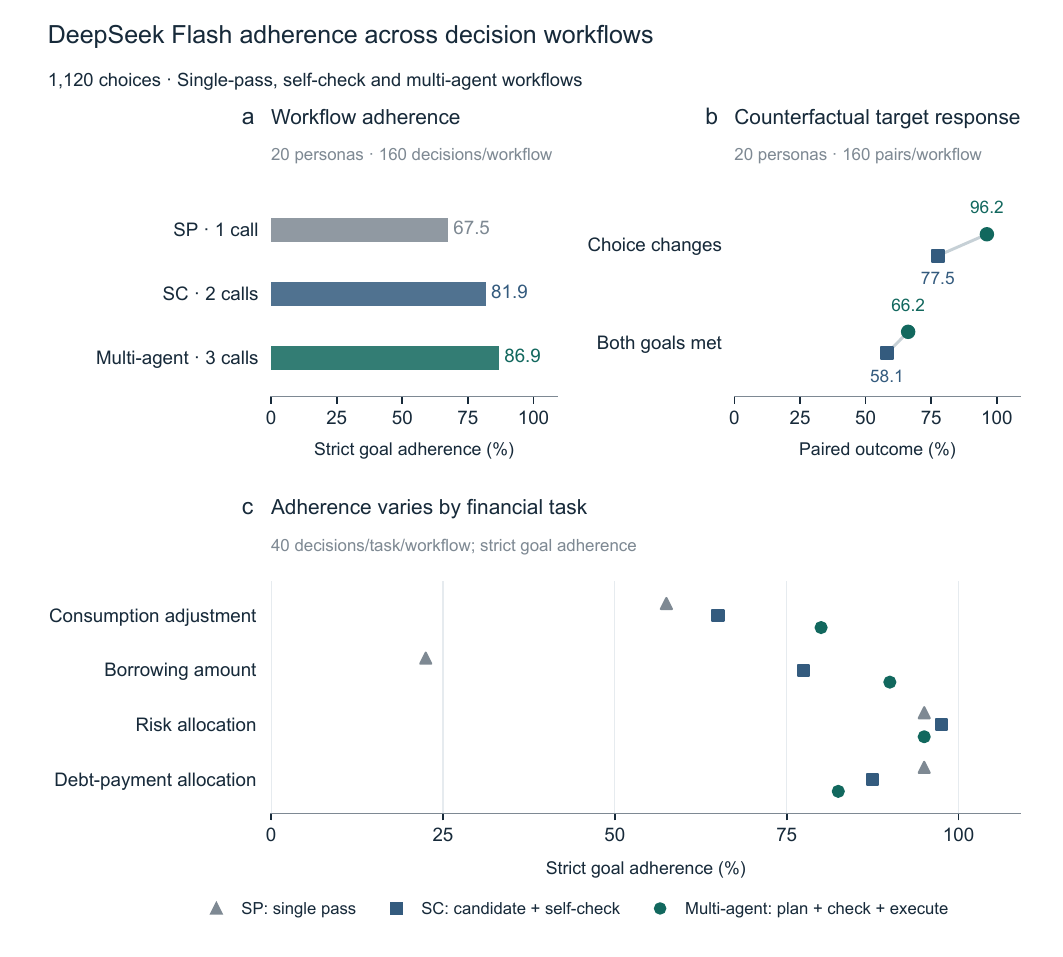}

\par\vspace{5pt}{\fontsize{8.5}{10.3}\selectfont\raggedright Supplementary Figure S7 | DeepSeek financial workflows and counterfactual goal response. DeepSeek Flash is the actor in both experiments. a, Strict adherence requires every hard financial constraint and persona goal. The workflow experiment uses 20 personas, four tasks and two conditions, yielding 160 choices for each of SP, SC and multi-agent workflow, with one, two and three model calls per choice. b, The counterfactual experiment changes one goal while holding the ledger and options fixed. It uses 20 constructed personas, four goal axes, two polarities and two repeats: 320 choices and 160 pairs per workflow. Choice response means the selected option changes; two-end attainment means both choices satisfy their respective goals. c, The four-task breakdown locates gains in consumption and borrowing decisions and shows the variation in risk and debt-payment tasks. Percentages are the deterministic-rule summaries; persona-clustered exploratory confidence intervals for paired differences are reported in the text.\par}

\end{minipage}

\end{center}

\begin{center}
\begin{minipage}{\textwidth}
\fontsize{8.5}{10.3}\selectfont
Table S23 | Financial workflow comparison
\par\vspace{4pt}
\begin{tabularx}{\textwidth}{@{}XXXXXX@{}}
\toprule
Workflow & Calls / choice & Choices & Strict adherence & Hard violation & Strong pass \\
\midrule
SP: single pass & 1 & 160 & 67.5\% & 23.1\% & 75.0\% \\
SC: candidate + self-check & 2 & 160 & 81.9\% & 14.4\% & 83.8\% \\
Multi-agent: plan + check + execute & 3 & 160 & 86.9\% & 6.9\% & 89.4\% \\
\bottomrule
\end{tabularx}
\par\vspace{3pt}{\fontsize{8}{9.5}\selectfont Actor: DeepSeek Flash; 20 personas, four tasks and two conditions. Strict adherence requires all hard and soft persona goals. Strong pass requires a valid action, no critical violation, and both valid judges rating persona adherence (PA) at least 4/5; PA = 4 permits one secondary goal to be unmet. Missing judge responses count as non-passes in the 160-choice denominator.}
\end{minipage}
\end{center}

\section*{Appendix S10. Matched downstream adaptation and training costs}

\subsection*{Data, checkpoints and candidate classification}

The 50,000-example training set is nested in the 200,000-example set. The fixed checkpoint-selection development set contains 20,000 examples and 829 question versions; the method-selection partition contains all 412,873 development examples and the same 829 versions. The final shared test contains 240,458 examples, 119 question versions and 8,275 respondent groups. Both accuracy and question-macro F1 retain the entire test denominator. Candidate-class F1 includes every defined answer class, including classes absent from predictions, and versions receive equal weight. All ten formal test versions have zero invalid predictions.

Qwen candidate contexts independently encode the full state, question and alternatives, with the candidate under evaluation marked explicitly. The final nonpadding representation is converted to FP32, L2-normalized and scored by one shared bias-free scalar projection followed by candidate softmax. Head-only training freezes the backbone and reuses cached features; its learning rate is 0.003, batch size 4,096 and early-stopping patience three epochs, from a maximum of twenty. LoRA rank 16, $\alpha=32$ and zero dropout updates adapters together with the same head under candidate-classification cross-entropy. AdamW uses weight decay 0.01, adapter/head rates of $10^{-4}/10^{-3}$, 3\% warm-up and linear decay to zero, gradient clipping 1.0 and seed 20261018. One pass uses eight ranks, microbatch one and accumulation four: global batch 32. Qwen3.5-4B-Base and Qwen3.5-9B are different upstream initializations. The complete development comparison selected LoRA for 9B; training budgets and methods were fixed before the final test.

Native SFT and its RLCD/CE continuations preserve ModernBERT-large, the upstream two-layer Transformer decision head, tokenizer and option-marker scoring interface. The two SFT parents selected their terminal 50,000- and 200,000-exposure checkpoints. Continued stages reset AdamW and the scheduler, use the same seed and input order, and allow one complete additional pass with global batch 768 over eight ranks. Encoder and head learning rates are $5\times10^{-6}$ and $2\times10^{-5}$, respectively; warm-up occupies 5\% of updates followed by cosine decay. Native RLCD uses four noisy samples, Gaussian logit noise annealed from 0.4 to 0.1 and spherical-reward weight 0.75. The general objective's ranked-probability weight is 1, but that term is inactive for the Choice-typed examples used here. The RLCD loss includes cross-entropy. It perturbs discriminative option logits using target-answer rewards; financial ledger outcomes do not enter training.

All native child checkpoints are selected on the same fixed development rule as their parents. RLCD200 and CE continuation each process the same additional 200,000-example pass, but their development-best checkpoints occur at different points: 200,000 and 50,688 additional exposures. The selected cumulative downstream exposures are therefore 400,000 and 250,688, respectively; cumulative unique examples remain 200,000 in both cases. Exposure counts cover the recorded downstream stages.

\subsection*{Training work and selected exposures}

Table S24 records every formal run. Elapsed times are trainer-reported wall times, including in-stage validation and saving; allocated GPU-hours multiply this duration by eight H100 GPUs. Head-only time includes frozen feature preparation and small-head fitting under the reported run allocation. Inference is performed on six A100 40GB GPUs for the four 4B adaptations and two SFT controls, and on eight H100 80GB GPUs for 9B, native continuations and the original native reference.

\begin{center}\begin{minipage}{\textwidth}\fontsize{8.2}{10}\selectfont
Table S24 | Training exposure and cost for all ten formal versions
\par\vspace{4pt}
\begin{tabularx}{\textwidth}{@{}>{\hsize=1.6\hsize\linewidth=\hsize\raggedright\arraybackslash}X*{6}{>{\hsize=.9\hsize\linewidth=\hsize\raggedright\arraybackslash}X}@{}}
\toprule
Version & Unique QA & Stage terminal exposures & Selected cumulative exposures & Stage updates & Elapsed (s) & Allocated GPU-hours \\
\midrule
YN-Lite-4B-head-50k & 50,000 & 250,000 & 100,000 & 65 & 466.85 & 1.037 \\
YN-Lite-4B-LORA-50k & 50,000 & 50,000 & 50,000 & 1,563 & 3,498.63 & 7.775 \\
YN-Lite-4B-head-200k & 200,000 & 800,000 & 200,000 & 196 & 999.44 & 2.221 \\
YN-Lite-4B & 200,000 & 200,000 & 200,000 & 6,250 & 10,819.22 & 24.043 \\
YN-master-9B & 200,000 & 200,000 & 200,000 & 6,250 & 13,401.91 & 29.782 \\
YN-mini-SFT-50k & 50,000 & 50,000 & 50,000 & 66 & 87.65 & 0.195 \\
YN-mini-SFT-200k & 200,000 & 200,000 & 200,000 & 261 & 329.85 & 0.733 \\
YN-mini-RLCD-50k & 50,000 & 50,000 & 100,000 & 66 & 87.87 & 0.195 \\
YN-mini-RLCD-200k & 200,000 & 200,000 & 400,000 & 261 & 326.01 & 0.724 \\
YN-mini-CE-200k & 200,000 & 200,000 & 250,688 & 261 & 328.27 & 0.729 \\
\bottomrule\end{tabularx}
\par\vspace{3pt}{\fontsize{8}{9.5}\selectfont Stage terminal counts exclude parent training. Selected cumulative counts include the native SFT parent where applicable; cumulative unique QA count each example once. Head50k and Head200k stop after 250,000 and 800,000 presentations, selecting 100,000 and 200,000, respectively. The native original reference receives no additional training and is not a formal trained run.}
\end{minipage}\end{center}

\subsection*{Paired native comparisons}

Each paired interval resamples 8,275 respondent groups 1,000 times, carrying all question records from a selected group together. Accuracy and F1 differences use the same full 240,458-record test and candidate-class definitions. The six native comparisons distinguish initial SFT adaptation, the added RLCD stage, and equal added-budget CE continuation. A seventh direct comparison carries the same 200k-example test pairing across YN-Lite-4B and YN-master-9B, with 200,000 selected exposures and 6,250 updates each. Its 4B-Base and 9B post-trained initializations retain different upstream histories. Table S25 gives pointwise percentile 95\% intervals without adjustment for multiple comparisons. Full-data YN-mini serves as a separate system reference, with different cumulative supervision and MLX FP16 inference.

\begin{center}\begin{minipage}{\textwidth}\fontsize{8.5}{10.3}\selectfont
Table S25 | Direct downstream comparisons
\par\vspace{4pt}
\begin{tabularx}{\textwidth}{@{}XXX@{}}\toprule
Target $-$ reference & Accuracy difference (percentage points), 95\% CI & Question-macro F1 difference (percentage points), 95\% CI \\
\midrule
YN-mini-SFT-50k $-$ original & +12.4970 [+12.2933, +12.7005] & +5.2516 [+4.8365, +5.7448] \\
YN-mini-SFT-200k $-$ original & +13.6781 [+13.4662, +13.8857] & +6.3953 [+5.9603, +6.9003] \\
YN-mini-RLCD-50k $-$ YN-mini-SFT-50k & +0.2316 [+0.1771, +0.2854] & -0.0671 [-0.1516, +0.0197] \\
YN-mini-RLCD-200k $-$ YN-mini-SFT-200k & +0.0516 [+0.0029, +0.1020] & +0.0554 [-0.0380, +0.1654] \\
YN-mini-RLCD-200k $-$ YN-mini-CE-200k & +0.0283 [-0.0175, +0.0752] & -0.1229 [-0.2073, -0.0330] \\
YN-mini-CE-200k $-$ YN-mini-SFT-200k & +0.0233 [-0.0288, +0.0731] & +0.1783 [+0.0762, +0.2770] \\
YN-master-9B $-$ YN-Lite-4B & -0.0516 [-0.0981, -0.0071] & -0.0721 [-0.1725, +0.0404] \\
\bottomrule\end{tabularx}
\end{minipage}\end{center}

\subsection*{Complete reference comparison}

Expanded-supervision YN-mini reaches 74.5544\% accuracy and 38.3746\% question-macro F1 under MLX FP16. Table S26 compares every trained variant against this reference on identical test records; staged supervision, checkpoint calibration and deployment precision differ between systems.

\begin{center}\begin{minipage}{\textwidth}\fontsize{8.5}{10.3}\selectfont
Table S26 | Every fixed-budget version minus YN-mini-MLX
\par\vspace{4pt}
\begin{tabularx}{\textwidth}{@{}XXX@{}}\toprule
Version $-$ YN-mini-MLX & Accuracy difference (percentage points), 95\% CI & Question-macro F1 difference (percentage points), 95\% CI \\
\midrule
YN-Lite-4B-head-50k & -7.2175 [-7.3973, -7.0389] & -7.3580 [-7.6125, -7.0713] \\
YN-Lite-4B-LORA-50k & -0.8255 [-0.9248, -0.7308] & -1.7520 [-1.9896, -1.4840] \\
YN-Lite-4B-head-200k & -7.2703 [-7.4509, -7.0849] & -7.4751 [-7.7300, -7.1973] \\
YN-Lite-4B & -0.4109 [-0.4977, -0.3248] & -1.6402 [-1.8938, -1.3596] \\
YN-master-9B & -0.4625 [-0.5455, -0.3733] & -1.7124 [-1.9451, -1.4417] \\
YN-mini-SFT-50k & -1.8602 [-1.9884, -1.7458] & -3.5234 [-3.7717, -3.2552] \\
YN-mini-SFT-200k & -0.6791 [-0.7745, -0.5885] & -2.3797 [-2.6138, -2.1158] \\
YN-mini-RLCD-50k & -1.6286 [-1.7503, -1.5133] & -3.5905 [-3.8261, -3.3252] \\
YN-mini-RLCD-200k & -0.6276 [-0.7120, -0.5414] & -2.3242 [-2.5683, -2.0443] \\
YN-mini-CE-200k & -0.6558 [-0.7453, -0.5602] & -2.2014 [-2.4396, -1.9297] \\
\bottomrule\end{tabularx}
\par\vspace{3pt}{\fontsize{8}{9.5}\selectfont Complete test: 240,458 records, 119 question versions, 8,275 respondent groups. Pointwise 95\% intervals use 1,000 respondent-group bootstrap draws; every defined candidate class enters F1. All ten trained variants have zero invalid decisions.}
\end{minipage}\end{center}

\section*{Appendix S11. Complete static task execution and language sensitivity}

The static task suite samples household--parameter combinations within the conditions in Table S27. Each task contains 2,842 baseline/intervention pairs. The overall pool has 207 financial parents; task-specific counts range from 22 to 207. Each language contains 45,472 paired inputs plus one additional, unpaired T1 intervention input. Both models execute every input in both languages, giving $45,473\times2\times2=181,892$ choices. The paired results use 2,842 pairs per task; the four additional T1 choices enter the completed-work count.

The English and Chinese inputs have identical execution specifications, action consequences and presentation orders. Both languages use the same frozen full-data YN-mini checkpoint and the same original Laya reference in MLX CUDA FP16. Tables S28 and S29 report English YN-mini and Laya; Tables S30 and S31 report their Chinese-input counterparts. Each cell is intervention minus baseline within a model and language. Pointwise intervals resample the financial parents represented in that task.

For this suite, monthly income $I$ and usual spending $C$ are annual financial-anchor amounts divided by twelve; $K$ is opening demand deposits. Assigned adjustable spending is an integer percentage from 10\% through 30\% of $C$, with the remainder essential. Existing bank-housing pools retain their recorded principal and monthly payment; dates and annual rates are assigned, with rate $\min(r_0,\lfloor60,000M_j/P_j\rfloor/10,000)$ for sampled $r_0\in\{0.02,0.03,0.04,0.05\}$ and pool-specific payment $M_j$ and principal $P_j$. Let $q\in\{10,000,20,000,50,000,100,000,174,000\}$ CNY, $P$ denote the largest opening pool's principal and $M$ its monthly payment. The quoted new loans in T2--T3 use annual rates of 3\%, 5\% or 8\% and terms of 12, 24 or 36 months, zero fees and repayment beginning one month after origination. Within each pair, household information and the other sampled parameters remain common.

\begin{center}\begin{minipage}{\textwidth}\fontsize{8}{9.7}\selectfont
Table S27 | Conditions and executed actions in the eight-task suite
\par\vspace{4pt}
\begin{tabularx}{\textwidth}{@{}>{\hsize=.40\hsize\linewidth=\hsize\raggedright\arraybackslash}X>{\hsize=1.20\hsize\linewidth=\hsize\raggedright\arraybackslash}X>{\hsize=1.20\hsize\linewidth=\hsize\raggedright\arraybackslash}X>{\hsize=1.20\hsize\linewidth=\hsize\raggedright\arraybackslash}X@{}}\toprule
Task / parents & Baseline $\to$ intervention & Sampled parameters & Action consequences \\
\midrule
T1 / 207 & Monthly income $I\to uI$ throughout twelve months, where $u\in\{0.65,0.66,\ldots,0.85\}$ & Usual spending and its essential/adjustable split remain fixed & Select normal, half-normal or zero adjustable spending for the ledger; essential spending remains due \\
\addlinespace[4pt]
T2 / 162 & Immediate net medical bill $b\to\min(K,1.5b)$ & $b=\min[K,\max(100,rK)]$ CNY, with $r$ an integer percentage from 15\% to 30\%; common loan-rate and term sets above. Realized bills are CNY 100--300,000 at baseline and 150--450,000 under intervention & Pay the bill from cash, borrow the full amount, or fund approximately half from each. The medical payment counts as consumption; the selected loan creates dated installments \\
\addlinespace[4pt]
T3 / 207 & Purchase goal and tied approved limit $q\to1.5q$ & Baseline goals CNY 10k, 20k, 50k, 100k or 174k; intervention goals 15k, 30k, 75k, 150k or 261k; common loan-rate and term sets above & Defer the whole purchase, finance half, or finance all. Borrowed funds purchase the corresponding amount immediately; deferral leaves the unpurchased amount outside routine shortfalls \\
\addlinespace[4pt]
T4 / 207 & Add 3 percentage points to both offered annual rates & Purchase amount $q$ fixed. Short term 12 or 24 months at 2\%, 3\% or 4\% initially; long term adds twelve months and 2 rate points. Intervention short rates are 5--7\%, long rates 7--9\%. The long-product fee is $\min(0.01q,0.1K)$ & Select the short product, the long product, or defer. Either accepted loan buys $q$ immediately; its fee is paid from cash, and its quoted installments enter the ledger \\
\bottomrule\end{tabularx}
\par\vspace{3pt}{\fontsize{7.7}{9.3}\selectfont Each row has 2,842 complete pairs. Parent counts identify the actual task-specific clusters used in the response intervals. Parameters are sampled combinations accepted under the stated action constraints. Amounts use integer cents, with the constructor's rounding. All choices are made once and executed over twelve months.}
\end{minipage}\end{center}

\begin{center}\begin{minipage}{\textwidth}\fontsize{8}{9.7}\selectfont
Table S27 (continued) | Conditions and executed actions in the eight-task suite
\par\vspace{4pt}
\begin{tabularx}{\textwidth}{@{}>{\hsize=.40\hsize\linewidth=\hsize\raggedright\arraybackslash}X>{\hsize=1.20\hsize\linewidth=\hsize\raggedright\arraybackslash}X>{\hsize=1.20\hsize\linewidth=\hsize\raggedright\arraybackslash}X>{\hsize=1.20\hsize\linewidth=\hsize\raggedright\arraybackslash}X@{}}\toprule
Task / parents & Baseline $\to$ intervention & Sampled parameters & Action consequences \\
\midrule
T5 / 161 & Widen the monthly risky-return support from $\{-15,-7.5,0,7.5,15\}\%$ to $\{-40,-20,0,20,40\}\%$ & Investable amount $\min(rK,q)$, where $r$ is an integer percentage from 10\% to 40\%; at least CNY 100. Returns are drawn from the five support points and are unknown at choice & Allocate 100\%, 75\%, 50\%, 25\% or 0\% to the risky product; the rest stays liquid at zero interest. The invested balance is revalued and liquidates at the start of month 6 \\
\addlinespace[4pt]
T6 / 169 & Reduce the first-month earmarked debt-payment budget from $L(0.7+0.2v)$ to $L(0.4+0.2v)$, with the same $v\in[0,1]$ & $L=\min(K+I-C,D_1,D_2)$; $D_1,D_2$ are the two current dues. Add an opening consumer loan of $q$ at 8\%, 10\% or 12\%, for 12, 24 or 36 months. The largest housing pool has opening arrears aged 15, 30 or 45 days & Pay the full budget to the housing pool, to the consumer loan, or proportionally to their dues. Other first-month cash is reserved and released next month; later debt payments follow the common priority \\
\addlinespace[4pt]
T7 / 149 & Monthly income $I\to uI$ for twelve months, with $u\in\{0.65,0.66,\ldots,0.85\}$; extension approval responds to current income & The same prepayment amount and extension quote enter both arms. Prepayment is capped at 5\% of $P$ and one-third of resources after spending and dues under reduced income. Extension doubles or triples the calculated remaining term, at the same rate and zero fee & Keep the contract, pay the current due plus the stated principal prepayment, or request extension. Approval requires the new payment to be at most 40\% of current income. Approved terms begin next month; existing arrears remain \\
\addlinespace[4pt]
T8 / 22 & Reduce the opening market-asset value from $A$ to $wA$, with $w\in\{0.70,0.71,\ldots,0.90\}$ & Initial cash and first-month income are zero after prepaid bills. A further bill $g=\min(0.7A/8,\max[100,rM])$ is due on day 15, with $r\in\{0.25,0.26,\ldots,0.75\}$ (step 0.01). Half the assets can be sold; settlement takes 1, 3, 5 or 10 days and the sale fee is 0\%, 0.25\%, 0.5\% or 1\%. Refinancing terms are 240, 300 or 360 months, with the assigned rate drawn from 2--4\% and capped at the old rate plus 1 point & Sell enough assets to pay the bill, refinance the named old pool and fund the bill, or combine a half-bill sale with refinancing. The bill and action amounts/contracts are common across asset-price arms. Refinancing first settles the named old debt; recorded payments include that settlement \\
\bottomrule\end{tabularx}
\par\vspace{3pt}{\fontsize{7.7}{9.3}\selectfont T6 draws $v$ uniformly and truncates the resulting cent budget; opening arrears are $\min(M,P/4)$. T8 uses the same integer-percentage draw for its bill rule in both arms and retains only pairs with executable asset-sale alternatives. After initial actions, income precedes selected household spending and dated bank payments; unfunded selected spending remains a shortfall. These conditions define the means in Tables S28--S31.}
\end{minipage}\end{center}

Static task execution chooses once and then runs its twelve-month ledger. T1 records unmet selected consumption targets; T3/T4 deferral of a proposed purchase is not counted as an unfunded routine need. T8 payments include refinancing settlement of existing debt.

Tables S28--S31 report all eight task families for both models and languages. Differences are the designated intervention minus baseline within each model. Monetary outcomes are CNY per constructed case over twelve months. The full aggregate source data retain paid debt, interest accrued, terminal debt, minimum cash and unfunded months with financial-parent-cluster intervals, together with paired YN-mini-minus-original-Laya responses and Chinese-minus-default sensitivity.
\begin{center}\begin{minipage}{\textwidth}\fontsize{8}{9.8}\selectfont
Table S28 | YN-mini: complete within-model task responses
\par\vspace{4pt}
\begin{tabularx}{\textwidth}{@{}>{\hsize=.4\hsize\linewidth=\hsize}X*{3}{>{\hsize=1.2\hsize\linewidth=\hsize\raggedright\arraybackslash}X}@{}}\toprule
Task & Consumption (CNY), 95\% CI & Closing cash (CNY), 95\% CI & Unfunded needs (CNY), 95\% CI \\
\midrule
T1 & -7,065.95 [-9,702.21, -4,909.79] & -26,969.82 [-31,722.80, -22,433.23] & +9,415.55 [+7,019.80, +12,333.77] \\
T2 & +5,119.61 [+3,651.29, +6,977.79] & -3,658.58 [-4,959.08, -2,550.75] & +544.93 [+303.66, +825.40] \\
T3 & +17,148.49 [+14,818.58, +19,601.23] & -3,439.10 [-4,376.02, -2,605.90] & +161.86 [+56.59, +295.36] \\
T4 & -1,052.95 [-1,779.68, -345.29] & +115.83 [-153.66, +399.32] & -1.94 [-16.32, +9.16] \\
T5 & -31.11 [-65.31, +1.78] & +46.02 [-116.34, +221.06] & +31.11 [-1.78, +65.31] \\
T6 & +34.06 [+14.30, +57.15] & -0.26 [-0.34, -0.17] & -34.06 [-57.15, -14.30] \\
T7 & -6,134.77 [-8,250.80, -4,228.10] & -33,424.05 [-39,676.36, -27,805.84] & +6,134.77 [+4,228.10, +8,250.80] \\
T8 & -0.00 [-0.00, +0.00] & +59.91 [-215.76, +361.55] & +0.00 [-0.00, +0.00] \\
\bottomrule\end{tabularx}
\par\vspace{3pt}{\fontsize{8}{9.5}\selectfont Each task has 2,842 complete pairs. Financial-parent cluster counts for T1--T8 are 207, 162, 207, 207, 161, 169, 149 and 22, respectively (Table S27). Pointwise 95\% intervals use 2,000 cluster bootstrap draws. Unfunded outcomes retain each task's definition.}
\end{minipage}\end{center}
\begin{center}\begin{minipage}{\textwidth}\fontsize{8}{9.8}\selectfont
Table S29 | original Laya: complete within-model task responses
\par\vspace{4pt}
\begin{tabularx}{\textwidth}{@{}>{\hsize=.4\hsize\linewidth=\hsize}X*{3}{>{\hsize=1.2\hsize\linewidth=\hsize\raggedright\arraybackslash}X}@{}}\toprule
Task & Consumption (CNY), 95\% CI & Closing cash (CNY), 95\% CI & Unfunded needs (CNY), 95\% CI \\
\midrule
T1 & -11,000.04 [-13,899.55, -8,548.88] & -22,952.29 [-27,537.90, -18,660.47] & +11,251.27 [+8,617.24, +14,360.57] \\
T2 & +5,285.73 [+3,839.13, +7,132.03] & -3,337.99 [-4,768.49, -2,297.21] & +378.81 [+198.14, +590.52] \\
T3 & +34,807.17 [+33,709.99, +35,959.66] & -5,003.76 [-6,000.29, -4,062.40] & +214.65 [+96.89, +357.22] \\
T4 & -922.61 [-1,555.46, -320.36] & -429.30 [-703.34, -166.70] & +2.84 [-17.86, +24.34] \\
T5 & -47.67 [-102.83, +6.51] & -30.78 [-213.92, +148.44] & +47.67 [-6.51, +102.83] \\
T6 & +34.06 [+14.30, +57.15] & -0.09 [-0.17, -0.02] & -34.06 [-57.15, -14.30] \\
T7 & -6,084.71 [-8,181.91, -4,174.70] & -34,092.61 [-40,371.37, -28,431.99] & +6,084.71 [+4,174.70, +8,181.91] \\
T8 & +0.00 [+0.00, +0.00] & +264.68 [+41.13, +565.57] & -0.00 [-0.00, +0.00] \\
\bottomrule\end{tabularx}
\par\vspace{3pt}{\fontsize{8}{9.5}\selectfont Each task has 2,842 complete pairs. Financial-parent cluster counts for T1--T8 are 207, 162, 207, 207, 161, 169, 149 and 22, respectively (Table S27). Pointwise 95\% intervals use 2,000 cluster bootstrap draws. Unfunded outcomes retain each task's definition.}
\end{minipage}\end{center}
\begin{center}\begin{minipage}{\textwidth}\fontsize{8}{9.8}\selectfont
Table S30 | Chinese YN-mini: complete within-model task responses
\par\vspace{4pt}
\begin{tabularx}{\textwidth}{@{}>{\hsize=.4\hsize\linewidth=\hsize}X*{3}{>{\hsize=1.2\hsize\linewidth=\hsize\raggedright\arraybackslash}X}@{}}\toprule
Task & Consumption (CNY), 95\% CI & Closing cash (CNY), 95\% CI & Unfunded needs (CNY), 95\% CI \\
\midrule
T1 & -8,950.22 [-11,508.39, -6,927.20] & -25,340.71 [-29,854.28, -20,990.65] & +7,634.45 [+5,701.08, +9,968.84] \\
T2 & +5,271.55 [+3,823.66, +7,162.49] & -3,702.98 [-5,075.09, -2,533.84] & +392.99 [+163.88, +663.76] \\
T3 & +1,194.03 [+638.59, +1,825.40] & -369.11 [-643.23, -156.06] & +19.55 [+1.23, +48.47] \\
T4 & -2,497.49 [-4,043.44, -960.37] & +319.62 [-300.39, +958.42] & +19.66 [-24.85, +72.90] \\
T5 & -14.59 [-31.68, +0.03] & -30.14 [-113.77, +55.04] & +14.59 [-0.03, +31.68] \\
T6 & +34.06 [+14.30, +57.15] & -0.15 [-0.26, -0.05] & -34.06 [-57.15, -14.30] \\
T7 & -6,320.23 [-8,358.10, -4,416.56] & -31,278.26 [-37,511.45, -25,610.21] & +6,320.23 [+4,416.56, +8,358.10] \\
T8 & +0.00 [-0.00, +0.00] & -244.78 [-659.67, +38.44] & -0.00 [-0.00, +0.00] \\
\bottomrule\end{tabularx}
\par\vspace{3pt}{\fontsize{8}{9.5}\selectfont Each task has 2,842 complete pairs. Financial-parent cluster counts for T1--T8 are 207, 162, 207, 207, 161, 169, 149 and 22, respectively (Table S27). Pointwise 95\% intervals use 2,000 cluster bootstrap draws. Unfunded outcomes retain each task's definition.}
\end{minipage}\end{center}
\begin{center}\begin{minipage}{\textwidth}\fontsize{8}{9.8}\selectfont
Table S31 | Chinese original Laya: complete within-model task responses
\par\vspace{4pt}
\begin{tabularx}{\textwidth}{@{}>{\hsize=.4\hsize\linewidth=\hsize}X*{3}{>{\hsize=1.2\hsize\linewidth=\hsize\raggedright\arraybackslash}X}@{}}\toprule
Task & Consumption (CNY), 95\% CI & Closing cash (CNY), 95\% CI & Unfunded needs (CNY), 95\% CI \\
\midrule
T1 & -9,941.04 [-12,557.41, -7,833.34] & -24,696.32 [-29,280.60, -20,341.76] & +8,015.96 [+5,983.57, +10,474.95] \\
T2 & +5,220.59 [+3,781.94, +7,062.02] & -3,866.51 [-5,555.35, -2,572.03] & +443.95 [+208.87, +729.75] \\
T3 & +32,567.83 [+29,697.63, +35,465.54] & -5,608.78 [-6,903.31, -4,381.07] & +348.78 [+112.69, +658.72] \\
T4 & +167.65 [-1,060.06, +1,372.52] & -545.27 [-930.22, -174.69] & -10.72 [-34.16, +8.95] \\
T5 & -64.40 [-129.52, -4.34] & +23.59 [-183.09, +239.37] & +64.40 [+4.34, +129.52] \\
T6 & +34.06 [+14.30, +57.15] & -0.15 [-0.26, -0.06] & -34.06 [-57.15, -14.30] \\
T7 & -6,397.08 [-8,451.63, -4,471.02] & -31,651.45 [-37,813.25, -25,990.92] & +6,397.08 [+4,471.02, +8,451.63] \\
T8 & +0.00 [-0.00, +0.00] & -7.87 [-164.27, +146.87] & +0.00 [-0.00, +0.00] \\
\bottomrule\end{tabularx}
\par\vspace{3pt}{\fontsize{8}{9.5}\selectfont Each task has 2,842 complete pairs. Financial-parent cluster counts for T1--T8 are 207, 162, 207, 207, 161, 169, 149 and 22, respectively (Table S27). Pointwise 95\% intervals use 2,000 cluster bootstrap draws. Unfunded outcomes retain each task's definition.}
\end{minipage}\end{center}
For the completed T3 grid, YN-mini's intervention response differs from original Laya by lower consumption, less additional debt and less erosion of cash: a smaller financed-purchase response. Increasing the purchase need within YN-mini still raises closing debt, debt payments, accrued interest and unmet routine needs. The Chinese response is materially smaller than the default response.

\subsection*{T3 purchase-funding application with twelve million synthetic household agents}

One prechosen T3 setting is executed on the complete 12M cohort using YN-mini MLX CUDA FP16. The input targets and contract are fixed as specified in M10. Table S32 gives both condition means and their paired differences, weighting all households equally. The full-panel difference in consumption combines the initial financed purchase with subsequent routine consumption; the default measures follow the dated bank-dues rule. Table S33 records the complete execution scope and measured work.

\begin{center}\begin{minipage}{\textwidth}\fontsize{8.1}{9.8}\selectfont
Table S32 | Complete financial outcomes for the 12M T3 application
\par\vspace{4pt}
\begin{tabularx}{\textwidth}{@{}>{\hsize=1.15\hsize\linewidth=\hsize}X>{\hsize=.62\hsize\linewidth=\hsize}X>{\hsize=.62\hsize\linewidth=\hsize}X>{\hsize=1.61\hsize\linewidth=\hsize}X@{}}\toprule
Measure per household & Baseline & Intervention & Paired difference [row-variation reference] \\
\midrule
Total consumption (CNY) & 68,960.33 & 71,847.52 & 2,887.18 [2,882.36, 2,892.00] \\
Closing cash (CNY) & 97,805.83 & 96,877.07 & -928.76 [-930.67, -926.85] \\
Debt payments (CNY) & 8,286.33 & 9,218.88 & 932.55 [930.64, 934.46] \\
Closing debt (CNY) & 84,569.06 & 86,641.51 & 2,072.45 [2,068.82, 2,076.08] \\
Accrued interest (CNY) & 5,238.03 & 5,352.06 & 114.03 [113.84, 114.22] \\
Unfunded routine needs (CNY) & 10,517.57 & 10,521.36 & 3.79 [3.72, 3.86] \\
Minimum cash (CNY) & 60,303.21 & 60,172.21 & -131.00 [-131.67, -130.33] \\
Closing noncash assets (CNY) & 0.00 & 0.00 & 0.00 [0.00, 0.00] \\
Opening-facility EAD (CNY) & 48,848.37 & 49,112.49 & 264.12 [259.29, 268.96] \\
New-facility EAD (CNY) & 174.50 & 1,159.84 & 985.34 [982.50, 988.18] \\
Total EAD (CNY) & 49,022.87 & 50,272.33 & 1,249.46 [1,243.65, 1,255.28] \\
Unfunded months (months) & 1.3296 & 1.3301 & 0.0006 [0.0005, 0.0006] \\
Any bank-facility default (\%) & 17.3485 & 19.0148 & 1.6663 [1.6590, 1.6736] \\
Opening-facility default (\%) & 16.8974 & 17.2058 & 0.3085 [0.3053, 0.3116] \\
New-facility default (\%) & 0.9225 & 4.1371 & 3.2146 [3.2046, 3.2246] \\
\bottomrule\end{tabularx}
\par\vspace{3pt}{\fontsize{8}{9.5}\selectfont Every mean uses the same 12,000,000 synthetic households. CNY rows are monetary means; default rows give percentages with percentage-point differences, and unfunded months give months per household. EAD denotes exposure at first default. The bracketed reference is paired mean $\pm1.96$ SE across synthetic rows, conditional on the constructed cohort. Exact panel means are primary; the reference measures synthetic-row variation rather than survey-cluster or national sampling uncertainty. M10 defines the twelve-month horizon and 24-month contract.}
\end{minipage}\end{center}

\begin{center}\begin{minipage}{\textwidth}\fontsize{8.2}{10}\selectfont
Table S33 | Completed work and measurement scopes in the 12M application
\par\vspace{4pt}
\begin{tabularx}{\textwidth}{@{}>{\hsize=.95\hsize\linewidth=\hsize}X>{\hsize=.95\hsize\linewidth=\hsize}X>{\hsize=1.10\hsize\linewidth=\hsize}X@{}}\toprule
Quantity & Measured value & Scope \\
\midrule
Synthetic households / choices & 12,000,000 / 24,000,000 & Two conditions per household; T3 \\
Monthly ledger states & 288,000,000 & Twelve accounting months per choice \\
GPU allocation & 8 H100 80GB + 6 A100 40GB & Two service pools, same YN-mini FP16 policy \\
Completed service batches & 1,641,533 & Unique recorded service batches \\
Input / generated output tokens & 12,719,522,472 / 0 & Actual encoded nonpadding input; discriminative outputs \\
Compressed private trace & 67,664,771,925 bytes & Full inputs, choices, usage and monthly ledgers \\
Common end-to-end wall & 14,356.9775 s & First nested-pilot preparation through final work-unit completion \\
End-to-end rates & 835.831 households/s; 1,671.661 choices/s; 885,947.092 input tokens/s & All denominators use the same common wall \\
Common batch-processing window & 14,348.8386 s & Preparation, tokenization, collation, forward calls and decoding \\
Summed forward-call wall & 111,369.7866 s & Sum across unique batches on parallel endpoints \\
Pure kernel time, padding, cache & Unmeasured & Separate from the reported service-call and end-to-end intervals \\
Recorded requests / responses & 24,000,000 / 24,000,000 & Complete-input indicator present for every request \\
HTTP / response-validation errors & 0 & Interrupted requests without recorded responses: 0 \\
Process/status finalization retries & 14 across 11 complete units & Existing complete summaries adopted, preserving saved choices \\
\bottomrule\end{tabularx}
\par\vspace{3pt}{\fontsize{8}{9.5}\selectfont The common end-to-end interval includes nested-pilot work, transfer waits, subsequent input preparation, CPU ledgers, compressed persistence and recovery waits. Service-call wall is a sum over parallel batches. The measured aggregate rates use the common interval rather than summed endpoint durations. Trace and input records remain private.}
\end{minipage}\end{center}

\section*{Appendix S12. Supporting two-million-household fixed-plan application}

This application samples two million constructed households uniformly without replacement from the 10M experimental bank. Each household selects one plan under unavailable and available credit, producing four million one-shot choices and 64 million monthly ledger records over months 0--15. Both conditions share the same initial household state and an income reduction of 20\% in months 0--2, with full recovery from month 3. The available product has a CNY 20,000 limit, 7.68\% annual rate, twelve installments due in months 1--12 and zero fees (Table S34). A selected plan governs the full ledger path.

The supporting application uses the delivered 36-field consumer-bank representation and the assigned old-debt contracts in Table S34. The 12M main application draws from the separately calibrated 500M resource and uses the opening-debt assignment specified in M10.

C0 adjusts spending to preserve cash; C1 uses reserves to sustain consumption. Credit options combine partial financing with half-normal adjustable spending or seek to sustain normal spending while preserving reserves. The ledger applies necessary spending, dated debt payments and adjustable spending with integer-cent cash constraints. Unfunded needs and unpaid obligations remain explicit. The selected adjustable-spending target defines the corresponding execution shortfall.

Across the full panel, available minus unavailable credit yields CNY 293.76 less cumulative consumption, CNY 311.50 more closing cash and CNY 17.74 more closing old debt (Table S35). Reduced consumption and lower actual debt payments jointly fund the cash difference. Within the realized fully funded plans, 90,667 households increase consumption and 220,427 preserve cash; these are outcome-defined subsets of the full cohort (Table S36). Cash reaches zero in at least one month for 317,879 households without available credit and 265,770 with it, comprising 62,326 avoided events and 10,217 added events.

\begin{center}
\begin{minipage}{\textwidth}
\fontsize{8.5}{10.3}\selectfont
Table S34 | Assigned environment and economic references
\par\vspace{4pt}
\begin{tabularx}{\textwidth}{@{}XXX@{}}
\toprule
Component & Assignment / rule & Reference and basis \\
\midrule
Simulated population & Survey-derived constructed households; 2M sampled from 10M & Actual bank: 59.98\% urban, 36.15\% rural, 3.87\% unclassified. \\
Income and horizon & -20\% in m0–m2; full recovery in m3; observe m0–m15 & Buffer-stock saving (Carroll, 1997) and consumption commitments (Chetty and Szeidl, 2007); references [22]–[23]. \\
Spending allocation & 70\% harder to adjust; 30\% adjustable & Assigned common spending split \\
Opening spendable cash & 39.1\% of assigned financial assets & Assigned common opening liquidity \\
Existing debt & Housing: 5.42\%, 240 remaining months; other: 7.68\%, 36 months; mixed 50:50 & March 2022 new-loan rates supply references; remaining terms, mix and zero opening arrears are assignments \\
Available-credit contract & CNY 20,000 limit; 7.68\%; 12 installments m1–m12; no fees & Rate reference and assigned limit, schedule and fee controls \\
Amount categories & Closed-bracket midpoint; open upper bracket lower bound \ensuremath{\times} 1.25 & Assigned representative amounts shared by both conditions \\
Missing amounts & Annual income 72,000; consumption 80\% of income; financial assets one month income; old debt 50,000 & Disclosed missing-value assignments; annual flows divided by 12 \\
Payment execution & Income, then harder-to-adjust spending, due debt, adjustable spending; integer cents; no overdraft & Deterministic cash constraint; unmet spending and unpaid debt retained \\
Risk definition & Cash-zero: month-end balance 0; default: ≥CNY 1 overdue >90 days; LGD 0.6 & Household cash boundary, contract event threshold and assigned conditional-loss proxy \\
\bottomrule
\end{tabularx}

\end{minipage}
\end{center}

\begin{center}
\begin{minipage}{\textwidth}
\fontsize{8.5}{10.3}\selectfont
Table S35 | Full-cohort financial results
\par\vspace{4pt}
\begin{tabularx}{\textwidth}{@{}XXXX@{}}
\toprule
Outcome & Credit unavailable & Credit available & Paired difference \\
\midrule
Consumption: shock, m0–m2 & 15,890.78 & 15,632.74 & -258.04 \\
Consumption: recovery/repayment, m3–m12 & 55,244.15 & 55,228.73 & -15.42 \\
Consumption: follow-up, m13–m15 & 16,482.21 & 16,461.91 & -20.30 \\
Consumption: total, m0–m15 & 87,617.15 & 87,323.39 & -293.76 \\
Harder-to-adjust consumption shortfall & 4,234.76 & 4,243.42 & +8.66 \\
Adjustable consumption shortfall & 788.33 & 286.29 & -502.05 \\
Use of opening cash & 3,708.86 & 3,364.57 & -344.29 \\
Closing cash & 71,866.27 & 72,177.77 & +311.50 \\
Old-debt principal paid & 3,397.29 & 3,385.53 & -11.76 \\
Old-debt interest paid & 3,103.16 & 3,097.18 & -5.98 \\
Old-debt interest accrued & 3,852.80 & 3,852.80 & +0.00 \\
Fees & 0.00 & 0.00 & +0.00 \\
Closing old debt & 50,001.71 & 50,019.45 & +17.74 \\
Closing old principal & 49,252.07 & 49,263.83 & +11.76 \\
Closing old arrears & 3,072.61 & 3,090.34 & +17.74 \\
Old-debt exposure at default & 16,655.83 & 16,664.71 & +8.88 \\
Old-debt conditional loss & 9,993.50 & 9,998.83 & +5.33 \\
Old-debt defaults: contracts & 59,565 & 59,938 & +373 \\
\bottomrule
\end{tabularx}
\par\vspace{3pt}{\fontsize{8}{9.5}\selectfont Amounts are CNY per household over the stated window; the contract-count row gives the total count across 2,000,000 households. Conditional losses use LGD 0.6. Adjustable shortfall is relative to each selected plan’s target.}
\end{minipage}
\end{center}

\begin{center}
\begin{minipage}{\textwidth}
\fontsize{8.5}{10.3}\selectfont
Table S36 | Fully funded realized adjustments
\par\vspace{4pt}
\begin{tabularx}{\textwidth}{@{}XXXXXX@{}}
\toprule
Realized group & Households / share & Shock consumption \ensuremath{\Delta} & Total consumption \ensuremath{\Delta} & Closing cash \ensuremath{\Delta} & Closing old debt \ensuremath{\Delta} \\
\midrule
Consumption protection & 90,667 (4.53335\%) & +2,023.69 & +2,401.49 & -2,401.49 & +0.00 \\
Cash preservation & 220,427 (11.02135\%) & -3,588.38 & -4,335.41 & +4,331.25 & -4.16 \\
Avoided-zero subset of cash preservation & 62,326 (3.11630\%) & -2,104.64 & -3,260.69 & +3,245.96 & -14.73 \\
Full cohort reference & 2,000,000 (100\%) & -258.04 & -293.76 & +311.50 & +17.74 \\
\bottomrule
\end{tabularx}
\par\vspace{3pt}{\fontsize{8}{9.5}\selectfont CNY per household; \ensuremath{\Delta} = credit available - unavailable. Consumption protection requires higher 16-month consumption, fully funded available-condition plan, positive month-end cash throughout and no added closing old debt. Cash preservation requires higher closing cash and a fully funded available-condition plan. The 62,326 avoided-zero households are nested in the cash-preservation group.}
\end{minipage}
\end{center}

\clearpage
Controls on 2,000 households from this sixteen-month application compare learned selections with fixed plans and separate information, menu and budget changes. Under the available full menu, YN-mini selects fully funded plans for 87.00\% of households, compared with 79.30\% for original Laya and 88.05\% for always selecting the restrained-consumption plan C0. Relative to C0, YN-mini executes CNY 1,146.60 more consumption (95\% interval 1,007.92--1,314.65) and retains CNY 1,115.34 less cash. A matched information/menu comparison gives CNY 199.88 more consumption when credit availability is stated with the common menu, followed by CNY 447.98 less when financing alternatives are added. In the matched recovery-versus-persistent-income comparison, both paths lose 20\% of income in months 0--2; one recovers from month 3 and the other remains lower through month 15. The CNY 4,043.31 consumption reduction comprises 3,998.78 from executing the recovery-selected plan under the changed budget and 44.53 from plan reselection.

\begin{center}\begin{minipage}{\textwidth}\fontsize{8.2}{10}\selectfont
Table S37 | Fixed-plan references in the supporting sixteen-month application
\par\vspace{4pt}
\begin{tabularx}{\textwidth}{@{}Xrrrr@{}}\toprule
Policy & Fully funded (\%) & Consumption & Closing cash & Closing debt \\
\midrule
Always C0 & 88.05 & 85,299.26 & 73,315.33 & 44,419.50 \\
Always C1 & 77.60 & 87,991.01 & 70,745.46 & 44,541.38 \\
YN-mini & 87.00 & 86,445.86 & 72,199.99 & 44,450.76 \\
Original Laya & 79.30 & 88,319.86 & 70,843.45 & 45,025.26 \\
\bottomrule\end{tabularx}
\par\vspace{3pt}{\fontsize{7.7}{9.3}\selectfont The same 2,000 households, original wording and order/IDs, available full menu and sixteen-month ledger enter every row. Monetary values are CNY per household. C0 restrains adjustable consumption during the initial shock; C1 targets regular consumption. A fully funded plan has no shortfall in selected consumption or scheduled debt payments. The menu contains at least one fully funded plan for 1,761 households. Fixed references execute existing plans without additional model calls.}
\end{minipage}\end{center}

\section*{Appendix S13. Consumption heterogeneity in the 12M application}

Tables S38--S40 report age, income, financial stocks and consumption components; Tables S41--S42 hold household composition common for the financial-stock comparisons. Table S43 adds the funding and purchase-choice channels for education, employment and household size, alongside the variation increments for all 18 additional profile attributes. Table S44 retains all 91 occupied categories, including unknown groups.

\begin{center}
\begin{minipage}{\textwidth}
\fontsize{8}{9.8}\selectfont
Table S38 | Consumption and financial channels by age and income
\par\vspace{4pt}
\begin{tabularx}{\textwidth}{@{}Xrrrrr@{}}
\toprule
Profile group & Households & Consumption $\Delta$ & Cash $\Delta$ & Debt $\Delta$ & Actual payments $\Delta$ \\
\midrule
\multicolumn{6}{l}{Decision-maker age} \\
{18--24} & 286,623 & 6,339.29 & -1,766.25 & 4,828.24 & 1,782.07 \\
{25--34} & 1,883,785 & 9,161.69 & -2,961.16 & 6,562.22 & 2,974.16 \\
{35--44} & 1,850,016 & 5,953.92 & -1,951.95 & 4,236.36 & 1,958.56 \\
{45--54} & 2,589,544 & 1,320.65 & -420.41 & 952.48 & 421.67 \\
{55--64} & 2,025,895 & 12.88 & -2.78 & 10.63 & 2.78 \\
{65+} & 2,364,307 & 0.00 & 0.00 & 0.00 & 0.00 \\
{Unknown} & 999,830 & 1,109.93 & -355.32 & 798.47 & 356.31 \\
\addlinespace[3pt]
\multicolumn{6}{l}{Annual household income (CNY)} \\
{[0, 20,000)} & 1,351,144 & 1,059.67 & -58.41 & 1,048.34 & 60.45 \\
{[20,000, 50,000)} & 2,310,535 & 2,085.36 & -407.45 & 1,765.10 & 414.37 \\
{[50,000, 100,000)} & 3,812,459 & 2,753.65 & -860.61 & 2,002.24 & 864.64 \\
{[100,000, 200,000)} & 3,080,612 & 3,427.51 & -1,206.00 & 2,355.02 & 1,209.24 \\
{[200,000, 500,000)} & 1,373,056 & 5,026.86 & -2,127.22 & 3,088.72 & 2,128.21 \\
{$\geq$500,000} & 72,194 & 6,052.72 & -2,877.92 & 3,397.27 & 2,877.92 \\
\addlinespace[3pt]
\bottomrule
\end{tabularx}
\par\vspace{3pt}{\fontsize{7.7}{9.3}\selectfont CNY per household over twelve months; $\Delta$ is 30k minus 20k goal and tied limit for the same household. Cash and debt are closing stocks; actual payments and consumption are cumulative. Each marginal partition retains all 12M households. Income and financial-asset intervals include their lower and exclude their upper endpoint; debt intervals exclude their lower and include their upper endpoint. Unknown denotes an unclassified profile field.}
\end{minipage}
\end{center}
\clearpage
\begin{center}
\begin{minipage}{\textwidth}
\fontsize{8}{9.8}\selectfont
Table S39 | Consumption and financial channels by financial stocks
\par\vspace{4pt}
\begin{tabularx}{\textwidth}{@{}Xrrrrr@{}}
\toprule
Profile group & Households & Consumption $\Delta$ & Cash $\Delta$ & Debt $\Delta$ & Actual payments $\Delta$ \\
\midrule
\multicolumn{6}{l}{Financial asset stock (CNY)} \\
{[0, 10,000)} & 3,743,151 & 2,580.88 & -670.55 & 2,015.27 & 670.55 \\
{[10,000, 50,000)} & 2,719,820 & 2,601.10 & -908.20 & 1,793.95 & 911.45 \\
{[50,000, 200,000)} & 1,859,166 & 2,485.13 & -1,076.97 & 1,501.57 & 1,093.86 \\
{[200,000, 500,000)} & 378,504 & 1,932.62 & -900.19 & 1,103.90 & 910.57 \\
{[500,000, 1,000,000)} & 58,469 & 2,400.45 & -1,137.38 & 1,351.73 & 1,155.82 \\
{[1,000,000, 2,000,000)} & 185,377 & 3,054.60 & -1,470.71 & 1,695.84 & 1,471.99 \\
{[2,000,000, 5,000,000)} & 111,221 & 3,085.48 & -1,488.70 & 1,709.77 & 1,488.70 \\
{[5,000,000, 10,000,000)} & 55,373 & 2,779.06 & -1,341.13 & 1,539.70 & 1,341.13 \\
{$\geq$10,000,000} & 18,683 & 2,759.73 & -1,331.80 & 1,528.99 & 1,331.80 \\
{Unknown} & 2,870,236 & 3,938.38 & -1,121.23 & 2,975.30 & 1,121.23 \\
\addlinespace[3pt]
\multicolumn{6}{l}{Debt principal stock (CNY)} \\
{No debt} & 6,774,791 & 2,163.34 & -751.92 & 1,495.58 & 755.74 \\
{(0, 10,000]} & 323,366 & 1,815.25 & -643.06 & 1,242.48 & 644.33 \\
{(10,000, 20,000]} & 117,583 & 3,449.65 & -1,211.95 & 2,371.62 & 1,214.62 \\
{(20,000, 50,000]} & 208,558 & 3,455.38 & -1,200.16 & 2,389.69 & 1,203.54 \\
{(50,000, 100,000]} & 254,046 & 3,202.63 & -1,091.81 & 2,236.02 & 1,096.06 \\
{(100,000, 200,000]} & 309,937 & 3,990.29 & -1,259.01 & 2,889.14 & 1,262.36 \\
{(200,000, 300,000]} & 208,232 & 3,619.08 & -1,055.29 & 2,708.95 & 1,056.88 \\
{(300,000, 500,000]} & 390,806 & 4,098.32 & -1,037.35 & 3,227.74 & 1,039.43 \\
{(500,000, 1,000,000]} & 353,945 & 3,035.16 & -627.62 & 2,533.57 & 628.94 \\
{(1,000,000, 2,000,000]} & 156,096 & 3,442.46 & -417.06 & 3,172.88 & 418.57 \\
{(2,000,000, 5,000,000]} & 33,255 & 2,970.19 & -111.48 & 2,990.43 & 112.12 \\
{>5,000,000} & 31 & 322.58 & 0.00 & 337.37 & 0.00 \\
{Unknown} & 2,869,354 & 4,238.15 & -1,347.54 & 3,059.09 & 1,352.50 \\
\addlinespace[3pt]
\bottomrule
\end{tabularx}
\par\vspace{3pt}{\fontsize{7.7}{9.3}\selectfont CNY per household over twelve months; $\Delta$ is 30k minus 20k goal and tied limit for the same household. Cash and debt are closing stocks; actual payments and consumption are cumulative. Each marginal partition retains all 12M households. Income and financial-asset intervals include their lower and exclude their upper endpoint; debt intervals exclude their lower and include their upper endpoint. Unknown denotes an unclassified profile field.}
\end{minipage}
\end{center}
\clearpage
\begin{center}
\begin{minipage}{\textwidth}
\fontsize{8}{9.8}\selectfont
Table S40 | Income-standardized age contrasts and consumption components
\par\vspace{4pt}
\begin{tabularx}{\textwidth}{@{}Xrrr@{}}
\toprule
Decision-maker age & Households & Raw consumption $\Delta$ & Income-standardized $\Delta$ \\
\midrule
{18--24} & 286,623 & 6,339.29 & 6,343.15 \\
{25--34} & 1,883,785 & 9,161.69 & 9,181.62 \\
{35--44} & 1,850,016 & 5,953.92 & 5,856.21 \\
{45--54} & 2,589,544 & 1,320.65 & 1,364.57 \\
{55--64} & 2,025,895 & 12.88 & 12.77 \\
{65+} & 2,364,307 & 0.00 & 0.00 \\
{Unknown} & 999,830 & 1,109.93 & 1,108.03 \\

\bottomrule
\end{tabularx}
\par\vspace{8pt}
\begin{tabularx}{\textwidth}{@{}Xrrrrr@{}}
\toprule
Annual household income (CNY) & Purchase $\Delta$ & Routine $\Delta$ & Interest $\Delta$ & Total consumption $\Delta$ & Age 25--34 consumption $\Delta$ \\
\midrule
{[0, 20,000)} & 1,061.72 & -2.04 & 47.07 & 1,059.67 & 9,945.53 \\
{[20,000, 50,000)} & 2,092.28 & -6.93 & 87.19 & 2,085.36 & 9,380.60 \\
{[50,000, 100,000)} & 2,757.68 & -4.03 & 109.20 & 2,753.65 & 8,892.69 \\
{[100,000, 200,000)} & 3,430.75 & -3.25 & 133.51 & 3,427.51 & 8,306.18 \\
{[200,000, 500,000)} & 5,027.86 & -0.99 & 189.08 & 5,026.86 & 10,714.27 \\
{$\geq$500,000} & 6,052.72 & 0.00 & 222.47 & 6,052.72 & 11,980.83 \\

\bottomrule
\end{tabularx}
\par\vspace{3pt}{\fontsize{7.7}{9.3}\selectfont CNY per household. Standardization weights each age group's six income-specific paired consumption means by the full panel's income distribution. All 42 age--income cells are occupied. The lower panel decomposes total consumption into the initial purchase and twelve-month routine spending; its final column shows the nonmonotonic income pattern within ages 25--34. The corresponding age-65+ contrasts are zero in every income bracket. Interest is cumulative accrued interest and enters the closing-cash/debt identity. Source aggregates retain the full group means and synthetic-row variability references.}
\end{minipage}
\end{center}

\begin{center}
\begin{minipage}{\textwidth}
\fontsize{8}{9.8}\selectfont
Table S41 | Financial-stock comparisons under a common age--income mixture
\par\vspace{4pt}
\begin{tabularx}{\textwidth}{@{}Xrrrrrr@{}}
\toprule
Financial-stock group & Households & \shortstack[r]{Raw\\consumption $\Delta$} & \shortstack[r]{Standardized\\consumption $\Delta$} & \shortstack[r]{Cash\\drawdown} & \shortstack[r]{Actual\\payments $\Delta$} & \shortstack[r]{Closing\\debt $\Delta$} \\
\midrule
Assets below 10,000 & 3,743,151 & 2,580.88 & 3,014.26 & 905.61 & 905.61 & 2,229.04 \\
Assets 1--2 million & 185,377 & 3,054.60 & 2,361.25 & 1,136.28 & 1,137.87 & 1,311.52 \\
No opening debt & 6,774,791 & 2,163.34 & 2,864.61 & 1,059.38 & 1,063.72 & 1,915.54 \\
Opening debt 2--5 million & 33,255 & 2,970.19 & 1,811.13 & 43.97 & 44.31 & 1,848.23 \\

\bottomrule
\end{tabularx}
\par\vspace{3pt}{\fontsize{7.7}{9.3}\selectfont Amounts are CNY per household over twelve months; $\Delta$ is 30k minus 20k purchase goal and tied approved limit. Cash drawdown is minus the closing-cash difference. The cash, payment and debt columns use the standardized mixture; the first consumption column gives each group's raw mean. Each reported group covers all 42 decision-maker-age by household-income cells and uses the same full-panel reference weights. Tables S38--S39 preserve every marginal group, and source aggregates give the category-level standardization.}
\end{minipage}
\end{center}

\begin{center}
\begin{minipage}{\textwidth}
\fontsize{8}{9.8}\selectfont
Table S42 | Profile groupings and variation in paired financial responses
\par\vspace{4pt}
\begin{tabularx}{\textwidth}{@{}Xrrr@{}}
\toprule
Marginal grouping & Consumption & Closing cash & Actual payments \\
\midrule
Decision-maker age & 16.1842\% & 10.7556\% & 10.8359\% \\
Annual household income & 1.6048\% & 3.0395\% & 3.0250\% \\
Financial asset stock & 0.5070\% & 0.3678\% & 0.3746\% \\
Debt principal stock & 1.1950\% & 0.6661\% & 0.6681\% \\

\bottomrule
\end{tabularx}
\par\vspace{8pt}
\begin{tabularx}{\textwidth}{@{}Xrrr@{}}
\toprule
Nested grouping & Consumption & Closing cash & Actual payments \\
\midrule
Age $\times$ income & 16.6635\% & 12.9679\% & 13.0142\% \\
Age $\times$ income $\times$ assets & 17.1740\% & 13.2838\% & 13.3584\% \\
Age $\times$ income $\times$ debt & 18.0637\% & 15.1056\% & 15.1615\% \\
All four attributes & 18.9879\% & 15.8103\% & 15.8912\% \\

\bottomrule
\end{tabularx}
\par\vspace{3pt}{\fontsize{7.7}{9.3}\selectfont Entries are the between-group sum of squares divided by the total sum of squares of the within-household paired response, expressed as percentages. Marginal rows group one attribute at a time; nested rows refine the common age--income partition. Every row uses all 12M households and retains unknown categories. The nested increments reported in the text subtract the first nested row. M10 defines the calculation from the saved paired sums and squared sums.}
\end{minipage}
\end{center}

\clearpage
\begin{center}\begin{minipage}{\textwidth}\fontsize{7.8}{9.5}\selectfont
Table S43 | Additional household attributes and the channels of consumption response
\par\vspace{4pt}
\begin{tabularx}{\textwidth}{@{}Xrrrrr@{}}\toprule
Standardized response contrast & Consumption & Closing debt & Cash drawdown & $-$Interest & Actual payments \\
\midrule
College minus primary or below & 3,422.87 & 2,555.05 & 1,004.78 & -136.97 & 1,009.48 \\

Employed minus not working & 2,031.84 & 1,494.71 & 618.00 & -80.88 & 621.37 \\

Three-person minus one-person & 2,130.99 & 1,530.96 & 684.12 & -84.09 & 686.76 \\

\bottomrule\end{tabularx}
\par\vspace{8pt}
\begin{tabularx}{\textwidth}{@{}Xrrrrr@{}}\toprule
Standardized response contrast & \shortstack[r]{Deferral to\\purchase} & \shortstack[r]{Same choice,\\larger amount} & \shortstack[r]{Other choice\\transitions} & Purchase & Routine spending \\
\midrule
College minus primary or below & 3,183.17 & 234.46 & 9.94 & 3,427.57 & -4.70 \\

Employed minus not working & 1,822.37 & 213.57 & -0.73 & 2,035.21 & -3.37 \\

Three-person minus one-person & 1,948.40 & 183.08 & 2.14 & 2,133.62 & -2.63 \\

\bottomrule\end{tabularx}
\par\vspace{8pt}
\begin{tabularx}{\textwidth}{@{}Xrrr@{}}\toprule
Attribute added to age $\times$ income & Consumption & Closing cash & Actual payments \\
\midrule
Household size & 1.8624 & 1.3688 & 1.3748 \\

Child under 18 & 0.1473 & 0.0895 & 0.0904 \\

Member aged 65+ & 3.1344 & 2.3232 & 2.3328 \\

Employment status & 2.8286 & 1.5554 & 1.5676 \\

Main income source & 0.2665 & 0.1712 & 0.1728 \\

Housing tenure & 0.0661 & 0.0598 & 0.0608 \\

Existing debt type & 1.3944 & 1.7465 & 1.7547 \\

Risk preference & 0.1288 & 0.0857 & 0.0864 \\

Financial literacy & 0.3280 & 0.1826 & 0.1838 \\

Investment activity & 1.9069 & 1.4829 & 1.4893 \\

Education & 2.5609 & 1.2520 & 1.2630 \\

Urban/rural residence & 0.1688 & 0.1151 & 0.1171 \\

Region & 0.3127 & 0.1924 & 0.1937 \\

Gender & 0.1231 & 0.1071 & 0.1075 \\

Marital status & 0.3800 & 0.2493 & 0.2509 \\

Income--expenditure status & 0.0709 & 0.1534 & 0.1461 \\

General trust & 0.4883 & 0.3271 & 0.3295 \\

Annual household spending (CNY) & 0.5958 & 4.8452 & 4.7294 \\

\bottomrule\end{tabularx}
\par\vspace{3pt}{\fontsize{7.5}{9.1}\selectfont The first two panels give CNY per household, comparing the 30k-minus-20k response between profile groups after weighting each to the full panel's 42 age--income shares. Closing debt plus cash drawdown minus interest equals consumption; purchase plus routine spending gives the same total. Table S44 gives group sizes. The lower panel gives the added between-group share of paired-response variation in percentage points, above the age--income shares of 16.6635\%, 12.9679\% and 13.0142\%, respectively. Each nested partition uses all 12M households, retaining unknown categories.}
\end{minipage}\end{center}
\clearpage
\begin{center}\begin{minipage}{\textwidth}\fontsize{7.7}{9.3}\selectfont
Table S44 | Complete additional profile groups
\par\vspace{4pt}
\begin{tabularx}{\textwidth}{@{}Xrrrrrr@{}}\toprule
Profile category & Households & \shortstack[r]{Raw\\consumption} & \shortstack[r]{Standardized\\consumption} & \shortstack[r]{Cash\\drawdown} & \shortstack[r]{Closing\\debt} & \shortstack[r]{Actual\\payments} \\
\midrule
\multicolumn{7}{l}{\textit{Household size}} \\

1 & 2,095,057 & 2,817.32 & 1,666.17 & 542.04 & 1,189.89 & 544.35 \\

2 & 3,035,740 & 1,996.56 & 3,046.54 & 982.41 & 2,184.43 & 986.50 \\

3 & 2,835,219 & 3,679.33 & 3,797.16 & 1,226.16 & 2,720.85 & 1,231.11 \\

4 & 2,143,688 & 3,729.14 & 3,346.76 & 1,066.37 & 2,412.70 & 1,070.95 \\

5 or more & 1,890,296 & 2,251.97 & 2,631.32 & 825.01 & 1,910.55 & 828.76 \\

\addlinespace[3pt]
\multicolumn{7}{l}{\textit{Child under 18}} \\

Yes & 2,201,358 & 3,498.94 & 3,026.77 & 961.32 & 2,185.18 & 965.44 \\

No & 9,413,213 & 2,693.79 & 2,799.27 & 903.67 & 2,006.11 & 907.30 \\

Unknown & 385,429 & 4,116.29 & 4,105.07 & 1,296.06 & 2,971.51 & 1,302.44 \\

\addlinespace[3pt]
\multicolumn{7}{l}{\textit{Member aged 65+}} \\

Yes & 4,202,185 & 124.72 & 368.65 & 122.69 & 260.46 & 122.91 \\

No & 7,412,774 & 4,333.31 & --- & --- & --- & --- \\

Unknown & 385,041 & 5,194.86 & 5,197.20 & 1,612.24 & 3,791.23 & 1,620.10 \\

\addlinespace[3pt]
\multicolumn{7}{l}{\textit{Employment status}} \\

Employed & 4,978,882 & 5,029.75 & 3,437.93 & 1,063.20 & 2,511.29 & 1,068.04 \\

Self-employed & 959,304 & 5,322.91 & 4,210.97 & 1,296.93 & 3,081.42 & 1,302.83 \\

Farming & 2,785,684 & 203.71 & 429.51 & 141.57 & 304.83 & 142.20 \\

Not working & 1,349,210 & 1,502.71 & 1,406.09 & 445.20 & 1,016.58 & 446.67 \\

Retired & 1,326,412 & 50.02 & --- & --- & --- & --- \\

Unknown & 600,508 & 3,057.55 & 3,058.30 & 975.44 & 2,203.81 & 979.75 \\

\addlinespace[3pt]
\multicolumn{7}{l}{\textit{Main income source}} \\

Wages & 7,513,822 & 3,426.27 & 2,910.63 & 930.86 & 2,094.82 & 934.70 \\

Business & 1,094,749 & 2,452.13 & 2,876.46 & 950.75 & 2,038.85 & 954.25 \\

Property & 176,410 & 2,012.11 & 2,133.60 & 674.25 & 1,543.87 & 676.18 \\

Transfers & 2,161,380 & 1,065.27 & 2,317.38 & 755.03 & 1,653.71 & 757.88 \\

Mixed & 262,085 & 2,590.17 & 2,908.93 & 932.82 & 2,091.09 & 936.70 \\

Unknown & 791,554 & 3,639.76 & 3,648.49 & 1,173.77 & 2,618.80 & 1,179.25 \\

\addlinespace[3pt]
\multicolumn{7}{l}{\textit{Housing tenure}} \\

Owned & 10,564,041 & 2,744.13 & 2,876.62 & 918.25 & 2,072.12 & 921.89 \\

Rented & 620,423 & 5,028.21 & 2,705.67 & 904.78 & 1,907.09 & 908.99 \\

Other & 745,321 & 3,142.92 & 3,260.14 & 1,100.48 & 2,287.40 & 1,105.76 \\

Unknown & 70,215 & 2,776.87 & 2,764.88 & 893.44 & 1,980.58 & 897.16 \\

\addlinespace[3pt]
\bottomrule\end{tabularx}
\par\vspace{3pt}{\fontsize{7.5}{9.1}\selectfont CNY per household over twelve months, 30k minus 20k goal and tied approved limit. All financial columns after raw consumption use the common 42-cell age--income mixture. Cash drawdown is minus the closing-cash difference. Dashes mark categories missing at least one reference cell. Each attribute partition contains all 12M households. College and graduate education remain separate categories. No stated risk preference preserves the source's none category; unknown is retained separately. Annual spending intervals include the lower and exclude the upper endpoint.}
\end{minipage}\end{center}
\clearpage
\begin{center}\begin{minipage}{\textwidth}\fontsize{7.7}{9.3}\selectfont
Table S44 (continued) | Complete additional profile groups
\par\vspace{4pt}
\begin{tabularx}{\textwidth}{@{}Xrrrrrr@{}}\toprule
Profile category & Households & \shortstack[r]{Raw\\consumption} & \shortstack[r]{Standardized\\consumption} & \shortstack[r]{Cash\\drawdown} & \shortstack[r]{Closing\\debt} & \shortstack[r]{Actual\\payments} \\
\midrule
\multicolumn{7}{l}{\textit{Existing debt type}} \\

Consumer credit & 540,837 & 4,100.82 & 2,840.90 & 538.76 & 2,422.49 & 540.98 \\

Consumer credit + Education & 45 & 5,111.11 & --- & --- & --- & --- \\

Education & 846 & 3,875.51 & --- & --- & --- & --- \\

Education + Medical & 188 & 5,053.19 & --- & --- & --- & --- \\

Housing & 1,687,505 & 2,954.00 & 1,880.61 & 525.89 & 1,429.72 & 527.30 \\

Housing + Consumer credit & 124,640 & 4,943.47 & 2,212.86 & 430.41 & 1,875.72 & 433.38 \\

Housing + Education & 121 & 5,867.77 & --- & --- & --- & --- \\

Housing + Medical & 101 & 7,326.73 & --- & --- & --- & --- \\

Medical & 1,572 & 2,830.79 & --- & --- & --- & --- \\

No debt & 6,774,791 & 2,163.34 & 2,864.61 & 1,059.38 & 1,915.54 & 1,063.72 \\

Unknown & 2,869,354 & 4,238.15 & 3,697.26 & 1,123.29 & 2,721.95 & 1,128.00 \\

\addlinespace[3pt]
\multicolumn{7}{l}{\textit{Risk preference}} \\

High & 330,309 & 3,363.65 & 2,702.59 & 866.92 & 1,942.46 & 870.56 \\

Slightly high & 464,118 & 5,337.10 & 2,949.84 & 928.94 & 2,137.76 & 933.02 \\

Average & 1,843,098 & 4,105.33 & 2,706.32 & 871.77 & 1,941.42 & 875.18 \\

Slightly low & 1,685,994 & 3,766.38 & 3,038.47 & 971.16 & 2,187.43 & 975.13 \\

No stated risk preference & 5,438,837 & 1,828.61 & 2,761.75 & 887.57 & 1,983.27 & 891.20 \\

Unknown & 2,237,644 & 3,215.87 & 3,213.28 & 1,038.19 & 2,301.90 & 1,042.60 \\

\addlinespace[3pt]
\multicolumn{7}{l}{\textit{Financial literacy}} \\

Both correct & 1,861,069 & 4,903.51 & 3,158.63 & 984.45 & 2,299.54 & 988.81 \\

Interest only correct & 855,268 & 3,400.22 & 2,889.27 & 932.34 & 2,070.97 & 936.31 \\

Inflation only correct & 2,071,454 & 3,938.31 & 3,403.68 & 1,081.06 & 2,457.31 & 1,085.71 \\

Neither correct & 7,088,293 & 1,982.16 & 2,555.86 & 830.61 & 1,826.03 & 833.88 \\

Unknown & 123,916 & 3,261.96 & 3,249.77 & 1,044.83 & 2,333.29 & 1,049.24 \\

\addlinespace[3pt]
\multicolumn{7}{l}{\textit{Investment activity}} \\

Funds & 300,995 & 7,084.99 & 3,862.86 & 1,164.25 & 2,852.60 & 1,169.03 \\

No investment activity & 8,810,256 & 2,207.84 & 2,344.71 & 744.64 & 1,692.87 & 747.73 \\

Other & 48,853 & 4,340.00 & --- & --- & --- & --- \\

Stocks & 267,266 & 5,390.89 & 4,484.23 & 1,375.62 & 3,286.96 & 1,381.82 \\

Stocks + Funds & 172,729 & 7,532.84 & --- & --- & --- & --- \\

Unknown & 2,399,901 & 4,211.87 & 4,212.95 & 1,354.36 & 3,024.95 & 1,360.32 \\

\addlinespace[3pt]
\multicolumn{7}{l}{\textit{Education}} \\

Primary or below & 3,765,905 & 196.85 & 398.12 & 127.34 & 286.53 & 127.85 \\

Junior secondary & 3,526,704 & 2,692.61 & 2,760.94 & 872.58 & 1,997.69 & 876.70 \\

Senior/vocational & 1,773,107 & 3,217.80 & 3,227.56 & 999.82 & 2,355.92 & 1,004.25 \\

College & 2,166,144 & 6,706.66 & 3,820.99 & 1,132.12 & 2,841.58 & 1,137.33 \\

Graduate & 167,396 & 10,080.55 & --- & --- & --- & --- \\

Unknown & 600,744 & 4,142.03 & 4,145.38 & 1,297.74 & 3,011.96 & 1,303.45 \\

\addlinespace[3pt]
\multicolumn{7}{l}{\textit{Urban/rural residence}} \\

Urban & 7,052,749 & 3,355.19 & 2,984.50 & 915.89 & 2,187.33 & 919.33 \\

Rural & 4,383,957 & 2,061.56 & 2,677.15 & 920.34 & 1,861.40 & 924.88 \\

Unknown & 563,294 & 3,453.09 & 3,505.62 & 1,209.81 & 2,432.67 & 1,215.95 \\

\addlinespace[3pt]
\bottomrule\end{tabularx}
\par\vspace{3pt}{\fontsize{7.5}{9.1}\selectfont CNY per household over twelve months, 30k minus 20k goal and tied approved limit. All financial columns after raw consumption use the common 42-cell age--income mixture. Cash drawdown is minus the closing-cash difference. Dashes mark categories missing at least one reference cell. Each attribute partition contains all 12M households. College and graduate education remain separate categories. No stated risk preference preserves the source's none category; unknown is retained separately. Annual spending intervals include the lower and exclude the upper endpoint.}
\end{minipage}\end{center}
\clearpage
\begin{center}\begin{minipage}{\textwidth}\fontsize{7.7}{9.3}\selectfont
Table S44 (continued) | Complete additional profile groups
\par\vspace{4pt}
\begin{tabularx}{\textwidth}{@{}Xrrrrrr@{}}\toprule
Profile category & Households & \shortstack[r]{Raw\\consumption} & \shortstack[r]{Standardized\\consumption} & \shortstack[r]{Cash\\drawdown} & \shortstack[r]{Closing\\debt} & \shortstack[r]{Actual\\payments} \\
\midrule
\multicolumn{7}{l}{\textit{Region}} \\

East & 4,504,091 & 3,199.82 & 2,795.16 & 912.23 & 1,993.08 & 915.69 \\

Central & 2,933,392 & 2,243.66 & 2,471.01 & 784.15 & 1,784.68 & 787.34 \\

West & 2,918,945 & 3,144.45 & 3,159.90 & 1,017.29 & 2,267.39 & 1,021.77 \\

Northeast & 1,643,572 & 2,722.05 & 3,353.68 & 1,071.89 & 2,414.35 & 1,076.04 \\

\addlinespace[3pt]
\multicolumn{7}{l}{\textit{Gender}} \\

Male & 6,422,653 & 2,822.42 & 3,071.77 & 994.34 & 2,198.64 & 998.34 \\

Female & 4,677,144 & 3,017.81 & 2,698.21 & 864.12 & 1,940.73 & 867.75 \\

Unknown & 900,203 & 2,670.57 & 2,726.76 & 859.91 & 1,974.82 & 863.17 \\

\addlinespace[3pt]
\multicolumn{7}{l}{\textit{Marital status}} \\

Married & 9,329,082 & 2,755.57 & 3,067.41 & 981.29 & 2,207.34 & 985.44 \\

Cohabiting & 55,067 & 3,940.78 & 3,141.31 & 1,000.41 & 2,265.21 & 1,004.25 \\

Previously married & 1,094,434 & 1,394.88 & 3,074.01 & 988.79 & 2,206.65 & 992.45 \\

Never married & 920,976 & 5,677.16 & 2,284.61 & 741.66 & 1,633.14 & 744.87 \\

Unknown & 600,441 & 3,276.07 & 3,285.44 & 1,064.18 & 2,350.86 & 1,068.88 \\

\addlinespace[3pt]
\multicolumn{7}{l}{\textit{Income--expenditure status}} \\

Surplus & 4,211,945 & 3,128.90 & 2,800.48 & 988.89 & 1,920.57 & 990.85 \\

Balance & 4,876,542 & 2,896.48 & 3,013.81 & 957.51 & 2,175.48 & 961.00 \\

Deficit & 2,818,405 & 2,499.10 & 2,900.60 & 789.89 & 2,227.97 & 797.00 \\

Unknown & 93,108 & 3,212.82 & 3,179.56 & 1,032.81 & 2,272.16 & 1,037.10 \\

\addlinespace[3pt]
\multicolumn{7}{l}{\textit{General trust}} \\

Trust & 6,567,125 & 2,748.88 & 2,937.96 & 938.60 & 2,115.51 & 942.53 \\

Depends & 1,342,964 & 3,101.72 & 2,577.88 & 827.80 & 1,851.95 & 831.10 \\

Distrust & 1,689,871 & 2,384.35 & 2,193.51 & 713.91 & 1,566.09 & 716.55 \\

Unknown & 2,400,040 & 3,499.61 & 3,503.40 & 1,134.43 & 2,507.15 & 1,139.15 \\

\addlinespace[3pt]
\multicolumn{7}{l}{\textit{Annual household spending (CNY)}} \\

Below 20,000 & 1,925,783 & 2,671.05 & 4,101.64 & 1,810.63 & 2,444.34 & 1,810.63 \\

20,000--50,000 & 3,651,505 & 2,589.10 & 2,975.96 & 1,223.33 & 1,865.44 & 1,224.12 \\

50,000--100,000 & 3,504,275 & 2,561.22 & 2,377.91 & 708.29 & 1,764.68 & 716.00 \\

100,000--200,000 & 1,571,594 & 3,624.69 & 2,858.00 & 412.46 & 2,568.33 & 431.92 \\

200,000--500,000 & 597,743 & 4,194.46 & 2,899.03 & 222.55 & 2,804.79 & 237.35 \\

500,000 or more & 31,471 & 4,793.84 & 3,274.04 & 99.28 & 3,323.30 & 121.33 \\

Unknown & 717,629 & 3,788.00 & 3,345.67 & 1,152.40 & 2,323.70 & 1,152.40 \\

\addlinespace[3pt]
\bottomrule\end{tabularx}
\par\vspace{3pt}{\fontsize{7.5}{9.1}\selectfont CNY per household over twelve months, 30k minus 20k goal and tied approved limit. All financial columns after raw consumption use the common 42-cell age--income mixture. Cash drawdown is minus the closing-cash difference. Dashes mark categories missing at least one reference cell. Each attribute partition contains all 12M households. College and graduate education remain separate categories. No stated risk preference preserves the source's none category; unknown is retained separately. Annual spending intervals include the lower and exclude the upper endpoint.}
\end{minipage}\end{center}

\clearpage
\clearpage
\section*{Supplementary sandbox example: Repeated household decisions}
\begin{figure}[H]
\centering
\includegraphics[width=\textwidth]{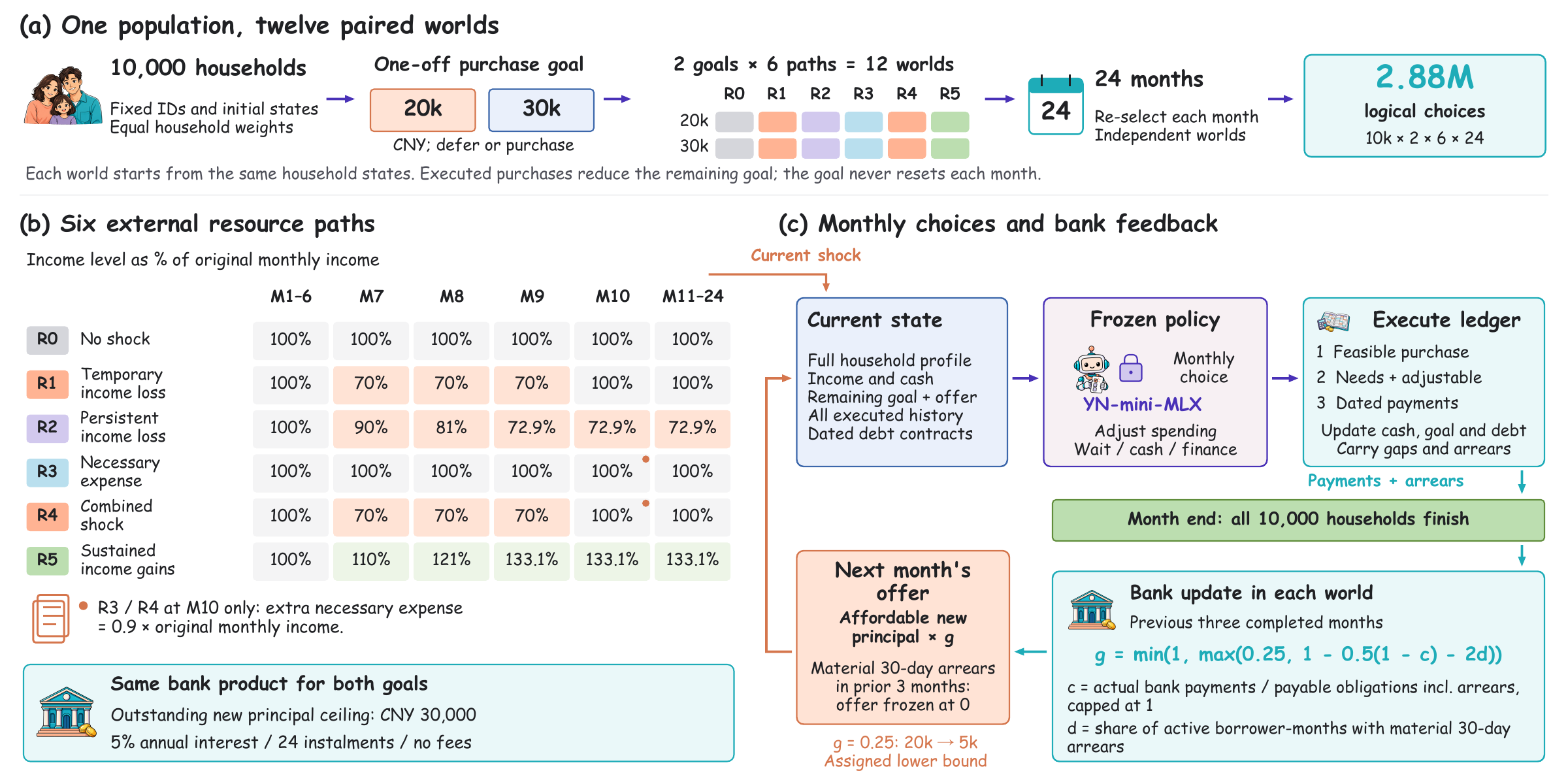}
\caption*{Supplementary Figure S8 | A repeated-decision sandbox with six assigned resource paths. The fixed panel faces CNY 20k and 30k purchase goals for 24 months, yielding 2.88M logical choices. R0 retains income; R1 applies a temporary loss; R2 a persistent loss; R3 a one-time necessary expense; R4 combines the temporary loss and expense; R5 raises income persistently. All worlds share a CNY 30k ceiling on outstanding new principal, 5\% annual interest, 24 installments and zero fees. Each world's lagged payment coverage and arrears determine its bank factor; household affordability and material arrears determine individual offers.}
\makeatletter\def\@currentlabel{S8}\phantomsection\label{figure:monthly-scenario}\makeatother
\end{figure}

This sandbox example illustrates how the decision interface can support repeated choices and bank feedback (Supplementary Figure~\ref{figure:monthly-scenario}). A fixed panel of 10,000 households faces two purchase goals and six assigned resource paths over 24 months. Purchases reduce the remaining goal; cash, dated obligations and arrears pass to the next decision. Payment records update the following month's bank offers.

\end{document}

%% file: household_experiment.tex
\section*{7. Learned decision policies for heterogeneous households}

Large heterogeneous-agent simulations require decision rules that respond to diverse household profiles and financial states. We compared fixed candidate-scoring policies through retrospective annual consumption prediction and information-timing tests. Standalone YN-Lite-4B and YN-master-9B chose directly from household information; advised variants also received a recommendation from a fully rational benchmark. Paired changes in consumption and liquid reserves then show how these policies allocate additional income.

\subsection*{7.1. A shared household problem with three decision rules}

\begin{figure}[H]
\centering
\includegraphics[width=\textwidth]{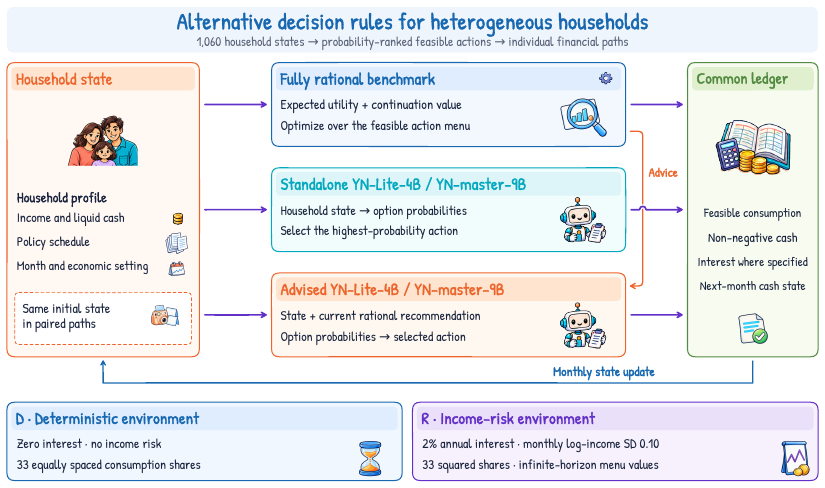}
\caption{\textbf{Standalone and advised learned policies share a household accounting interface.} The fully rational benchmark optimizes consumption and saving. Standalone policies map household information directly to candidate probabilities; advised policies additionally receive a rational recommendation evaluated at their own current cash state. Argmax selects a feasible action and the ledger advances the household. Standalone results are available in D; rational and advised comparisons are available in both D and R, using the same households and annual evaluation target.}
\label{hh:fig:scenario}
\end{figure}

We studied 1,060 urban households with positive income, zero observed pre-policy debt and the financial information required to initialize the simulation. Demographic and financial profiles, liquid resources and an announced income path defined each household's state. The policy setting was China's wage-income-tax reform effective 1 October 2018, which raised the monthly basic deduction from CNY 3,500 to 5,000 and revised the wage-income tax brackets~\cite{ref44}. The experiment treats September as the announcement month and October as the first payment month for a permanent tax-saving income component. Each rule generated a baseline and a policy path over September--December, giving 2,120 paths and 8,480 monthly decisions. A common ledger updated cash after every decision (Figure~\ref{hh:fig:scenario}). Monthly regular income was pre-policy annual income divided by twelve; initial cash carried pre-policy liquid balances forward as the September proxy. These fixed inputs define a controlled retrospective household problem. The policies were fixed before outcome scoring; separation of these households from all prior model-development records has not been established.

The deterministic specification, D, used zero interest, constant income and 33 equally spaced consumption shares $s_k=k/32$. The income-risk specification, R, used a 2\% annual real interest rate, uncertain future income and shares $s_k=(k/32)^2$, concentrating options at low consumption rates. Both used $k=0,\ldots,32$. Standalone models and both advised models were evaluated in D; the rational benchmark and both advised models were also evaluated in R. Each specification retains its recorded price convention: D amounts are nominal, whereas R amounts are converted from reference-year prices by dividing by 0.964311. This conversion accounts for the reported income components of 221.12 and 229.31 yuan. D and R compare complete specifications that change prices, interest, risk and the menu together.

The household experiment used the head-only 200k YN-Lite-4B variant: Qwen3.5-4B-Base remained frozen while its decision head was trained. This differs from the LoRA-plus-head 4B checkpoint selected for the benchmark in Section 4.3. YN-master-9B used the selected LoRA-adapted Qwen3.5-9B. Each interface returned probabilities over all 33 candidates; the highest-probability feasible candidate determined consumption. In advised runs, the rational recommendation was recomputed each month at the learned policy's own cash state.

\clearpage
\subsection*{7.2. Standalone policies predicted annual consumption with distinct distributional errors}

We evaluated the forecasts against observed annual consumption, using fixed household weights and 2018 nominal yuan. Annual predictions combined eight-twelfths of pre-policy annual consumption with the four simulated months. This shared prefix is an assigned pre-announcement consumption level, not an observed eight-month outcome:
\begin{equation}
 \widehat C_i=\frac{8}{12}C_i^{\mathrm{pre}}+\sum_{t=9}^{12}c_{it}^{\mathrm{policy}}.
 \label{hh:eq:annual}
\end{equation}
Income and net wealth, each divided by the square root of household size, defined separate weighted tertiles before model evaluation. For normalized weights $w_i$, group mass $W_g=\sum_{i\in g}w_i$ and observed annual consumption $C_i$, group fit was
\begin{equation}
 D=\left(\sum_{g=1}^{3}W_g\mu_g^2\right)^{1/2},\qquad
 \mu_g=\sum_{i\in g}\frac{w_i}{W_g}\left[\log(1+\widehat C_i)-\log(1+C_i)\right].
 \label{hh:eq:group}
\end{equation}
MAE and RMSE summarized household-level amount errors; income and wealth $D$ summarized the weighted root-mean-square of the group mean log errors, including overall prediction bias. Table~\ref{hh:tab:group-mae} additionally reports within-tertile MAE, averaging absolute household errors separately in each income and wealth group.

\begin{table}[H]
\centering\footnotesize
\setlength{\tabcolsep}{3.5pt}
\begin{tabularx}{\textwidth}{@{}Xrrrrr@{}}
\toprule
Decision rule & MAE & RMSE & Predicted mean & Income $D$ & Wealth $D$\\
\midrule
Rational benchmark (D) & 53,128.42 & 139,410.00 & 92,840.11 & 0.19283 & 0.07147\\
Standalone YN-Lite-4B (D) & 53,989.67 & 136,065.36 & 90,543.88 & 0.16533 & 0.08471\\
Standalone YN-master-9B (D) & 51,312.38 & 135,018.99 & 79,745.82 & 0.20688 & 0.16664\\
Advised YN-Lite-4B (D) & 54,899.68 & 135,972.02 & 92,809.03 & 0.17723 & 0.07424\\
Advised YN-master-9B (D) & 53,103.68 & 139,384.72 & 93,083.54 & 0.19104 & 0.07129\\
\midrule
Rational benchmark (R) & 54,650.51 & 140,118.28 & 98,300.12 & 0.20730 & 0.10311\\
Advised YN-Lite-4B (R) & 54,601.82 & 140,107.74 & 96,818.73 & 0.19621 & 0.08966\\
Advised YN-master-9B (R) & 54,636.46 & 140,109.03 & 98,276.71 & 0.20714 & 0.10279\\
\midrule
Continuous rational reference (D) & 52,880.93 & 138,521.14 & 91,940.42 & 0.18880 & 0.06272\\
Continuous rational reference (R) & 54,588.50 & 139,751.83 & 98,168.47 & 0.20689 & 0.10189\\
\midrule
Standalone YN-mini (D) & 72,429.94 & 173,019.10 & 116,041.44 & 0.23321 & 0.17079\\
Advised YN-mini (D) & 67,301.29 & 166,334.61 & 105,796.50 & 0.17193 & 0.11323\\
\midrule Pre-policy consumption carried forward & 52,420.72 & 138,162.11 & -- & -- & --\\
\bottomrule
\end{tabularx}
\caption{\textbf{Annual consumption forecasts for 1,060 households.} Amounts are household-weighted 2018 nominal yuan; the observed mean is 99,300.73 yuan. D denotes the deterministic specification and R the income-risk specification. Comparisons within each specification share the economic environment and menu. Continuous solutions and the 421M variants are included as references. Carry-forward predicts each household's pre-policy consumption without a simulated decision path.}
\label{hh:tab:complete}
\end{table}

Standalone YN-master-9B had the lowest annual MAE and RMSE among the evaluated systems. Against the deterministic rational benchmark, its MAE fell by 3.42\% and RMSE by 3.15\%. Relative to carrying pre-policy consumption forward, MAE was 2.11\% lower and RMSE 2.27\% lower. Its predicted population mean was 79,746 yuan, compared with an observed 99,301 yuan, and its income- and wealth-group distances exceeded the benchmark's. The mean signed error was negative in all six tertiles. Its MAE advantage was concentrated in the highest income group (83,282 versus 90,874 yuan), while the low- and middle-income group errors were higher (Table~\ref{hh:tab:group-mae}). Lower individual amount error therefore coexisted with a systematic downward bias.

\begin{figure}[H]
\centering
\includegraphics[width=\textwidth]{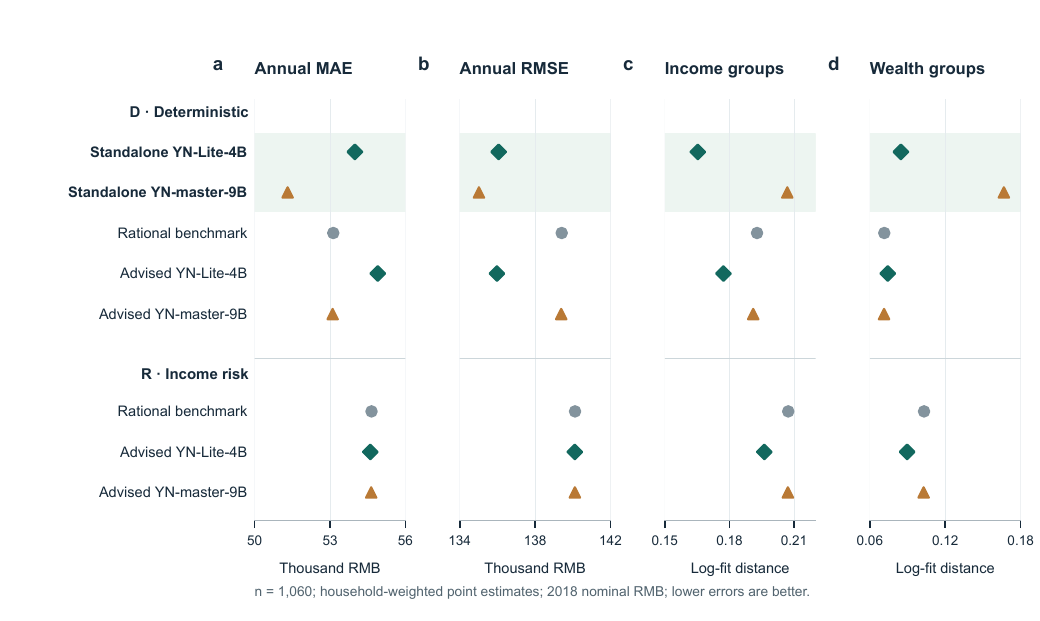}
\caption{\textbf{Standalone policies and rational advice produce different prediction profiles.} Annual MAE, RMSE and income- and wealth-group log-fit distances for the eight principal systems. The two specifications are separated: D uses deterministic income and an equally spaced menu; R includes income risk and a squared-share menu. All forecasts use the same households, weights, price units and observed annual target. Lower values indicate closer fit.}
\label{hh:fig:results}
\end{figure}

Standalone YN-Lite-4B had lower mean log prediction bias across income tertiles. Its income-group distance was 14.26\% lower than the deterministic rational benchmark's and its RMSE was 2.40\% lower, whereas MAE was 1.62\% higher and wealth-group distance was 18.52\% higher. Figure~\ref{hh:fig:results} shows these differences in individual and group prediction errors.

Advice moved the 9B forecast toward the rational benchmark. Under D, its predicted mean rose from 79,746 to 93,084 yuan and wealth $D$ fell from 0.16664 to 0.07129, while MAE rose to 53,103.68 yuan. Under R, advised YN-master-9B closely tracked the benchmark. Advised YN-Lite-4B reduced income and wealth $D$ by 5.35\% and 13.04\% relative to the R benchmark, with an MAE reduction of 0.089\%.

\clearpage
\subsection*{7.3. Household-specific optimization provides a common rational baseline}

Following Kaplan et al.'s consumption--saving household approach~\cite{ref32}, the R benchmark used CRRA utility $u(c)=c^{1-\gamma}/(1-\gamma)$, $\gamma=2$, monthly discounting $\beta_m=0.96^{1/12}$ and return $R_m=1.02^{1/12}$. Future income multipliers were independent mean-one lognormal draws with log standard deviation $0.10$; recorded paths realized their conditional mean. For resources $m$, regular income $\bar y_i$ and policy component $g_{i,t+1}$, the rational policy solved
\begin{equation}
 V_{it}(m)=\max_{s\in\mathcal S}\left\{u(sm)+\beta_m\mathbb{E}\left[V_{i,t+1}\!\left(R_m(1-s)m+\bar y_i\varepsilon_{t+1}+g_{i,t+1}\right)\right]\right\}.
 \label{hh:eq:bellman}
\end{equation}
The solver optimized all 33 shares on a 4,096-point resource grid with seven income-risk quadrature nodes and infinite-horizon same-menu continuation. A continuous endogenous-grid solution~\cite{ref43} supplied the numerical reference. The D benchmark selected $\beta_m$ over $[0.8,0.999999]$ by minimizing the household-weighted mean squared error in pre-policy monthly $\log(1+c)$; the solution reached the upper boundary. This calibration used only pre-policy consumption. D used exact continuous continuation after four menu periods.

Income and wealth groups defined evaluation strata; optimization used each household's income, cash and policy component with common preferences. R parameters were fixed before outcome scoring. Within each specification, the rational economic agent and learned policies faced the same menu, income schedule, cash constraint and accounting equations.

\subsection*{7.4. Future-information tests reveal when model decisions change}

\begin{figure}[H]
\centering
\includegraphics[width=\textwidth]{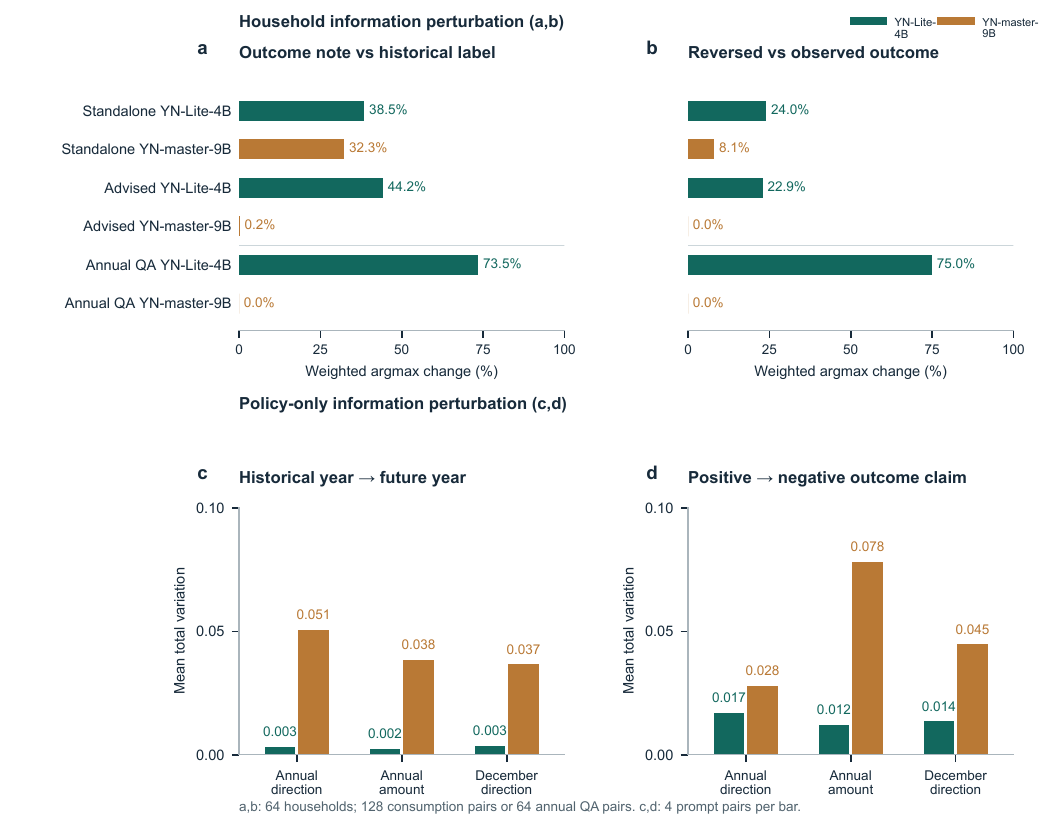}
\caption{\textbf{Decision sensitivity depends on both the model and the information task.} \textbf{a,b,} The outcome note supplied two adjusted weighted-regression slopes linking the observed pre-policy-to-outcome-year change in real $\log(1+C)$ to predicted fourth-quarter tax-income gains in the full 1,060-household sample: $-0.05375229$ for high-liquidity households and $-0.05404250$ for low-liquidity households, per 1,000 yuan of exposure. These are descriptive associations conditional on household covariates. Panel a adds this note to the historical-label input; panel b reverses both signs. Consumption comparisons contain 128 pairs (64 households in two income paths); direct annual-direction questions contain 64. Rates use household weights. \textbf{c,d,} Policy-only tests compare historical and future year labels, and positive and negative synthetic retrospective claims. Total variation is $\tfrac12\sum_k|p_k-q_k|$; each cell averages four wording/order pairs.}
\label{hh:fig:information}
\end{figure}

The primary annual forecasts used anonymous policy inputs without later outcome notes. We tested sensitivity to added information in two separate designs. A fixed 64-household panel held September states and menus constant while adding a historical policy label, two later consumption--exposure regression slopes from the full 1,060-household sample, or the same slopes with reversed signs (Figure~\ref{hh:fig:information}). It covered standalone/advised consumption choices and direct annual-consumption and December-cash questions. The 120 policy-only inputs omitted household profiles and varied an anonymous description, a historical year, a future-year relabeling, and synthetic positive or negative retrospective claims. Each model answered three question types under five information conditions, two equivalent wordings and two option orders.

\clearpage

Observed later statistics altered standalone consumption choices in 57 of 128 comparisons for YN-Lite-4B and 38 of 128 for YN-master-9B. With rational advice, the corresponding counts were 45 and 1. Reversing the statistics changed 32 standalone 4B choices and 7 standalone 9B choices. Direct annual-direction questions were especially responsive for 4B: the observed note changed 48 of 64 answers, and reversing it changed 50. The 9B annual-direction argmax remained unchanged in both comparisons, although its mean weighted probability shifts were 0.04624 and 0.05035 in total variation.

In policy-only questions, YN-Lite-4B selected ``cannot determine'' in all 60 inputs. YN-master-9B selected an increase in all 40 direction questions; for annual magnitude, it selected ``cannot determine'' in 19 of 20 cases and a change within $\pm0.1\%$ once. Relabeling the policy year from 2018 to 2031 left the selected choices unchanged in all 12 paired tests per model, while moving the 9B annual-direction distribution by mean TV 0.05075 and the December-direction distribution by 0.03659. Reversing the synthetic retrospective claim changed its annual-magnitude distribution by 0.07819 and flipped one of four choices. Equivalent wording and reversed option order each changed one of 30 paired 9B choices and no 4B choices. These interventions measure decision sensitivity to supplied dates and outcome claims; they do not measure what the training corpus contains.

\clearpage
\subsection*{7.5. Interpretation: An alternative policy module with distinct consumption--cash behavior}

These results support learned decision rules as another route to large heterogeneous-agent simulation. A shared trained policy maps household profiles and budgets to feasible actions, while the simulator updates financial states. The contrast between 9B's lower individual error and 4B's smaller income-tertile mean log errors makes both levels of evaluation relevant as the simulated population becomes more heterogeneous.

The standalone 9B underprediction arises from lower selected consumption and greater cash retention. At identical September states, it chose a lower consumption share than the deterministic rational benchmark for 65.8\% of household weight; mean selected shares were 25.8\% and 38.4\%. Its four-month consumption totaled 30,141 yuan versus 43,235 yuan, leaving 13,094 yuan more in December cash. Added to the common 49,605-yuan eight-month prefix, this produced the 79,746-yuan annual forecast, 19.7\% below the observed mean. Rational advice raised four-month consumption to 43,478 yuan and narrowed the annual gap.

Paired policy increments compare two independently simulated September--December paths for each household and decision rule. Both start from the same initial state: the baseline retains the original tax schedule, while the policy path includes the tax-cut income component announced in September and paid from October. Each path updates consumption and cash monthly. We subtract baseline from policy within each household, then average these paired differences with household weights.

\begin{figure}[H]
\centering
\includegraphics[width=\textwidth]{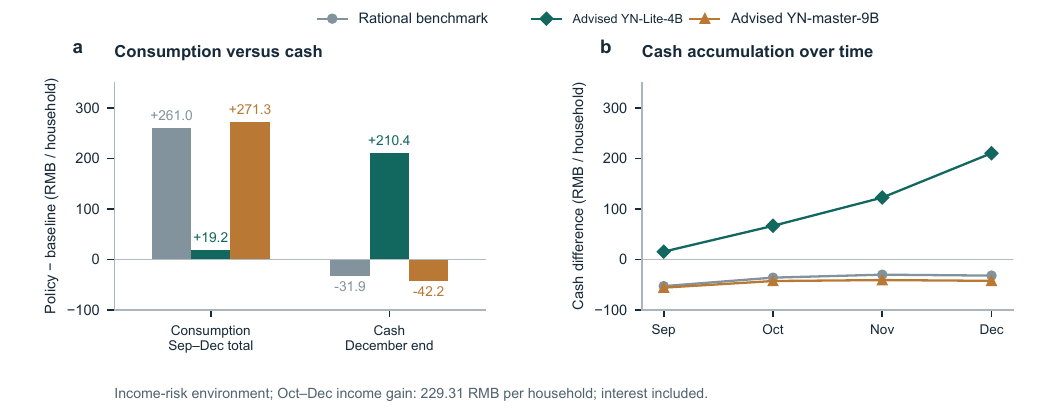}
\caption{\textbf{Conditional income gains are allocated differently between consumption and cash.} The income-risk specification compares the fully rational benchmark with advised 4B and 9B. All changes are policy minus baseline: consumption is summed over September--December, whereas cash is the December-end stock. The right panel tracks month-end cash differences. Amounts are household-weighted 2018 nominal yuan. Table~\ref{hh:tab:channels} gives all decision rules.}
\label{hh:fig:allocation}
\end{figure}

In R, the income component provided 229.31 yuan over October--December. Advised YN-Lite-4B allocated 19.23 yuan to additional consumption and 210.42 yuan to closing cash, retaining about 92\% of the component. Advised YN-master-9B increased consumption by 271.30 yuan and reduced cash by 42.22 yuan, close to the rational benchmark's allocation (Figure~\ref{hh:fig:allocation}). In D, standalone YN-Lite-4B increased consumption by 502.78 yuan, whereas standalone YN-master-9B reduced it by 904.38 yuan and accumulated 1,125.51 yuan in cash. Thus the 9B policy's favorable annual prediction score accompanied a negative simulated consumption response to the income component.

The common budget connects these responses through
\begin{equation}
 \Delta C_{9:12}+\Delta a_{13}=\sum_{t=9}^{12}\Delta g_t+\Delta I_{9:12}.
 \label{hh:eq:budget}
\end{equation}
The component totals were 221.12 yuan in D and 229.31 yuan in R in common price units.

\clearpage
\subsection*{7.6. Complete group, information and consumption--cash results}
\begingroup
\setlength{\intextsep}{5pt}
\captionsetup{font=footnotesize,skip=3pt}
\begin{table}[H]
\centering\footnotesize
\setlength{\tabcolsep}{3pt}
\begin{tabularx}{\textwidth}{@{}Xrrrrrr@{}}
\toprule
 & \multicolumn{3}{c}{Income tertile} & \multicolumn{3}{c}{Wealth tertile}\\
\cmidrule(lr){2-4}\cmidrule(l){5-7}
Decision rule & Low & Middle & High & Low & Middle & High\\
\midrule
Rational benchmark (D) & 28,023 & 40,908 & 90,874 & 34,949 & 60,060 & 64,540\\
Standalone YN-Lite-4B (D) & 29,435 & 45,778 & 87,109 & 35,864 & 62,477 & 63,795\\
Standalone YN-master-9B (D) & 29,564 & 41,443 & 83,282 & 34,856 & 59,530 & 59,704\\
Advised YN-Lite-4B (D) & 30,348 & 44,303 & 90,438 & 37,097 & 65,836 & 61,933\\
Advised YN-master-9B (D) & 27,807 & 41,050 & 90,873 & 34,765 & 60,093 & 64,619\\
\midrule
Rational benchmark (R) & 27,192 & 41,772 & 95,438 & 35,625 & 60,765 & 67,731\\
Advised YN-Lite-4B (R) & 27,377 & 41,993 & 94,880 & 35,701 & 60,677 & 67,595\\
Advised YN-master-9B (R) & 27,189 & 41,779 & 95,392 & 35,619 & 60,770 & 67,690\\
\midrule
Continuous rational reference (D) & 28,029 & 41,297 & 89,724 & 34,882 & 60,060 & 63,864\\
Continuous rational reference (R) & 27,315 & 41,795 & 95,103 & 35,639 & 60,722 & 67,573\\
\midrule
Standalone YN-mini (D) & 32,668 & 56,906 & 128,321 & 38,960 & 67,461 & 111,137\\
Advised YN-mini (D) & 30,327 & 51,342 & 120,821 & 36,892 & 64,061 & 101,197\\
\midrule
Pre-policy consumption carried forward & 31,040 & 43,808 & 82,744 & 35,499 & 61,191 & 60,728\\
\bottomrule
\end{tabularx}
\caption{\textbf{Annual MAE within income and wealth tertiles.} Household-weighted 2018 nominal yuan, rounded to the nearest yuan. Each cell is $\mathrm{MAE}_g=\sum_{i\in g}(w_i/W_g)|\widehat C_i-C_i|$. Income groups contain 337/338/385 households; wealth groups contain 351/308/401. Group weights reconstruct each rule's overall MAE.}
\label{hh:tab:group-mae}
\end{table}

\begin{table}[H]
\centering\footnotesize
\begin{tabularx}{\textwidth}{@{}lXrrr@{}}
\toprule
Model & Fixed-household task & Historical label & Outcome note & Reversed note\\
\midrule
4B & Standalone consumption & 42/128 & 57/128 & 32/128\\
4B & Advised consumption & 36/128 & 45/128 & 30/128\\
4B & Annual consumption direction & 14/64 & 48/64 & 50/64\\
4B & December cash direction & 21/64 & 20/64 & 37/64\\
\midrule
9B & Standalone consumption & 16/128 & 38/128 & 7/128\\
9B & Advised consumption & 1/128 & 1/128 & 0/128\\
9B & Annual consumption direction & 0/64 & 0/64 & 0/64\\
9B & December cash direction & 17/64 & 7/64 & 1/64\\
\midrule
421M & Standalone consumption & 20/128 & 8/128 & 4/128\\
421M & Advised consumption & 13/128 & 11/128 & 0/128\\
421M & Annual consumption direction & 0/64 & 0/64 & 0/64\\
421M & December cash direction & 0/64 & 0/64 & 0/64\\
\bottomrule
\end{tabularx}
\caption{\textbf{Changed choices in the fixed-household information test.} Entries are unweighted flips/pairs. Historical label is compared with the anonymous input; observed outcome information is compared with the historical label; the reversed note is compared with the observed note. Consumption tasks include both paths at the same initial household state. The 421M reference used the same 64-household design.}
\label{hh:tab:information}
\end{table}
\begin{table}[H]
\centering\footnotesize
\begin{tabularx}{\textwidth}{@{}Xrrr@{}}
\toprule
Decision rule & $\Delta C_{9:12}$ & $\Delta a_{13}$ & $\Delta I_{9:12}$\\
\midrule
Rational benchmark (D) & +249.58 & -28.46 & +0.00\\
Standalone YN-Lite-4B (D) & +502.78 & -281.66 & +0.00\\
Standalone YN-master-9B (D) & -904.38 & +1,125.51 & +0.00\\
Advised YN-Lite-4B (D) & -645.67 & +866.79 & +0.00\\
Advised YN-master-9B (D) & +193.22 & +27.90 & +0.00\\
\midrule
Rational benchmark (R) & +260.97 & -31.86 & -0.20\\
Advised YN-Lite-4B (R) & +19.23 & +210.42 & +0.34\\
Advised YN-master-9B (R) & +271.30 & -42.22 & -0.23\\
\midrule
Continuous rational reference (D) & +292.58 & -71.46 & +0.00\\
Continuous rational reference (R) & +308.42 & -79.50 & -0.38\\
\midrule
Standalone YN-mini (D) & -2,006.08 & +2,227.20 & +0.00\\
Advised YN-mini (D) & -8,962.40 & +9,183.53 & +0.00\\
\bottomrule
\end{tabularx}
\caption{\textbf{Paired consumption and cash changes.} Policy minus baseline, in household-weighted 2018 nominal yuan. Cash is the December-end stock; consumption and interest are four-month totals. D and R retain their original economic specifications and component paths.}
\label{hh:tab:channels}
\end{table}
\endgroup